\documentclass[
aps,
pra,
reprint,
]{revtex4-2} 
\usepackage{graphicx}
\usepackage[utf8]{inputenc}
\usepackage{hyperref}
\usepackage{amsmath}
\usepackage{amsthm}
\usepackage{amssymb}

\usepackage{algorithmic}
\usepackage{algorithm}
\usepackage{tikz}
\usetikzlibrary{quantikz2}
\usepackage{svg}
\usepackage{orcidlink}
\usepackage{comment}

\begin{document}
\title{Adaptive quantum metrology with large dynamic range using short one-axis twists}

\author{Tyler G. Thurtell\,\orcidlink{0000-0001-8731-6160}}\email{tthurtll@unm.edu}

\author{Akimasa Miyake}\email{amiyake@unm.edu}

\affiliation{
Department of Physics and Astronomy,
Center for Quantum Information and Control, 
Quantum New Mexico Institute,
University of New Mexico, Albuquerque, NM 87106, USA
}

\date{\today}

\begin{abstract}
Phase estimation with potentially large phase values, i.e., with large dynamic range, has many applications in quantum metrology, for example to atomic clocks. A recently proposed phase estimation scheme approaches the Heisenberg scaling in this global setting using sequences of increasingly squeezed Gaussian states as probes and adaptively chosen, potentially mid-circuit, measurements. In this work, we first observe that the pattern of increase in the squeezing of the probes is applicable even to states with some non-Gaussian features. We then propose an experimentally feasible version of this phase estimation scheme, based on the alternating application of one-axis twist (OAT) operations and rotations. Our protocols are explicitly described in terms of multiple OAT angles whose durations decrease polynomially with system size and spin-squeezing parameters that decay as  $N^{-\mu}$, with $\mu>2/3$ in most cases. Using numerical computation of the system-size dependence $N^{-\nu}$ of the Bayesian mean-squared error of an estimator, we show that these states are suitable for use in the phase estimation scheme, and highlight the protocols to achieve $\nu=17/9$ and $53/27$ using two and three OAT operations respectively in the last adaptation stage. We also analyze the limited non-Gaussianity of the resulting probe states and discuss the role of non-Gaussianity in this protocol more generally. Finally, we analyze how robust these protocols are with respect to imperfections such as particle number fluctuations and coherent control fluctuations.
\end{abstract}

\maketitle

\section{Introduction} 
Quantum metrology uses coherence and entanglement to improve sensing precision and accuracy~\cite{Giovannetti_2011,Ma_2011, Toth_2014, Ludlow_2015, Degen_2017, Pezze_2018}. The set of platforms that make excellent sensors for various physical quantities is quite diverse and includes, for example: trapped ions~\cite{Steinel_23,Pelzer_24,Hausser_25}, neutral atoms in tweezer arrays~\cite{Madjarov_19, Norcia_2019,Young_20,Shaw_24} or optical lattices~\cite{Katori_03,Takamoto_15,Al_Masoudi_15}, Bose-Einstein condensates~\cite{Wildermuth2006,Mao2023}, nitrogen-vacancy (NV) centers in diamond~\cite{Levine1973, Barry2020}, and optical interferometers~\cite{Caves1981,Tse2019}. 

In order to make the use of delicate quantum mechanical resources desirable, the estimation error exhibited by the quantum sensor may need to scale more favorably with the sum of time that all particles are exposed to the signal, which we denote $T$, than alternative methods. Typically, protocols that do not utilize quantum phenomena exhibit the error scaling one would expect from the central limit theorem,
i.e., $\sim T^{-1}$. On the other hand, when quantum resources are used, scaling of up to $\sim T^{-2}$ can be achieved. If each particle is exposed to the signal for the same amount of time, then the system size $N$ is proportional to $T$ and we focus on the $N$-dependence of the estimation error~\cite{Bollinger_1996,Giovannetti_2006}. A system size dependence of $\sim N^{-1}$ is referred to as standard quantum limit (SQL) scaling while a system size dependence of $\sim N^{-2}$ is referred to as Heisenberg limit (HL) scaling. 

The paradigmatic example in quantum metrology is quantum phase estimation~\cite{Toth_2014,Pezze_2018}. Here, the probes are pseudo-spins that are rotated about the \textit{z}-axis by an unknown angle, also referred to as an unknown phase, and the goal is to estimate this angle. Problems that phase estimation applies to include atomic clock stabilization~\cite{Levine_99,Udem_02,Kessler2014,Komar_2014,Ludlow_2015,Fraas_16,Colombo_22}, magnetometry~\cite{Degen_2017}, searches for new fundamental physics~\cite{Derevianko_2014,Safronova2018,Roussy_2023,Ye_2024}, the alignment of reference frames~\cite{Rudolph2003}, gravitational wave detection~\cite{Kolkowitz2016}, and geodesy~\cite{Mehlstäubler_2018,Grotti2018}. Phase estimation is also an essential piece of many quantum algorithms~\cite{Nielsen_00}.

One approach to entanglement enhanced phase estimation is spin squeezing~\cite{Wineland_1992, Kitagawa_1993, Wineland_1994}. Spin squeezed states have been produced, for example, via measurement back action~\cite{Kuzmich_1998, Kuzmich_2000, Smith_2006, Appel_2009, Shah_2010, Wasilewski_2010, Bohnet_2014, Hosten2016, Bao_2020}, one-axis twisting~\cite{Henkel2010,Leroux2010, Bohnet_2016, Braverman2019, Pogorelov2021,Gil2014,Colombo_2022,Li_2022}, twisting-and-turing type dynamics~\cite{Gross_2010, Riedel_2010} and two-axis counter-twisting~\cite{Luo2025}. 

Care is needed when applying these strategies if they must succeed over a sizable range of values of $\phi$~\cite{Andr__2004,Ma_2011, Pezze_2018}. A decrease in the variance of one angular momentum component, say of $J_{y}$, leads to an increase in the variance of the perpendicular angular momentum components. Since a large value of $\phi$ causes a measurement of $J_{y}$ to rotate into a measurement of an operator with some component of $J_{x}$, the associated measurement variance will be large if the state is highly squeezed due to the increased variance of $J_{x}$. This phenomena is deeply tied to the fact that the phase space of the collection of pseudo-spins is spherical. In the case of a bosonic mode, for which the phase space is a plane. If we denote the quadrature operators by $X$ and $P,$ a displacement according to $\exp(-i\phi P)$ sends $X'= X+\phi$. The variance of $X'$ is equal to the variance of $X$ so there is no consequence squeezing the observable $X$ as much as possible.   

\begin{figure*}
\includesvg[width=1.0\linewidth]{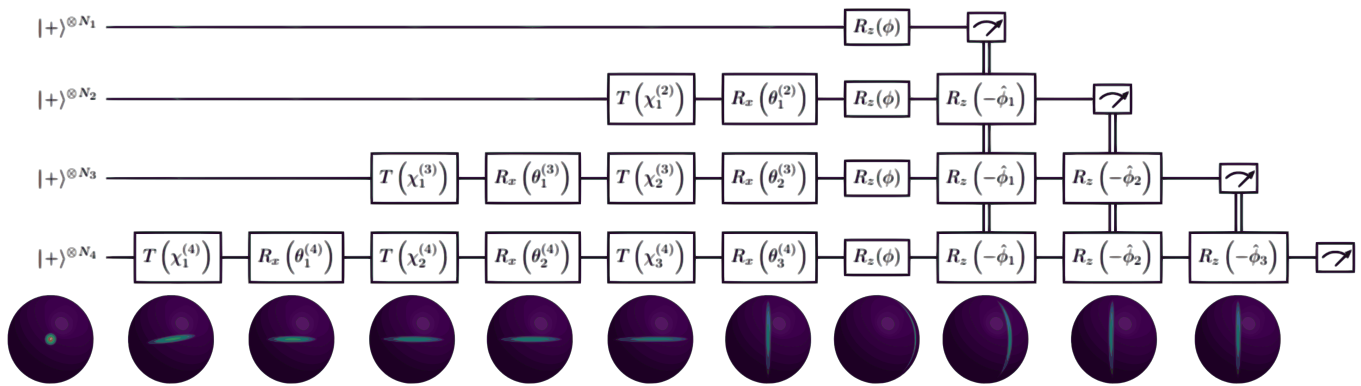}
\caption{In the adaptive protocols we consider, the spin-1/2 particles that make up the sensor are divided up into  separate ensembles. The ensembles then undergo separate squeezing protocols where different ensembles undergo a different number of one-axis twisting operations, denoted by $T(\chi_{j}^{(k)})$, interspersed with $x$-rotations, denoted by $R_{x}(\theta_{j}^{(k)})$. After the squeezing, the ensembles undergo a unknown rotation about the $z$-axis, denoted by $R_{z}(\phi)$, by the angle to be estimated. Finally, the ensembles are measured one at a time from smallest, and least squeezed, to largest with larger ensembles being counter rotated by the estimate of the residual phase made by the previous ensembles, i.e. by $\hat{\phi}_{j}$'s. The quasi-probability distributions are the Q-distributions (on a logarithmic scale) associated with the fourth, largest, ensemble at each step of the protocol.}
\label{fig:scheme}
\end{figure*}

One workaround is to perform a more complicated measurement~\cite{Kaubruegger2021,Marciniak2022, Kaubruegger2023,Thurtell2024,kielinski2025, Liu_2025}. In this work we focus on an alternative strategy which we refer to as the adaptive approach~\cite{rosenband2013,Borregaard2013,Pezze_2018,Shaw2024}. Here, the probes are divided into separable ensembles with differing sizes and differing amounts of squeezing. The ensembles are not entangled with each other and undergo the unknown rotation separately. Despite the lack of entanglement, the division into ensembles is useful because it allows for a broadening of the dynamic range. The less squeezed ensembles are measured first, giving preliminary estimates of $\phi$. These preliminary estimates are used to adapt the measurements on the more squeezed ensembles. Finally, all of the measurement outcomes are combined to produce a final estimate. An important algorithm of this type using squeezed states was proposed by Pezz\`e and Smerzi~\cite{Pezze2020,Pezze2021}, with related work done by Huang and Moore~\cite{Huang_2008}, based on a family of states they call Gaussian spin-squeezed (GSS) states. In this algorithm, each ensemble is prepared in a spin-squeezed state where the amount of squeezing has a different dependence on the system size. In this work, we describe simple and experimentally friendly state preparation protocols for states similar to GSS states with the amounts of squeezing required by this algorithm.

Often the unknown rotation is produced by a fluctuating signal, for example, during atomic clock stabilization~\cite{Ludlow_2015}. In this case, the ensembles must undergo the unknown rotation at the same time and the measurements used in the adaptive procedure become mid-circuit measurements in which part of the system must be projectively measured while the rest of the system remains in coherent superposition. Fortunately, this capability has recently been demonstrated experimentally, e.g. in ion traps~\cite{Schindler2011, Negnevitsky_2018,Ryan_Anderson2021,Egan2021} and neutral atom arrays~\cite{Deist2022,Singh_2023,Graham2023,Lis2023,Norcia2023,Huie2023}.

The manuscript is organized as follows. We discuss every step of a procedure to achieve improved $N$-dependence of the estimation error across a large range of values of $\phi$ using an adaptive procedure. First, we argue that the pattern of spin-squeezing required for the squeezing based phase estimation algorithm~\cite{Pezze2020,Pezze2021,Huang_2008} should also be utilized for a more general family of states. We then propose a procedure based on OAT operations that prepares states with the requisite amount of squeezing. Our procedure uses several one-axis twists per ensemble, each of which has a duration that decreases with system size. This builds on Carrasco et al.'s~\cite{Carrasco2022} result that just a few one-axis twists can produce extreme spin-squeezing. Then we show how to combine these state preparations in an adaptive phase estimation algorithm. A schematic representing the total procedure is shown in Fig.~\ref{fig:scheme} and the quantities needed to understand the protocol are summarized in Table~\ref{tab:phase_est}.

\section{Background}
\label{sec:back}

\begin{table*}
\begin{tabular}{ c c c c }
\hline\hline
Quantity & \quad\quad & Symbol & Scaling exponent \\
\hline
Prior standard deviation (dynamic range) & \quad\quad & $\sigma$ & \\
Total number of ensembles & \quad\quad & $M=j+1$ &  \\
Number of particles in ensemble $k$ & \quad\quad & $N_{k}$ & \\
Measurement outcome from ensemble $k$ & \quad\quad & $m_{k}$ & \\
Residual phase estimate associated with ensemble $k$ & \quad\quad & $\hat{\phi}_{k}=\frac{2m_{k}}{N_{k}}$ & \\
Estimation error & \quad\quad & $\Delta\phi^{2}$ & $\nu=2-\frac{1}{3^{j}}$ \\
Squeezing of $k$th ensemble & \quad\quad & $\xi^{2}_{k}$ & $\mu=1-\frac{1}{3^{k-1}}$ \\
$m$th twisting angle & \quad\quad & $\chi_{m}^{(k)}$ & $\gamma_{m}=\frac{2}{3^{m}}$ \\
$n$th rotation angle ($n<k-1$) & \quad\quad & $\theta_{n}^{(k)}$ & $\alpha_{n}=1-\frac{2}{3^{n}}$ \\
$k$th shifted final rotation for ensemble $k$ & \quad\quad & $\eta_{k}=\frac{\pi}{2}-|\theta_{k-1}^{(k)}|$ & $\beta_{k}=1-\frac{2}{3^{k-1}}$ \\
\hline\hline
\end{tabular}
\caption{For easy reference, we collect all of the quantities relevant to understand the phase estimation algorithm studied in Sec.~\ref{sec:phase_est}. The primary goal is to obtain a small value of the estimation error $\Delta\phi^{2}$ even when the prior standard deviation $\sigma$ is not negligible. To achieve this, the spins are divided up into $M=j+1$ ensembles with the $k$th ensemble prepared with a squeezing parameter $\xi^{2}_{k}$. This squeezing is achieved via a sequence of OAT operations with angles $\chi_{m}^{(k)}$ and global rotations with angles $\theta_{n}^{(k)}$. In the description of angles, the superscripts indicate the ensemble undergoing the operation and the subscript indicates which operation of that type on the specified ensemble we are considering with lower numbered operations occurring first. The final rotation for each ensemble as a slightly different form in order to align the squeezing along the correct direction. The system size dependence of many of these quantities is well described by a scaling exponent as described in Sec.~\ref{sec:back}. In all cases except for $\Delta\phi^{2}$, the exponent is associated with the number of spins in the specified ensemble $N_{k}$. For $\Delta\phi^{2}$, it is associated with the total number of particles across all ensembles $\sum_{k}N_{k}$. Recall that the $\ell$th twisting or rotation angle is specified by the number of particles in the ensemble but the functional form does not dependent on which ensemble we are considering. This is the reason that $\gamma_{m}$ and $\alpha_{n}$ do not depend on $k$.}
\label{tab:phase_est}
\end{table*}

We consider ensembles of many spin-1/2 particles, or qubits, so that the total state space is $\left(\mathbb{C}^{2}\right)^{\otimes N}$. We denote Pauli operators on the $j$th subsystem by $\sigma_{\mu}^{(j)}$ where $\mu=x,y,z$. Symmetric, or collective, spin operators are then defined by
$J_{\mu}=\frac{1}{2}\sum_{j=1}^{N}\sigma_{\mu}^{(j)}$. The total angular momentum squared is then $J^{2}=J_{x}^{2}+J_{y}^{2}+J_{z}^{2}$. We will refer to unitaries of the form 
\begin{equation}
R_{\mu}(\theta)\equiv e^{-i\theta J_{\mu}}
\end{equation}
as global rotations. The Hilbert space of $N$ spin-1/2 particles has dimension $2^{N}$ but the space of states that are invariant under permutations of the particles is much smaller with dimension $N+1$. This space, called the symmetric subspace, is exactly the space of $J^{2}$ eigenstates with eigenvalue $\frac{N}{2}\left(\frac{N}{2}+1\right)$. Fixing an eigenvalue of any one of the $J_{\mu}$ then completely specifies a state in the symmetric subspace. We will denote these states by $|J_{\mu}=m\rangle$ and refer to them as Dicke states. This reduction in the dimension of the accessible state space makes it possible to simulate much larger system sizes than would otherwise be feasible.

In phase estimation, these spins undergo a rotation about the $z$-axis by an unknown phase $\phi$. This imprints a phase dependence on the state of the probes
\begin{equation}
\rho_{\phi}=R_{z}(\phi)\rho_{0}R_{z}(\phi)^{\dag}.
\end{equation}
Note that the permutation invariance of $J_{z}$ implies that if $\rho_{0}$ is in the symmetric subspace then $\rho_{\phi}$ will also be in the symmetric subspace.

The goal of global phase estimation is to produce an estimate of the unknown phase $\phi$. We will evaluate protocols for estimating $\phi$ via the averaged estimation error, which is also referred to as the Bayesian mean squared error. Let $\Pi_{\hat{\phi}}$ be the measurement operator associated with making the estimate $\hat{\phi}$. Then the averaged estimation error is defined as
\begin{equation}
\Delta\phi^{2}\equiv\int d\phi\sum_{\hat{\phi}}\left(\phi-\hat{\phi}\right)^{2}\textrm{tr}(\rho_{\phi}\Pi_{\hat{\phi}})p(\phi),
\end{equation}
where $p(\phi)$ is a prior distribution. Throughout this paper we will denote the variance of this prior distribution by $\sigma$.  As alluded to above, we will often be interested in how $\Delta\phi^{2}$ depends on the system size $N$. It is particularly important to understand the form this dependence takes when $N$ is asymptotically large. Unentangled probe states exhibit standard quantum limit scaling $\Delta\phi^2\sim1/N$ at best. On the other hand, if arbitrary probe states are considered, then Heisenberg scaling $\Delta\phi^2\sim1/N^{2}$ may be achievable. Since we are interested in intermediate scalings in this work, we define the scaling exponent $\nu$ to satisfy  $\Delta\phi^2\sim1/N^{\nu}$.

Recall that we take $\phi$ to be real valued. Many authors restrict $\phi$ to the range $-\pi$ to $\pi$. We choose not to do this here for two reasons: (1) the source of the signal, e.g. an external magnetic field, will not generically be defined modulo $2\pi$ and (2) for the prior distributions we will be most interested the probability of $|\phi|>\pi$ will be negligible. If $J_{y}$ is measured and $\phi$ is small, the error of the resulting estimate is approximately 
\begin{equation}
\label{eq:xi_err}
\Delta\phi^{2}\approx\frac{\xi^{2}}{N},
\end{equation}
where 
\begin{equation}
\xi^{2}=\frac{N\left(\Delta J_{y}^{2}\right)_{\rho_{0}}}{\langle J_{x}\rangle^{2}_{\rho_{0}}},
\end{equation}
is the Wineland squeezing parameter~\cite{Wineland_1992}. The expectation values in this formula are evaluated with respect to the state immediately before the $R_{z}(\phi)$ rotation, which we denote by $\rho_{0}$. States with $\xi^{2}<1$ are referred to as spin-squeezed. Provided $\phi$ is small, probe states that exhibit $\xi^{2}<1/N^{\mu}$ achieve $\nu\approx\mu+1$. Throughout this paper, $\mu$ will be reserved for the exponent that dictates the asymptotic system size dependence of $\xi^{2}$. It turns out that all spin-squeezed states are entangled, i.e. $\xi^{2}$ is an entanglement witness. 

One-axis twisting (OAT)~\cite{Kitagawa_1993} dynamics, defined by
\begin{equation}
T(\chi)\equiv e^{-i\chi J_{z}^{2}},
\end{equation}
produces spin-squeezing. The OAT dynamics can be understood as an all-to-all Ising type interaction between the underlying pseudo-spins or as \textit{z}-dependent sheering action on the phase space where points undergo a rotation about the \textit{z}-axis by an amount that depends on their position relative to the equator. In particular, the state
\begin{equation}
|\psi_{0}\rangle=R_{x}(\theta)T(\chi)\big|J_{x}=\frac{N}{2}\big\rangle
\end{equation}
can exhibit $\xi^{2}\sim1/N^{2/3}$~\cite{Kitagawa_1993}, when
\begin{align}
\chi&\sim\frac{1}{N^{2/3}}, \\
\frac{\pi}{2}-\theta&\sim\frac{1}{N^{1/3}}.
\end{align}
Importantly, while these states can achieve values of $\mu>0$, they fall short of achieving Heisenberg scaling with $\mu=1$.

Numerical studies~\cite{Carrasco2022} have shown that states of the form 
\begin{equation*}
|\psi_{0}\rangle=R_{x}(\theta_{2})T(\chi_{2})R_{x}(\theta_{1})T(\chi_{1})\big|J_{x}=\frac{N}{2}\big\rangle
\end{equation*}
can overcome this $\mu=2/3$ barrier even the number of twists is kept quite small. Inspired by this development, we consider squeezing protocols of the form
\begin{equation}
|\psi_{0}\rangle=\prod_{k=1}^{L}\left[R_{x}(\theta_{k})T(\chi_{k})\right]\big|J_{x}=\frac{N}{2}\big\rangle,
\end{equation}
where, as for all operator products in this paper, operators labeled by smaller values of the dummy index $k$ values act first. Here $L$ denotes the depth of the state preparation procedure.

Naturally, we will be interested in the form of the circuit parameters $\{\chi_{k},\theta_{k}\}$ which give rise to a targeted amount of spin-squeezing. It turns out that the $N$-dependence of $\chi_{k}$ and $\theta_{k\neq L}$ is captured by polynomially decaying functions to which we can associate a scaling exponent in the same way that we did for $\Delta\phi^{2}$ and $\xi^{2}$. The exponent for $\chi_{k}$ will be denoted $\gamma_{k}$ while the exponent for $\theta_{k}$ we will denote by $\alpha_{k}$. For example, we have
\begin{equation}
|\chi_{k}|\sim\frac{1}{N^{\gamma_{k}}}.
\end{equation}
It is also useful to define $\eta\equiv\frac{\pi}{2}-|\theta_{L}|$ which can also be associated with a scaling exponent that will be denoted by $\beta$.

When the value of $\phi$ is no longer guaranteed to be small, a more appropriate approximation to the estimation error can be obtained by replacing the spin-squeezing parameter in Eq.~\eqref{eq:xi_err} by its post-rotation value. The states we consider in this work, this rotated squeezing parameter is given by
\begin{equation}
\Xi^{2}(\phi)\approx\frac{N}{\langle J_{x}\rangle_{\rho_{0}}^{2}}\left[\left(\Delta J_{y}^{2}\right)_{\rho_{0}}+\phi^{2}\left(\Delta J_{x}^{2}\right)_{\rho_{0}}\right].
\end{equation}
to the lowest order in $\phi$. Since we are usually interested in the average performance of protocols over the prior distribution $p(\phi)$, it makes sense to average this quantity over the prior as well to obtain
\begin{equation}
\label{eq:rot_avg}
\overline{\Xi^{2}}\approx\frac{N}{\langle J_{x}\rangle_{\rho_{0}}^{2}}\left[\left(\Delta J_{y}^{2}\right)_{\rho_{0}}+\sigma^{2}\left(\Delta J_{x}^{2}\right)_{\rho_{0}}\right].
\end{equation}
The second term is larger for more squeezed states; limiting how small the error can become by squeezing. Squeezing based phase estimation algorithms provide a workaround by using $M$ ensembles with variable squeezing. Less squeezed ensembles give rough estimates $\hat{\phi}_{j}$ and more squeezed ensembles are rotated by $\exp\left(i\hat{\phi}_{j}J_{z}\right)$ so that they effectively undergo a smaller rotation. In other words, the
angle that this ensemble is rotated by is described by a distribution with an effectively smaller standard deviation $\sigma$. Previous works~\cite{Pezze2020,Pezze2021,Huang_2008} studied such a protocol based on the Gaussian spin-squeezed (GSS) states
\begin{equation}
|\textrm{GSS}(s)\rangle_{y}=\frac{1}{\sqrt{\mathcal{N}}}\sum_{m=-N/2}^{N/2}e^{-m^{2}/(s^{2}N)}|J_{y}=m\rangle.
\end{equation}
These states have the convienent property that $s^{2}\approx\xi^{2}$. In this protocol, the $j$th ensmeble is prepared in a GSS state with
\begin{equation}
\label{eq:s_gss}
s^{2}\sim\frac{1}{N^{1-\left(\frac{1}{3}\right)^{j-1}}}.
\end{equation}
This works because the $j$th ensemble effectively undergoes a rotation by an angle described by a distribution that satisfies
\begin{equation}
\label{eq:sig_gss}
\sigma^{2}\sim\frac{1}{N^{2-\left(\frac{1}{3}\right)^{j-2}}}.
\end{equation}
This protocol achieves a system size dependence of the estimation error characterized by $\nu=2-1/3^{M-1}$ so Heisenberg scaling is approached as the number $M$ of ensembles tends to infinity. These $N$-dependences were derived by first computing spin-moments of the GSS states and then minimizing Eq.~\eqref{eq:rot_avg} over $s$ and $N$. In the next section, we will see the same pattern emerge via a different type of argument. Pezz\`e and Smerzi do not explicitly discuss how to prepare these states but suggest that there are likely many ways doing so thanks to their Gaussian nature. We discuss in detail a procedure that yield states similar to these.

Note that Pezz\`e and Smerzi~\cite{Pezze2021} also considered a protocol using one-axis twists in a twist-untwist protocol~\cite{Leibfried2004,Leibfried2005,Davis2016,Frowis2016,Marci2016,Nolan2017,Colombo_2022,Scharnagl_2023,Li_2023}. The physics of such protocols differs significantly from that of the protocols we consider. In these protocols, the variance of the final measured observable is not decreased, rather the rate of change of this observable with respect to $\phi$ is increased. This is similar to GHZ based schemes for which ensemble based protocols have also been studied.

\section{Optimal squeezing beyond GSS states}
\label{sec:opt}

In this section, we argue that the sequence of squeezing parameters described by Eq.~\eqref{eq:s_gss} holds for a broader class of states than GSS states. First, we consider an analogous semi-classical estimation problem and see the emergence of the same pattern. Then we will show that the classical reasoning can be extended to a class of quantum states which we refer to as weakly non-Gaussian states.

The reasoning pursued in this section is of a rather different nature than the reasoning that led to the original proposals of this sequence of squeezing parameter $N$-dependences. Previous works~\cite{Pezze2020,Pezze2021} computed the spin-moment expectation values for a particular family of states and minimized Eq.~\eqref{eq:rot_avg} over the states in this family. In this section on the other hand, the essential inputs in deriving this pattern are (1) the geometric relationship between the spin-moments and (2) the uncertainty principle
\begin{equation}
\label{eq:heis_uncer}
    \Delta J_{y}\Delta J_{z}\geq\frac{1}{2}|\langle J_{x}\rangle|.
\end{equation}
This serves to illustrate the generality of this pattern. However, a spin-moment analysis additionally gives information about the constant factors that appear in the squeezing parameters and estimation errors. These constants are not easily extractible with the approach in this section. Additionally, the spin-moment analysis is more easily related to the state preparation circuit parameters that define the family of states that we will be most interested in. For these reasons, in the next Sec.~\ref{sec:oat_gss}, we return to a spin-moment based analysis, albeit one where we have to make some reasonable but uncontrolled approximations, but due to the analysis in this section we will apriori expect to see this pattern fallout of our analysis.

\subsection{Analogous semi-classical problem}
Suppose that a sphere of radius $r$ undergoes a rotation about the \textit{z}-axis by an unknown angle and our goal is to estimate the rotation angle. We are allowed do this this by observing the \textit{y}-value of a marked point on the sphere after the rotation. The catch is that we are not allowed to know the initial location of the marked point. Instead, the point is chosen at random from a probability distribution.
These distributions are required to satisfy a constraint analogous to Eq.~(\ref{eq:heis_uncer}) 
\begin{equation}
\label{eq:semi_uncer}
\delta_{y}\delta_{z}\geq\frac{1}{2}r,
\end{equation}
where $\delta_{\mu}$ is the standard deviation of the marginal distribution along the $\mu$-axis. To see why this is analogous to the uncertainty principle, note that the radius of the sphere that constitutes the SU(2) phase space is given by $\sqrt{\langle J^{2}\rangle}=\sqrt{\frac{N}{2}\left(\frac{N}{2}+1\right)}\approx\frac{N}{2}$ which also coincides with the maximal value of $|\langle J_{x}\rangle|$. 

Assuming the rotation angle is small enough that we can work to only lowest order in $\phi$, the rotation amounts to a map from the random variables $\{x,y,z\}$ to $\{x_{f},y_{f},z\}$ given by
\begin{align}
x_{f}&\approx x-\phi y \\
y_{f}&\approx y+\phi x.
\end{align}
Then the variance of the distribution for $y_{f}$ is 
\begin{equation}
\overline{\delta_{y_{f}}^{2}}\approx\delta_{y}^{2}+\phi^{2}\delta_{x}^{2}+2\phi\textrm{Cov}(x,y),
\end{equation}
where for classical random variables the covariance is defined by
\begin{equation}
\textrm{Cov}(x,y)=\mathbb{E}[xy]-\mathbb{E}[x]\mathbb{E}[y].
\end{equation}
Averaging over $\phi$ for a prior with mean zero and standard deviation $\sigma$ gives
\begin{equation}
\delta_{y_{f}}^{2}\approx\delta_{y}^{2}+\sigma^{2}\delta_{x}^{2}.
\end{equation}
Since we are considering points on a sphere the standard deviations $\{\delta_{x},\delta_{y},\delta_{x}\}$ are not all independent. They are linked via the relation $x=\sqrt{r^{2}-y^{2}-z^{2}}$. We will assume, as is natural given the effect of the rotation on $y$ and the relationship between $\phi$, $y$, and the radius of the sphere $r$, that the distribution is somewhat localized in both the $y$ and $z$ directions but that it is more localized in the $y$ direction, \textit{i.e.} 
\begin{equation}
\label{eq:taylor_assumps}
\mathbb{E}\left[\left(\frac{y}{r}\right)^{2}\right]\ll \mathbb{E}\left[\left(\frac{z}{r}\right)^{2}\right]\ll1,
\end{equation}
we can approximate the mean of $x$ by
\begin{equation}
\begin{split}
\mathbb{E}[x]\approx r&-\frac{\mathbb{E}[y^{2}]+\mathbb{E}[z^{2}]}{2r} \\
&-\frac{\mathbb{E}[y^{4}]+\mathbb{E}[z^{4}]+2\mathbb{E}[y^{2}z^{2}]}{8r^{3}}.
\end{split}
\end{equation}
We also have the simple relation $\mathbb{E}[x^{2}]=r^{2}-\mathbb{E}[y^{2}]-\mathbb{E}[z^{2}]$. 
From this expression we can see the importance of fourth moments in this setting. Then the variance of the post-rotation \textit{y} variable is
\begin{equation}
\begin{split}
\overline{\delta_{y_{f}}^{2}}\approx\delta_{y}^{2}+\frac{\sigma^{2}}{4r^{2}}(\mathbb{E}[y^{4}]-\mathbb{E}[y^{2}]^{2}&+\mathbb{E}[z^{4}]-\mathbb{E}[z^{2}]^{2} \\
&+2\textrm{Cov}(y^{2},z^{2}))
\end{split}
\end{equation}
We can now optimize this expression with respect to the expectation values that appear. It is technically possible for the expression in parentheses to vanish exactly. 
Situations like this are interesting and might suggest useful probe states but they won't be our main focus here. Instead, we consider independent $y$ and $z$ distributions with continuous support. We will also assume that the probability distributions are even in $y$ and $z$, \textit{i.e.} $\mathbb{E}[y]=\mathbb{E}[z]=0$. Then the covariance term here vanishes. It is useful to write the averaged variance of $y_{f}$ in terms of the kurtosis which for classical random variables is defined as
\begin{equation}
\textrm{Kurt}[R]\equiv\mathbb{E}\left[\left(\frac{R-\mathbb{E}[R]}{\delta_{R}}\right)^{4}\right].
\end{equation}
For a random variable with a Gaussian outcome distribution, the kurtosis is always given by exactly 3. The kurtosis minus 3 is called the excess kurtosis. A positive excess kurtosis is indicative of a distribution with heavier tails than a Gaussian distribution with the same mean and variance would have while a negative excess kurtosis similarly indicates a distribution with lighter tails. The variance of the post-rotation $y$ variable is then
\begin{equation}
\begin{split}
\overline{\delta_{y_{f}}^{2}}\approx\mathbb{E}[y^{2}]+\frac{\sigma^{2}}{4r^{2}}(&\mathbb{E}[y^{2}]^{2}(\textrm{Kurt}[y]-1) \\
&+\mathbb{E}[z^{2}]^{2}(\textrm{Kurt}[z]-1)).
\end{split}
\end{equation}
Now suppose that we consider only probability distributions with kurtoses that are bounded from below by $W_{1,y/z}$ and from above by $W_{2,y/z}$. This results in upper and lower bounds on $\overline{\delta_{y_{f}}^{2}}$ that have essentially the same form, namely
\begin{equation}
\begin{split}
\overline{\delta_{y_{f}}^{2}}&\sim\mathbb{E}[y^{2}]+\frac{\sigma^{2}}{4r^{2}}\left(W_{y}\mathbb{E}[y^{2}]^{2}+W_{z}\mathbb{E}[z^{2}]^{2}\right) \\
&\approx\mathbb{E}[y^{2}]+\frac{\sigma^{2}}{4r^{2}}\left(W_{z}\mathbb{E}[z^{2}]^{2}\right),
\end{split}
\end{equation}
for some non-zero constants $W_{y}$ and $W_{z}$. Accordingly, we will focus on minimizing this expression. Assuming the inequality in Eq.~\eqref{eq:semi_uncer} is saturated gives
\begin{equation}
\begin{split}
    \delta_{y_{f}}^{2}
    &\sim\mathbb{E}[y^{2}]+\frac{\sigma^{2}}{64r^{2}}\left(\frac{r^{4}W_{z}}{\mathbb{E}[y^{2}]^{2}}\right).
\end{split}
\end{equation}
Minimizing this with respect to $\mathbb{E}[y^{2}]$ gives
\begin{equation}
\mathbb{E}[y^{2}]=\left(\frac{\sigma^{2}r^{2}W_{z}}{32}\right)^{1/3}.
\end{equation}
Error propagation then tells us that the variance in the our estimate of $\phi$ will go as
\begin{equation}
\delta_{\phi}^{2}\sim\left(\frac{\sigma^{2}W_{z}}{32r^{4}}\right)^{1/3}.
\end{equation}
From here we can recover the pattern utilized by the GSS state protocol. Suppose that $\sigma\sim\frac{1}{\sqrt{r}}$ (this arises naturally in the quantum case if an unentangled state is used to obtain a preliminary estimate of the phase value), then the optimal $y$ variance satisfies $\mathbb{E}[y^{2}]\sim r^{1/3}$ leading to $\delta_{\phi}^{2}\sim1/r^{5/3}$. If prior to observing the mark on the sphere we are allowed to rotate the sphere ourselves, then $\delta_{\phi}^{2}$ can play the role of the residual phase uncertainty, \textit{i.e.} of $\sigma$, for the next estimation round. Then for the next round the optimal choice is $\mathbb{E}[y^{2}]\sim r^{1/9}$ leading to $\delta_{\phi}^{2}\sim1/r^{17/9}$. This is the pattern that appears in the GSS state protocol with $r\delta_{\phi}^{2}$ playing the role of $s^{2}=\xi^{2}$. In other words, the $r$-dependence exhibited by $r\delta_{\phi}^{2}$ matches the $N$-dependence in Eq.~\eqref{eq:s_gss}.

\subsection{Weakly non-Gaussian states}

Now we will briefly describe a version of this argument for a family of quantum states that we refer to as weakly non-Gaussian states. Since the argument is so close to the semi-classical one we will only sketch it and comment on the few points where it differs from the above argument. We refer the reader to Appendix~\ref{app:opt} for details. In particular, we make three key assumptions about the states under consideration. First, we consider states that are symmetrically oriented in the $x$-direction
\begin{align}
\langle J_{y}\rangle=\langle &J_{z}\rangle=0 \\
\langle J_{x}\rangle&\approx\frac{N}{2} \\
\textrm{Cov}(J_{y},&J_{z})=0,
\end{align}
where the covariance is defined analogously to the classical case. Second, we assume that the states are not maximally delocalized in either the $J_{y}$ or $J_{z}$ bases but that the state is much more concentrated in $J_{y}$ basis
\begin{equation}
\bigg\langle\frac{J_{y}^{2}}{N^{2}}\bigg\rangle\ll \left\langle\frac{J_{z}^{2}}{N^{2}}\right\rangle\ll1 .
\end{equation}
In this context, $A\ll B$ than should be understood to mean that $A$ decreases more quickly with $N$ than $B$ does. Our final condition is a bound on how much the fourth moments of $J_{y}$ and $J_{z}$ can deviate from the values they would take if the outcome distributions were exactly Gaussian. We consider families of probe states such that $\textrm{Kurt}[J_{y}]$ and $\textrm{Kurt}[J_{z}]$ are constants, i.e. numbers independent of $N$, as in the semi-classical case. The kurtoses are defined analogously to the classical case. The argument is now almost identical to the semi-classical one except that we can no longer assume $\textrm{Cov}(J_{y}^{2},J_{z}^{2})=0$. However, it is still the case that
\begin{equation}
\begin{split}
\textrm{Cov}(J_{y}^{2},J_{z}^{2})\leq&\Delta J_{z}^{2}\Delta J_{y}^{2} \\
&\times\sqrt{(\textrm{Kurt}[J_{y}]-1)(\textrm{Kurt}[J_{z}]-1)}.
\end{split}
\end{equation}
Since we are assuming that $\Delta J_{y}^{2}\ll\Delta J_{z}^{2}$ and that the kurtoses essentially behave as constants, we see that the covariance will be asymptotically negligible compared to a term that goes as $\Delta J_{y}^{4}$. Then we proceed to minimize Eq.~\eqref{eq:rot_avg} with respect to $\Delta J_{y}^{2}$. The minimizing value can then be related to a value of the squeezing parameter $\xi^{2}$. The minimizing value of the squeezing parameter in this case becomes
\begin{equation}
\xi^{2}\sim\left(\frac{4\sigma^{2}W_{z}}{N^{4}}\right)^{1/3},
\end{equation}
where $W_{z}$ is a constant that plays a role analogous to the one it played in the semi-classical problem. This squeezing parameter displays the same relationship between the $N$-dependence of $\sigma^{2}$ and of the $N$-dependence of $\xi^{2}$ as indicated by Eqs.~\eqref{eq:s_gss} and\eqref{eq:sig_gss}. Interestingly, this calculation also suggests that non-Gaussianity can be a resource in this context. In particular, if $\textrm{Kurt}[J_{z}]-1\leq W_{2,z}<3$ then the optimal value of $\xi^{2}$ will be smaller than it would if all distributions were Gaussian but the first and second moments were unchanged. This in turn should result in a smaller value of $\Delta\phi^{2}$. 

\section{Twisted Gaussian states}
\label{sec:oat_gss}

We now consider how to prepare states with the desired amount of squeezing using sequences of OAT operations and global rotations. We begin by studying the effect of OAT operations on GSS states.  

Suppose we begin with the state $|\textrm{GSS}(s)\rangle_{z}$ and apply an OAT operation with twisting angle $\chi$. The result is the state
\begin{equation*}
T(\chi)|\textrm{GSS}(s)\rangle_{z}=\frac{1}{\sqrt{\mathcal{N}}}\sum_{m=-N/2}^{N/2}e^{-m^{2}/s^{2}N-i\chi m^{2}}|J_{z}=m\rangle.
\end{equation*}
We show in Appendix~\ref{app:tgs} that the expectation value of $J_{x}$ in this state is approximated as
\begin{equation}
\langle J_{x}\rangle\approx\frac{N}{2}e^{-\frac{1}{2s^{2}N}-\frac{1}{2}\chi^{2}s^{2}N}
\end{equation}
and the maximal/minimal variances for an observable of the form
\begin{equation}
J(\theta)=\sin\theta J_{z}-\cos\theta J_{y}
\end{equation}
are given by 
\begin{equation}
V_{\pm}\approx\frac{N}{8}\left[2s^{2}+A(\chi)\pm\sqrt{A(\chi)^{2}+B(\chi)^{2}}\right],
\end{equation}
where
\begin{align}
A(\chi)&=\frac{N}{2}\left(1-e^{-\frac{2}{s^{2}N}-2\chi^{2}s^{2}N}\right)-s^{2}, \\
B(\chi)&=2\chi s^{2}Ne^{-\frac{1}{2s^{2}N}-\frac{1}{2}\chi^{2}s^{2}N}.
\end{align}
The value of $\theta$ that specifies the direction along which $V_{-}$ is achieved is
\begin{equation}
\theta_{*}\approx\frac{1}{2}\textrm{arctan}\left(\frac{B(\chi)}{A(\chi)}\right).
\end{equation}
If we assume that 
\begin{equation}
\label{eq:assump}
\frac{1}{N}<\frac{1}{s^{2}N}\ll\chi^{2}s^{2}N\ll1
\end{equation}
when $N$ is large, then the expression for the minimal variance simplifies to
\begin{equation}
\begin{split}
\label{eq:v_min}
V_{-}&\approx \frac{N}{4}\left[s^{2}-\frac{\chi^{2}s^{4}N}{\sinh\left(\frac{1}{s^{2}N}+\chi^{2}s^{2}N\right)}\right] \\
&\approx\frac{s^{2}N}{4}\left[\frac{1}{\chi^{2}s^{4}N^{2}}+\frac{1}{6}\chi^{4}s^{4}N^{2}\right].
\end{split}
\end{equation}
In Appendix~\ref{app:tgs}, we argue that this is the regime of greatest interest for spin-squeezing. The optimal choice of the twisting angle $\chi_{*}$ is then
\begin{equation}
\chi_{*}\approx\frac{3^{1/6}}{(s^{2}N)^{2/3}}
\end{equation}
which gives an associated minimal variance $(V_{-})_{*}$ of
\begin{equation}
(V_{-})_{*}=\frac{(9s^{2}N)^{1/3}}{8}.
\end{equation}

\subsection{Unrotated squeezing parameter}
\label{sec:squee}

The squeezing parameter is then approximately
\begin{equation}
\xi^{2}\approx\frac{(9s^{2})^{1/3}}{2N^{2/3}}.
\end{equation}
Recall that the for the initial GSS state the value of the squeezing parameter was roughly $s^{2}$. So if $N$-dependence the initial squeezing parameter was
\begin{equation}
s^{2}\sim\frac{1}{N^{1-\left(\frac{1}{3}\right)^{j-1}}}
\end{equation}
then the $N$-dependence of the post-twist squeezed state is
\begin{equation}
\xi^{2}\sim\frac{1}{N^{1-\left(\frac{1}{3}\right)^{j}}}.
\end{equation}
So a single OAT operation takes a state with squeezing appropriate for the $j$th ensemble to a state with squeezing appropriate for the $(j+1)$th ensemble.

We emphasize that these expressions are all approximate generalizations of the original expressions obtained by Kitagawa and Ueda~\cite{Kitagawa_1993} for the spin moments obtained by applying an OAT operation to $|J_{x}=N/2\rangle$. The origin of this connection is the simple approximation of the state with all spin polarized along the \textit{z}-axis as
\begin{equation}
\big|J_{z}=\frac{N}{2}\big\rangle\approx\frac{1}{\sqrt{\mathcal{N}}}\sum_{m=-N/2}^{N/2}e^{-m^{2}/N}|J_{z}=m\rangle,
\end{equation}
which can be simply obtained from the central limit theorem. Now notice that the right hand side is the state $|\textrm{GSS}(1)\rangle_{z}$. 

\subsection{Rotated squeezing parameter}

We now show the behavior observed in the previous section applies to the rotated squeezing parameters as well. For this we additionally need an approximation of the variance of $J_{x}$ which is also obtained in Appendix~\ref{app:tgs}. There we show that 
\begin{equation}
\Delta J_{x}^{2}\approx\frac{N^{2}}{8}\left(\frac{1}{s^{2}N}+\chi^{2}s^{2}N\right)^{2}.
\end{equation}
Under the same assumptions as in the previous section, we then find that the rotated squeezing parameter is given by
\begin{equation}
\begin{split}
\overline{\Xi^{2}}\approx\frac{1}{\chi^{2}s^{2}N^{2}}&+\frac{1}{6}\chi^{4}s^{6}N^{2} \\
&+\sigma^{2}\frac{\chi^{4}s^{4}N^{3}}{2}+\frac{\sigma^{2}}{2s^{4}N}+\sigma^{2}\chi^{2}N.
\end{split}
\end{equation}
The optimal value of the twisting angle then becomes
\begin{equation}
\chi_{*}\approx\left[\frac{1}{\left(\frac{1}{3}s^{2}+N\sigma^{2}\right)s^{6}N^{4}}\right]^{1/6},
\end{equation}
where we have assumed $\chi^{2}s^{2}N\gg\frac{1}{s^{2}N}$. This means that the optimal value of the rotated squeezing parameter is
\begin{equation}
\begin{split}
\label{eq:xi_star}
\overline{\Xi^{2}}_{*}\approx&\frac{(9s^{2}+27N\sigma^{2})^{1/3}}{2N^{2/3}}+\frac{\sigma^{2}}{2s^{4}N} \\
&+\frac{\sigma^{2}}{(\frac{1}{3}s^{2}+N\sigma^{2})^{1/3}s^{2}N^{1/3}}.
\end{split}
\end{equation}
The rotated squeezing parameter of the initial GSS state is approximated by
\begin{equation}
\label{eq:xi_knot}
\overline{\Xi^{2}}_{0}\approx s^{2}+\frac{\sigma^{2}}{2s^{4}N}.
\end{equation}
If $s^{2}\ll\sigma^{2}/2s^{4}N$ in Eq.~\ref{eq:xi_knot} then, since we assume $s^{2}\gg1/N$ (see Eq.~\ref{eq:assump}), the post-twist value of the rotated squeezing parameter in Eq.~\ref{eq:xi_star} is $\overline{\Xi^{2}}_{*}\approx \sigma^{2}/2s^{4}N$ and the $N$-dependence is not changed by the twist. However, if $s^{2}\gg \sigma^{2}/2s^{4}N$ in Eq.~\ref{eq:xi_knot} then other terms in Eq.~\ref{eq:xi_star} can dominate. Since $N\sigma^{2}\ll s^{2}$ implies that $\Xi^{2}_{0}\approx s^{2}$ in Eq.~\ref{eq:xi_knot}, we see that as long as this condition is satisfied the $N$-dependence of the rotated squeezing parameter shifts in the same way as the $N$-dependence of the unrotated squeezing parameter in the last section until it hits a floor set by the second term in $\overline{\Xi^{2}}_{0}$.

\section{Preparation of squeezed states}

\subsection{Procedure suggested by squeezing parameters}

\begin{figure*}
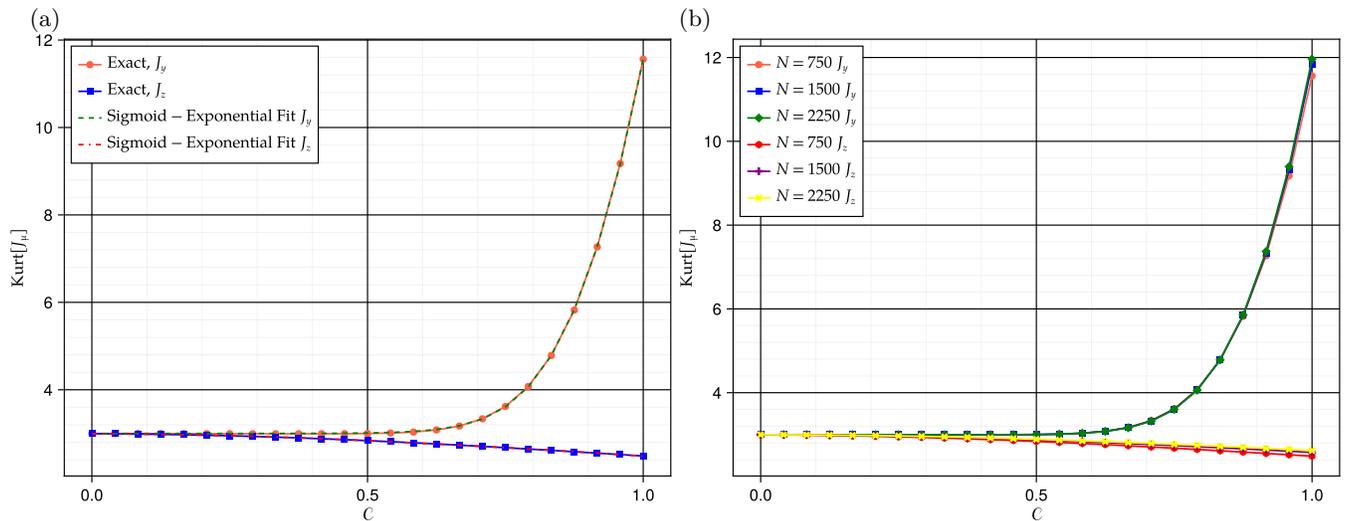

\centering
(a) \hspace{8cm} (b) \hspace{8cm} \ \\
\includesvg[width=0.49\linewidth]{kurtfit3.svg}
\includesvg[width=0.49\linewidth]{kurtN2.svg}
\caption{Here we show the dependences of the kurtoses of $J_{y}$ and $J_{z}$ on the value of $\mathcal{C}$. In (a) the orange dots are the numerically obtained kurtosis for the $J_{y}$ observable as a function of $\mathcal{C}$, the factor by which we shrink the twisting angle for the single twist case. The blue squares display the numerically obtain value of $\textrm{Kurt}[J_{z}]$. The green dashed line is a fit of $\textrm{Kurt}[J_{y}]$ to a sigmoid-exponential function of the form Eq.~\eqref{eq:sig_exp} while the red dash-dotted line is a fit of $\textrm{Kurt}[J_{z}]$ to the same functional form. The fits are of a fairly high quality with  $P_{4}\approx0.84$ for the $\textrm{Kurt}[J_{y}]$ indicating a turning-on behavior around this point. In (b) we plot for $N=750$ (orange circles and red hexagons), $N=1500$ (blue squares and purple crosses), and $N=2250$ (green diamonds and yellow $x$-crosses) the kurtosis of $J_{y}$ and $J_{z}$ respectively with respect to $\mathcal{C}$. These curves lie nearly on top of  one another indicating that the turning-on is approximately independent of system size.}
\label{fig:kurt}
\end{figure*}

These considerations suggest a state preparation procedure for appropriately squeezed states by applying OAT dynamics interspersed with global rotations. Suppose that
\begin{equation}
\sigma^{2}\sim\frac{1}{N^{2-\left(\frac{1}{3}\right)^{j-2}}}
\end{equation}
which is the expected $N$-dependence of the standard deviation of the effective phase seen by the $j$th ensemble (assuming $j\neq 1$). The optimal GSS state to prepare would then be one with
\begin{equation}
\label{eq:s_targ}
s^{2}\approx\frac{1}{N^{1-\left(\frac{1}{3}\right)^{j-1}}}.
\end{equation}
Then a state with the desired squeezing could be prepared by applying OAT dynamics to a GSS state with
\begin{equation}
s'^{2}\sim\frac{1}{N^{1-\left(\frac{1}{3}\right)^{j-2}}}.
\end{equation}
To develop a strategy for the preparation of this state, we assume that the effect of the OAT dynamics is to change the variances and covariances of the spin-moments, but not to produce a non-Gaussian state. In other words, we suppose that for our purposes
\begin{equation}
\label{eq:key_approx}
R_{x}(\theta)T(\chi)|\textrm{GSS}(s)\rangle_{z}\approx|\textrm{GSS}(\xi)\rangle_{z}
\end{equation}
with $\xi^{2}$ and $\theta$ taking the values specified in Sec.~\ref{sec:squee}. In this way beginning from a GSS state with $s^{2}$ we use OAT dynamics and a global rotation to prepare a GSS state with $s^{2}\approx N^{-2/3}$ then one with $N^{-8/9}$ and so on until one satisfying Eq.~\eqref{eq:s_targ} is obtained after $j-1$ OAT operations. In order for the correct angular momentum component to be squeezed at the end of the state preparation the final rotation angle is taken to be $\frac{\pi}{2}-\theta_{*}$ rather than $\theta_{*}$. More explicitly, this suggests that the probe states should have the form
\begin{equation}
|\psi_{0}\rangle=\prod_{k=1}^{j-1}\left[R_{x}(\theta_{k})T(\chi_{k})\right]\big|J_{x}=\frac{N}{2}\big\rangle,
\end{equation}
where
\begin{equation}
\chi_{k}\approx\frac{2^{1-\frac{1}{2}\gamma_{k}}}{3^{\frac{1}{2}-\gamma_{k}}N^{\gamma_{k}}},
\end{equation}
with
\begin{equation}
\gamma_{k}=\frac{2}{3^{k}},
\end{equation}
and $\theta_{k}=-\frac{1}{2}\textrm{arctan}\left[\frac{B(\chi_{k})}{A(\chi_{k})}\right]$. In Appendix~\ref{app:tgs}, we show that the asymptotic behavior of $\theta_{k}$ is
\begin{equation}
\theta_{k}\approx\frac{1}{N\chi_{k}},
\end{equation}
so that 
\begin{equation}
\alpha_{k}=1-\frac{2}{3^{k}}.
\end{equation}
Finally, the last rotation angle $\theta_{j-1}$ is shifted by $\frac{\pi}{2}$, i.e. $\theta_{j-1}=\frac{\pi}{2}-\frac{1}{2}\textrm{arctan}\left(\frac{B(\chi_{j-1})}{A(\chi_{j-1})}\right)$, in order for the squeezing to be oriented along the desired direction but is otherwise the same as the other rotation angles so that the scaling exponent associated with $\eta_{j-1}=\frac{\pi}{2}-\theta_{j-1}$ is  
\begin{equation}
\beta_{j-1}=1-\frac{2}{3^{j-1}}.
\end{equation}
Note that the magnitudes of all twisting angles used decease polynomially with system size. 

\subsection{Heuristic modifications to the state preparation}

\begin{figure*}
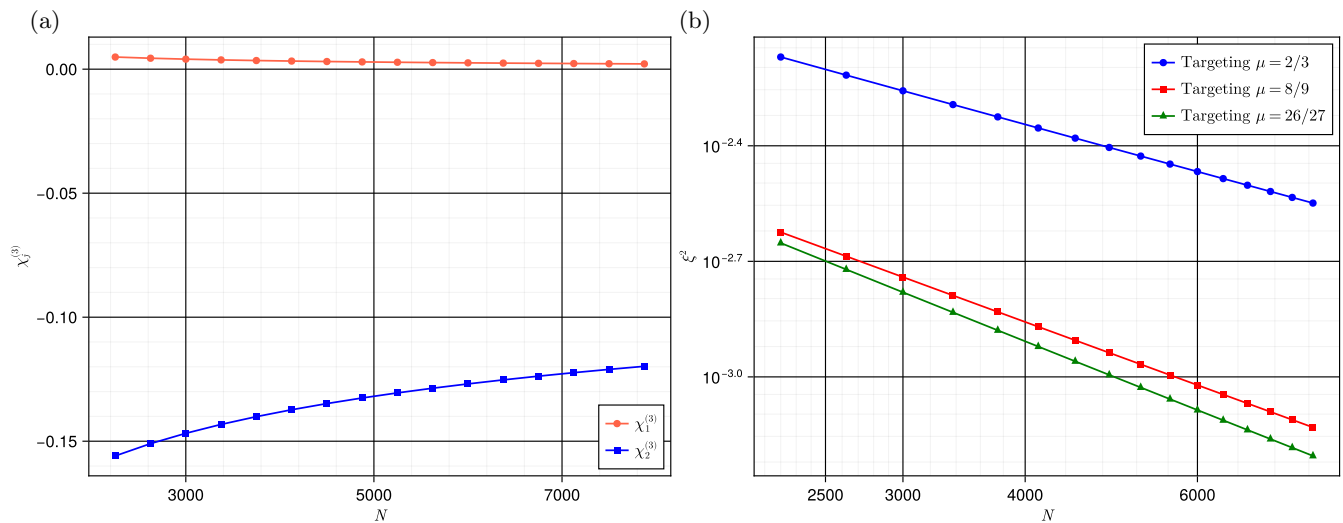

(a) \hspace{8cm} (b) \hspace{8cm} \ \\
\includesvg[width=0.49\linewidth]{chi.svg}
\includesvg[width=0.49\linewidth]{xi_new.svg}
\caption{(a) Example system size dependence of the two twisting parameters in our protocol targeting $\mu=8/9$ where the orange circles correspond to $\chi_{1}$ and the blue squares correspond to $\chi_{2}$. The magnitudes of both parameters approximately follow a simple algebraic decay. (b) Examples the  squeezing protocols considered here yielding a spin-squeezing parameter that decreases faster as the system size is increased than is possible with a single twist. The blue circles result from a single twist and correspond to the protocols proposed in~\cite{Kitagawa_1993}. The red squares correspond to the a two-twist protocol which achieves $\mu\approx8/9$. and the green triangles correspond to the three-twist protocol that achieves $\mu\approx26/27$.}
\label{fig:xi_collect}
\end{figure*}

Unfortunately, the states produced by this procedure do not turn out to be exactly GSS states. To mitigate the effects of this we find that several heuristic measures are necessary. First, we note that the above analysis predicts that nothing should change if the pair $(\chi_{k},\theta_{k})$ is replaced with $(-\chi_{k},-\theta_{k})$. However, through numerical experiments, we found that it is desirable for $\chi_{k}$ and $\chi_{k+1}$ to have opposite signs. We also found that reducing the size of all but the last twisting angle by a multiplicative constant, which we will denote $\mathcal{C}$, and adjusting the rotation angles accordingly greatly improves the performance of the multi-twist protocols without effecting the $N$-dependence predicted by the above analysis. In particular, this modifies the state preparation procedure to 
\begin{equation}
|\psi_{0}\rangle=\prod_{k=1}^{j-1}\left\{R_{x}[(-1)^{k+1}\tilde{\theta}_{k}]T[(-1)^{k+1}\mathcal{C}_{k}\chi_{k}]\right\}\big|J_{x}=\frac{N}{2}\big\rangle,
\end{equation}
where $\tilde{\theta}_{k\neq j-1}=-\frac{1}{2}\textrm{arctan}\left[\frac{B(\mathcal{C}_{k}\chi_{k})}{A(\mathcal{C}_{k}\chi_{k})}\right]$, $\tilde{\theta}_{j-1}$ is shifted by $\frac{\pi}{2}$ in the same way as described above, $\mathcal{C}_{k\neq j-1}=\mathcal{C}$ with $\mathcal{C}$ a constant fixed throughout the procedure, and $\mathcal{C}_{j-1}=1$.

The need for the $\mathcal{C}$ parameter is closely related to the non-Gaussianity of the state, as can be seen by examining the dependence of the kurtosis of the $J_{y}$ measurement outcome distribution vs the value of $\mathcal{C}$.
In Fig.~\ref{fig:kurt} we plot the kurtosis of $J_{y}$ and $J_{z}$ vs the value of $\mathcal{C}$ when the state $|J_{x}=\frac{N}{2}\rangle$ undergoes one-axis twists followed by an $x$-rotation chosen as described above. The kurtosis of $J_{y}$ displays a clear ``turning-on" behavior. We make this more quantitative by fitting the curve to a function of the form
\begin{equation}
\label{eq:sig_exp}
K(\mathcal{C})=3+\frac{P_{1}e^{P_{2}\mathcal{C}}}{1+e^{-P_{3}(\mathcal{C}-P_{4})}}.
\end{equation}
This is the dashed line in Fig.~\ref{fig:kurt}(a) which we refer to as the sigmoid-exponential fit. The denominator, i.e. the sigmoid, is responsible for the turning-on behavior. In particular, the turning-on occurs at roughly $P_{4}$. For this example, we find that the fit converges to a value of $P_{4}\approx0.84$. This suggests that we should choose values of $\mathcal{C}$ that are significantly less than this. In Fig.~\ref{fig:kurt}(b) we demonstrate that this behavior is remarkably constant across system sizes suggesting that the value of $\mathcal{C}$ chosen may not need to depend on system size. In practice, we find that values of $\mathcal{C}$ ranging from $0.35$ to $0.7$ work well. Additional numerical results relating to the kurtosis associated with these protocols are shown in Appendix~\ref{app:ng} and more details of the numerical techniques can be found in Appendix~\ref{app:numer}.

\subsection{Numerical simulation of state preparation}\label{sec:2t}

In Fig.~\ref{fig:xi_collect}a, we display the $N$-dependence of $\chi_{1}$ and $\chi_{2}$ for the three-ensemble state preparation protocol targeting $\mu=8/9$ with $\mathcal{C}=0.7$. In order to improve the finite system size performance we fix $\chi_{j}$ by numerically finding the solution to 
\begin{equation}
\label{eq:state1}
\chi_{j}^{2}\xi_{j}^{2}N-\tanh\left(\frac{1}{\xi^{2}_{j}N}+\chi_{j}^{2}\xi_{j}^{2}N\right)=0,
\end{equation}
which is obtained by differentiating Eq.~\eqref{eq:v_min}. Here $\xi^{2}_{1}=1$ and $\xi^{2}_{2}$ is the value of $\xi^{2}$ that would be obtained if the protocol were terminated after the first twist.
In Fig.~\ref{fig:xi_collect}b, we show the computed squeezing parameters for the cases where the targeted values of $\mu$ are $\{\frac{2}{3},\frac{8}{9},\frac{26}{27}\}$. Again with $\mathcal{C}=0.7$. The change in the $N$-dependence as the number of periods of OAT dynamics is increased can be seen by comparison of the three curves. We note that convergence to the asymptotically expected scaling can occur at quite large system sizes and perfect convergence is not achieved here. Remarkably, at smaller system sizes we actually observed faster decreases in the value of $\xi^{2}$ as the system size is increased. 

It is notable that alternative protocols using the same in or similar resources to achieve the same squeezing displayed here may exist. If they exist, alternative protocols may have their own setting dependent advantages and disadvantages but the protocol we describe here should be broadly useful due to the role it can play in the phase estimation algorithm we describe in the next section.

\section{One-axis twist adaptive phase estimation}
\label{sec:phase_est}

Let the prior distribution be normal with mean zero and standard deviation $\sigma$. As noted above, we allow $\phi$ to take any real value but focus on values of $\sigma$ for which phase wraps are unlikely. If the desired scaling exponent of $\Delta\phi^{2}$ is $\nu=2-1/3^{j}$, then the $N$ particles are divided up into $j+1$ ensembles. The number of particles in ensemble $j$, denoted $N_{j}$, is taken to increase exponentially with $j$. The particles are then squeezed via the method described in the previous section with the $k$th ensemble undergoing the version of the protocol leading to a squeezing exponent of $\mu=1-1/3^{k-1}$. In particular, we fix $\chi_{j}$ for the $k$th ensemble by numerically solving
\begin{equation}
\label{eq:state3}
\begin{split}
0=\chi_{j}^{2}\xi_{j}^{2}N_{k}&-\tanh\left(\frac{1}{\xi_{j}^{2}N_{k}}+\chi_{j}^{2}\xi_{j}^{2}N_{k}\right) \\
&+N_{k-1}\overline{\Xi^{2}}_{k-1}L(\chi_{k}),
\end{split}
\end{equation} 
where
\begin{equation}
\begin{split}
L(\chi_{j})\equiv&\frac{N_{k}\sinh\left(\frac{1}{\xi_{j}^{2}N_{k}}+\chi^{2}_{j}\xi^{2}_{j}N_{k}\right)}{\xi_{j}^{2}\coth\left(\frac{1}{\xi_{j}^{2}N_{k}}+\chi^{2}_{j}\xi^{2}_{j}N_{k}\right)} \\
&\times\left(e^{-\frac{1}{\xi^{2}_{j}N_{k}}-\chi^{2}_{j}\xi^{2}_{j}N_{k}}-e^{-\frac{2}{\xi^{2}_{j}N_{k}}-2\chi^{2}_{j}\xi^{2}_{j}N_{k}}\right).
\end{split}
\end{equation}
This equation is obtained from Eq.~\eqref{eq:state2} and reduces to Eq.~\eqref{eq:state1} whenever the final term is negligible compared to the first two which should occur for all but the last twist on each ensemble. Here $\xi_{j}^{2}$ is the value of $\xi^{2}$ that would be obtained for ensemble $k$ if the protocol were terminated after $j-1$ twists fixed recursively in this way and 
\begin{equation}
\overline{\Xi^{2}}_{k-1}\approx\frac{N}{\langle J_{x}\rangle_{\rho_{0},k-1}^{2}}\left[\left(\Delta J_{y}^{2}\right)_{\rho_{0,k-1}}+\overline{\Xi^{2}}_{k-2}\left(\Delta J_{x}^{2}\right)_{\rho_{0,k-1}}\right]
\end{equation}
with $\overline{\Xi^{2}}_{k-2}$ fixed recursively and where $\rho_{0,k-1}$ is the pre-squeezing state of ensemble $k-1$. The initial conditions of the recursion are $\xi^{2}_{1}=1$ and $\overline{\Xi^{2}}_{0}=\sigma^{2}$. After the state preparation, each particle undergoes the unknown rotation. The smallest, and least squeezed, ensemble is then measured in the eigenbasis of $J_{y}$ yielding a measurement outcome $m$. The associated preliminary estimate of $\phi$ is then simply $2m/N_{1}$. The relevant quantities to understand this protocol, along with their scaling exponents if applicable, are summarized in Table~\ref{tab:phase_est}.

We numerically study this protocol by evaluating the estimation error
\begin{equation}
\begin{split}
\label{eq:monte}
\Delta\phi^{2}=&\int d\phi\sum_{\hat{\phi}_{1}}\sum_{\hat\phi_{2}}\dots\sum_{\hat{\phi}_{M}}p(\phi)p(\hat{\phi}_{1}|\phi)p(\hat{\phi}_{2}|\phi,\hat{\phi}_{1}) \\
&\times\cdots p(\hat{\phi}_{M}|\phi,\hat{\phi}_{1},\hat{\phi}_{2}\cdots\hat{\phi}_{M-1})(\phi-\hat{\phi})^{2},
\end{split}
\end{equation}
with $\hat{\phi}=\sum\hat{\phi}_{j}$. Details of numerical techniques used to evaluate the errors can be found in Appendix~\ref{app:numer}. There are two cases where we make additional approximations when evaluating the error. The first case is when the number of particles in the ensembles needs to be made very large due to slow convergence with system size. The second case is when the particles are divided into a large number of ensembles as is required for this protocol to approach the Heisenberg limit. In both cases, we resort to Monte Carlo summation to evaluate the sums in the expression for the error. After computing the estimation error for a large number of system sizes $N$ and a large number of prior standard deviations $\sigma$, we numerically evaluate the scaling exponent $\nu$ as a function of $\sigma$.

\begin{figure*}
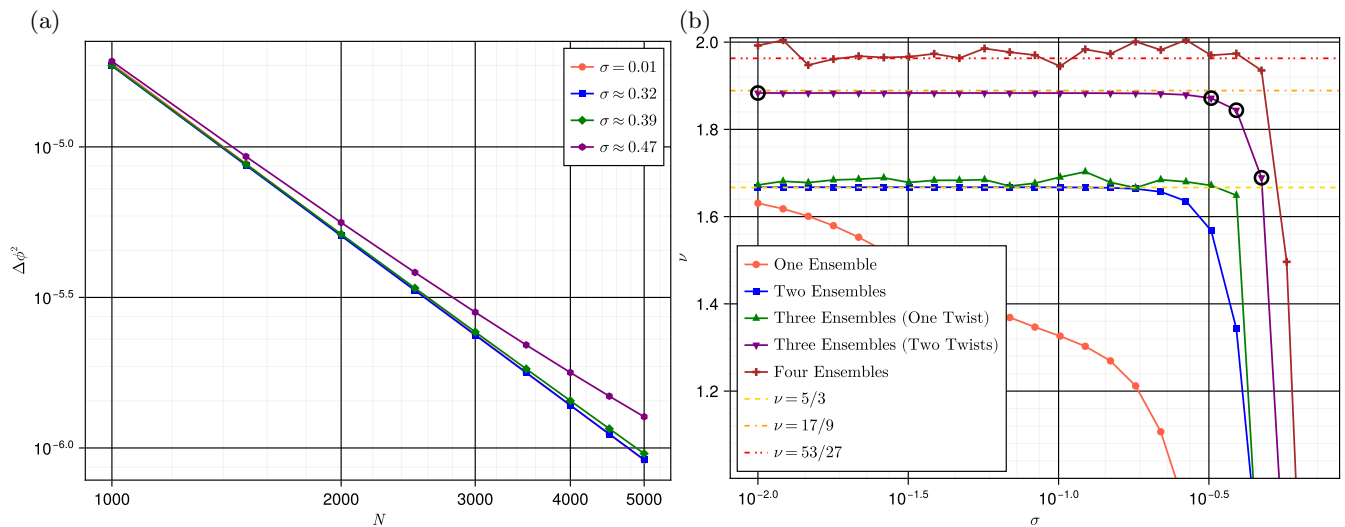

(a) \hspace{8cm} (b) \hspace{8cm} \ \\
\includesvg[width=0.49\linewidth]{ex_err.svg}
\includesvg[width=0.49\linewidth]{nu_v_sigma_new.svg}
\caption{Performance of adaptive phase estimation using one-axis twists. (a) Plot of the averaged estimation errors obtained by this protocol as a function of system size by the three-ensemble, two-twist protocol for various values of the prior standard deviation $\sigma$ where we target $\nu\approx17/9$. (b) Plot of the scaling exponents $\nu$ obtained as a function of the prior standard deviation $\sigma$. The dark orange circles correspond to the behavior exhibited by a protocol that uses one ensemble squeezed by a single twist. The scaling exponent rapidly decays in this case. Note that for this curve the exact values depend on the system size the fit is performed at but the qualitative behavior is consistent. The blue circles correspond to a protocol that uses two ensembles with the first being unsqueezed and the second being squeezed with a single twist. The green upward triangles correspond to a protocol that uses three ensembles where the first ensemble is unsqueezed and the second and third ensembles are squeezed with a single twist. The purple downward triangles correspond to a protocol that uses three ensembles but with the third ensemble undergoing two twists. The brown crosses correspond to a protocol that uses four ensembles where ensembles one through four are squeezed with zero to three twists respectively. The dashed lines correspond to the expected values for our two, three, and four ensemble protocols. The gold dashed line is at $\nu=5/3$, the orange dot-dashed line is at $\nu=17/9$, and the red dot-dot-dashed line is at $\nu=53/27$. The black circles around the purple downward triangles indicate the points associated with the curves in (a). indicate the points associated with the curves in subfigure (a). The numerical simulations give results that approximately agree with those suggested by the theory of the twisted Gaussian states. The ``Three Ensembles (One Twist)" and ``Four Ensembles" curves exhibit additional fluctuations due to the use of Monte Carlo integration to evaluate several of the sums in Eq.~\eqref{eq:monte} while for the other curves we are able to evaluate the sums exactly.}
\label{fig:nu_collect}
\end{figure*}

Fig.~\ref{fig:nu_collect}a shows the resulting averaged estimation errors as a function of the total number of probes for the case of three ensembles at various values of $\sigma$. For $\sigma\lesssim0.3$, as $\sigma$ is increased there is a constant increase in $\Delta\phi^{2}$ but the scaling remains roughly constant. Above this value of $\sigma$, we see a decrease in the scaling exponent $\nu$.

The scaling exponents associated with $\Delta\phi^{2}$ for various values of the $\sigma$ are shown in Fig.~\ref{fig:nu_collect}b. Regardless of how many ensembles are used in a given protocol, the scaling exponents plotted reflect the scaling with respect to the total number of particles in all ensembles combined. The dark orange circles correspond to the use of a single large squeezed state produced by a single one-axis twist followed by a global rotation about the $x$-axis with the twisting and rotation angles chosen to minimize $\Delta\phi$ for a given value of $\sigma$. At small values of $\sigma$, the one-ensemble protocol displays a scaling of nearly $\nu\approx5/3$. However, as $\sigma$ increases, the value of $\nu$ rapidly decreases. Note that the actual values of $\nu$ obtained for a protocol of this type is highly sensitive to the total number of particles used but qualitative behavior appears consistent. This sensitivity does not appear to occur for the other types of protocol considered here. 

The blue squares correspond to a two-ensemble protocol where the first ensemble is not squeezed and the second ensemble is squeezed using single-twist squeezing protocol described in the previous section. The ensemble displays a value of $\nu$ close to $5/3$ for a wider range of values of $\sigma$ before eventually displaying a fall-off at quite large values of $\sigma$.

The green upward triangles correspond to a three-ensemble protocol in which the first ensemble is not squeezed and the other ensembles are prepared in states squeezed via single-twist protocols. We again observe a maximum scaling exponent of $\nu\approx5/3$. The fall-off as $\sigma$ increases is present but occurs at larger values of $\sigma$ than in the two ensemble protocol. Since this type of protocol falls outside of the theory discussed in section~\ref{sec:oat_gss} we numerically optimize the circuit parameters for the third ensemble in this protocol. We also note that this protocol seems to converge to its asymptotic scaling only for much larger system sizes than the other protocols considered here. Due to these large system sizes, we use Monte Carlo integration to evaluate the third sum in Eq.~\eqref{eq:monte}. This results in some additional noise on the values of $\nu$ obtained. With only single-twist state preparation, increasing the number of ensembles cannot increase the scaling exponent $\nu$ but can lead to an increase in the values of $\sigma$ for which the protocol is effective.

The purple downward triangles correspond to a three-ensemble protocol where the first ensemble is not squeezed, the second ensemble utilizes single-twist state preparation, and the third ensemble utilizes the two-twist squeezing protocol described above. We find that $\mathcal{C}=0.35$ works well here. A scaling exponent of $\nu\approx17/9$ is achieved over a wide range of values of $\sigma$. Additionally, we also observe that the fall-off with increasing $\sigma$ occurs at larger values. 

Finally, the brown crosses correspond to the four-ensemble protocol discussed above where the first ensemble is unentangled, the second ensemble is squeezed with a single-twist protocol, the third ensemble is squeezed with a two-twist protocol, and the fourth ensemble is squeezed with a  three-twist protocol. Here we use $\mathcal{C}=0.7$. The resulting protocol gives values close to the expected value of  $\nu\approx53/27$ for a large range of $\sigma$ values. In particular, this protocol achieves nearly Heisenberg limited scaling across this large range of values. The relatively large fluctuations in this curve are due the Monte Carlo evaluation of the third and fourth sums in Eq.~\eqref{eq:monte}. We use Monte Carlo evaluation in this case because evaluation of the averaged estimation error via this method is exponentially expensive in the number of ensembles. We also observe the falloff of $\nu$ with $\sigma$ continuing to be pushed to larger values.

\section{Robustness to imperfections}
In this section we study the robustness of our protocols to various types of imperfections that may occur when utilizing this phase estimation protocol in experimental settings.

\subsection{Particle number uncertainty}

\begin{figure}
\includesvg[width=1.0\linewidth]{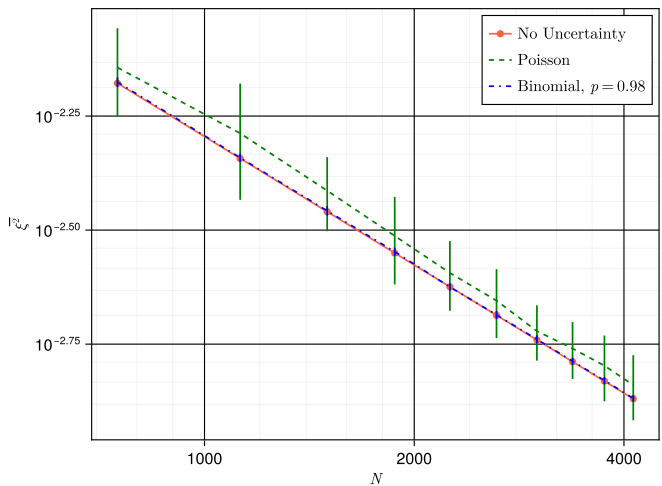}
\caption{The average squeezing parameter produced vs. the anticipated system size for various different particle number distributions in the case where $\mu\approx8/9$ is targeted. The orange circles correspond to the case where there is no particle number uncertainty, the green dashed curve corresponds to the case of a Poisson distribution, and the blue dash-dot curve corresponds to a binomial distribution with $p=0.98$. The vertical bars denote a single standard deviation from the mean for each distribution. We find that the noiseless squeezing parameter remains within the error bars for all system sizes.}
\label{fig:Nfluct}
\end{figure}

First, we consider the effect of allowing the number of particles in the probe to be uncertain. In particular, we consider the effect of fixing the twisting and rotation angles according to an anticipated particle number $N$ but allowing the actual particle number $N_{s}$ to be sampled from a probability distribution $P(N_{s})$. In Fig.~\ref{fig:Nfluct}, we show the value of the squeezing parameter averaged over the particle number distribution
\begin{equation}
\label{eq:xi_bar}
\overline{\xi^{2}}\equiv\sum_{N_{s}}P(N_{s})\xi^{2}(N_{s})
\end{equation}
for a Poisson distribution
\begin{equation}
            P(N_{s})=\frac{N^{N_{s}}e^{-N}}{N_{s}!},
\end{equation}
and a binomial distribution
\begin{equation}
P(N_{s})={\lceil N/p\rceil\choose N_{s}}p^{N_{s}}(1-p)^{N-N_{s}},
\end{equation}
where the number of trials is fixed so that the average number of particles is $N$. We consider the Poisson distribution because for some experimental platforms loading of qubits is a stochastic process in which many potential qubits are lost. For example, this is the case when a cloud of atoms is first cooled in a magneto-optic trap (MOT). The binomial distribution may describe platforms for which site-by-site loading is typical. Tweezer arrays provide an example of this type of loading. For all of the system sizes considered, the noiseless results are within one standard deviation of the averaged squeezing parameter. This is a result of the fact that the circuit parameters used in this protocol do not fluctuate too rapidly as the system size is changed.

\subsection{Feedback control fluctuations}

\begin{figure}
\includesvg[width=1.0\linewidth]{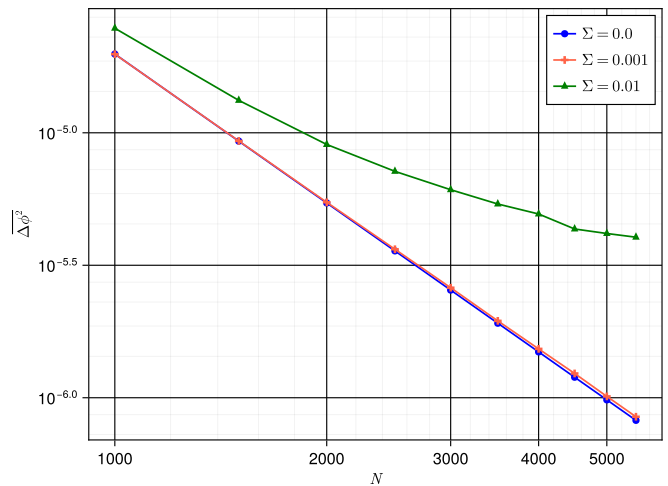}
\caption{Plot of the estimation error averaged over the distribution of feedback control fluctuations which is taken to be normal with standard deviation $\Sigma$. The blue circles correspond to the noiseless $\Sigma=0$ case, the orange crosses correspond to $\Sigma=0.001$, and the green triangles correspond to $\sigma=0.01$. These curves are for the three ensemble protocol, in which the final ensemble utilizes the two-twist squeezing protocol, with $\sigma\approx0.32$. In the case that $\Sigma$ is much less than the noiseless estimation error we observe good agreement with the noiseless case but the system size dependence changes qualitatively when this is no longer the case.}
\label{fig:pfluc}
\end{figure}

In the phase estimation protocol discussed above, after the measurement of ensemble $j$, all later ensembles are counter rotated by $-\hat{\phi}_{j}$. Here we consider the effect of instead counter rotating by $-\tilde{\phi}_{j}$ where
\begin{equation}
\tilde{\phi}_{j}=\hat{\phi}_{j}+r_{j},
\end{equation}
where $r_{j}$ is a normally distributed random variable with mean zero and standard deviation $\Sigma$. The results are shown in Fig.~\ref{fig:pfluc}.

In this figure we plot the value of $\Delta\phi^{2}$ squared averaged over the distribution for all of the $r_{j}$, denoted $\overline{\Delta\phi^{2}}$, for the three ensemble protocol vs. system size for $\sigma\approx0.34$. When $\Sigma=0.001$, we find that the feedback control fluctuations do not significantly effect the value of $\overline{\Delta\phi^{2}}$ up to several thousand spins. However, when $\Sigma=0.01$ we see a clear increase in the estimation error as well as a qualitative shift the $N$-dependence of the estimation error. In general, we expect that these control fluctuations will have a large impact only if $\Sigma\gtrsim\Delta\phi$. The reason is that this is roughly the point at which we begin to expect $r_{j}$ and $\hat{\phi}_{j}$ to become similar in magnitude for the second to last ensemble.

\subsection{Cavity induced contrast loss}
Certain cavity-QED schemes for implementing OAT interactions between neutral atoms rely inextricably on the atoms scattering photons out of the cavity. Such schemes result in additional contrast loss beyond that expected from the unitary OAT dynamics alone~\cite{Li_2022, Carrasco2022}. The resulting effective squeezing parameter is
\begin{equation}
\tilde{\xi^{2}}\equiv\frac{\xi^{2}}{C_{\textrm{SC}}},
\end{equation}
where
\begin{equation}
C_{\textrm{SC}}=\exp\left[-2\Gamma\sqrt{N}\sum_{j}|\chi_{j}|\right]
\end{equation}
and $\Gamma$ controls the amount of contrast loss and depends on the details of the experimental setup. Note that for the protocols we described above
\begin{equation}
\sqrt{N}\sum_{j}|\chi_{j}|\sim N^{\omega}, 
\end{equation}
where
\begin{equation}
\omega=\frac{1}{2}-\frac{2}{3^{j-1}}.
\end{equation}
This is an increasing function of $N$ whenever $j>2$. This suggests that we should not expect that the protocols we described above for more than two ensembles to lead to the desired $N$-dependence of the estimation error in the presence of this additional contrast loss.

\begin{figure}
\includesvg[width=1.0\linewidth]{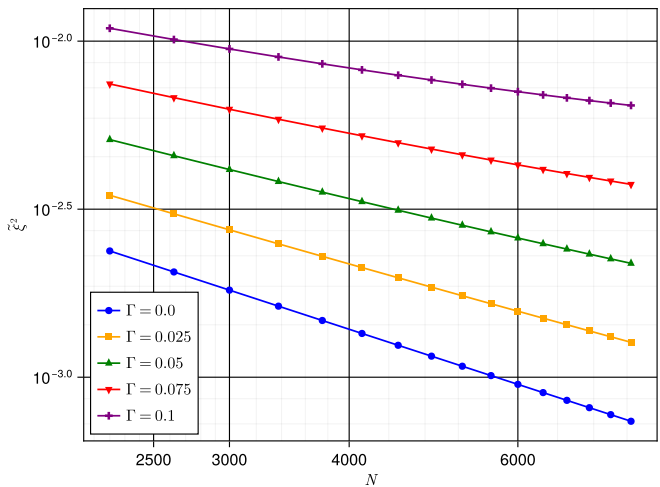}
\caption{The squeezing parameter adjusted for cavity induced contrast loss vs. system size for various values of the contrast loss strength $\Gamma$ with $\Gamma=0$ for the blue curves, $\Gamma=0.1$ for the orange squares, $\Gamma=0.2$ for the green upward triangles, $\Gamma=0.3$ for the red downward triangles, and $\Gamma=0.4$ for the purple crosses. The squeezing protocol considered here is the two-twist one that targets $\mu=8/9$. This squeezing protocol is used the phase estimation schemes that utilize three or more twists. Qualitatively, the orange squares still appear similar to a slower polynomial decay while the curves corresponding to larger values of $\Gamma$ clearly show the influence of the exponential dependence on system size $N$.}
\label{fig:xi_tild}
\end{figure}

Indeed we do observe this effect in the protocols described above. This is displayed in Fig.~\ref{fig:xi_tild} where we show $\tilde{\xi}^{2}$ vs. $N$ for the two-twist protocol utilized by the phase estimation algorithms with at least three ensembles. Here, we take $\mathcal{C}=0.7$. As a point of reference, the blue circles represent the noiseless case. Even for small values of $\Gamma$, an apparent decrease in the scaling exponents. For larger values of $\Gamma$ the behavior no longer appears qualitatively to be a polynomial decay. In other words, we begin to see the effect of the exponential dependence of $\tilde{\xi}^{2}$ on $N$. It may be possible to mitigate this effect to some extent by choosing a smaller value of $\mathcal{C}$. Thus could be a major obstacle for approaches to squeezing that rely on this type of scattering. Fortunately, we still see reasonable decrease of $\tilde{\xi}^{2}$ as system size is increased.  

\section{Conclusion}
We have analyzed a phase estimation algorithm based on probes prepared via multiple one-axis twist (OAT) operations adaptively and argued that the general structure of the algorithm is fairly general. We showed that the utilized states achieve squeezing parameter system size dependences characterized by scaling exponents satisfying $2/3\leq\mu\leq1$ and lead to similarly enhanced system size dependences of the averaged estimation error when used in a phase estimation algorithm. These protocols achieve a targeted scaling of the averaged estimation error $\Delta\phi^{2}$ across a wide range of prior standard deviations $\sigma$. Additionally, they do this using only resources that have already been experimentally demonstrated.

Some questions about these protocols remain open. First, prior work has numerically found protocol to prepare states with extreme spin-squeezing using only a two-twist protocol~\cite{Carrasco2022}. This suggests that two-twist state preparation might be possible for all ensembles but we do not have a good understanding of how such protocols would work at present. Second, it would be useful to have a less ad hoc understanding of the role of the parameter $\mathcal{C}$ and of how non-Gaussian features develop throughout the state preparation. In particular, we fix $\mathcal{C}$ to values that seem to work well based on heuristics but it would be useful to have a rigorous theory describing the optimal value of $\mathcal{C}$. A reasonable route to such a theory seems to be understanding the error in the approximation in Eq.~(\ref{eq:key_approx}). This seems especially important if interest develops in many ensemble protocols where many OAT operations are needed. A third question is the circuit level noise robustness of these protocols. It would be especially good to understand the effect of noise during the one-axis twists. Finally, many platforms cannot implement all-to-all interactions and it would be useful to understand how the multi-twist state preparation protocols should be modified for use when the interaction strengths between particles decay polynomially with the separation between the particles. Overall, we think that one-axis twist based phase estimation algorithms may provide a promising path to the experimental realization of entanglement enhanced phase estimation.

\section*{Acknowledgments}
This work is supported by the National Science Foundation QLCI Q-SEnSE (Grant No. OMA- 2016244), and STAQ (Grants No.PHY-2325080). We would like to thank Luca Pezz\`e for a helpful personal communication. We thank the UNM Center for Advanced Research Computing, supported in part by the National Science Foundation, for providing the high performance computing resources used in this work.

\bibliography{references}

@article{Ma_2011,
title = {Quantum spin squeezing},
journal = {Physics Reports},
volume = {509},
number = {2},
pages = {89-165},
year = {2011},
issn = {0370-1573},
doi = {https://doi.org/10.1016/j.physrep.2011.08.003},
url = {https://www.sciencedirect.com/science/article/pii/S0370157311002201},
author = {Jian Ma and Xiaoguang Wang and C.P. Sun and Franco Nori}
}

@article{Toth_2014,
doi = {10.1088/1751-8113/47/42/424006},
url = {https://dx.doi.org/10.1088/1751-8113/47/42/424006},
year = {2014},
month = {oct},
publisher = {IOP Publishing},
volume = {47},
number = {42},
pages = {424006},
author = {Géza Tóth and Iagoba Apellaniz},
title = {Quantum metrology from a quantum information science perspective},
journal = {Journal of Physics A: Mathematical and Theoretical}
}

@article{Ludlow_2015,
  title = {Optical atomic clocks},
  author = {Ludlow, Andrew D. and Boyd, Martin M. and Ye, Jun and Peik, E. and Schmidt, P. O.},
  journal = {Rev. Mod. Phys.},
  volume = {87},
  issue = {2},
  pages = {637--701},
  numpages = {65},
  year = {2015},
  month = {Jun},
  publisher = {American Physical Society},
  doi = {10.1103/RevModPhys.87.637},
  url = {https://link.aps.org/doi/10.1103/RevModPhys.87.637}
}

@article{Degen_2017,
  title = {Quantum sensing},
  author = {Degen, C. L. and Reinhard, F. and Cappellaro, P.},
  journal = {Rev. Mod. Phys.},
  volume = {89},
  issue = {3},
  pages = {035002},
  numpages = {39},
  year = {2017},
  month = {Jul},
  publisher = {American Physical Society},
  doi = {10.1103/RevModPhys.89.035002},
  url = {https://link.aps.org/doi/10.1103/RevModPhys.89.035002}
}

@article{Pezze_2018,
  title = {Quantum metrology with nonclassical states of atomic ensembles},
  author = {Pezz\`e, Luca and Smerzi, Augusto and Oberthaler, Markus K. and Schmied, Roman and Treutlein, Philipp},
  journal = {Rev. Mod. Phys.},
  volume = {90},
  issue = {3},
  pages = {035005},
  numpages = {70},
  year = {2018},
  month = {Sep},
  publisher = {American Physical Society},
  doi = {10.1103/RevModPhys.90.035005},
  url = {https://link.aps.org/doi/10.1103/RevModPhys.90.035005}
}

@article{Wineland_1992,
  title = {Spin squeezing and reduced quantum noise in spectroscopy},
  author = {Wineland, D. J. and Bollinger, J. J. and Itano, W. M. and Moore, F. L. and Heinzen, D. J.},
  journal = {Phys. Rev. A},
  volume = {46},
  issue = {11},
  pages = {R6797--R6800},
  numpages = {0},
  year = {1992},
  month = {Dec},
  publisher = {American Physical Society},
  doi = {10.1103/PhysRevA.46.R6797},
  url = {https://link.aps.org/doi/10.1103/PhysRevA.46.R6797}
}

@article{Kitagawa_1993,
  title = {Squeezed spin states},
  author = {Kitagawa, Masahiro and Ueda, Masahito},
  journal = {Phys. Rev. A},
  volume = {47},
  issue = {6},
  pages = {5138--5143},
  numpages = {0},
  year = {1993},
  month = {Jun},
  publisher = {American Physical Society},
  doi = {10.1103/PhysRevA.47.5138},
  url = {https://link.aps.org/doi/10.1103/PhysRevA.47.5138}
}

@article{Wineland_1994,
  title = {Squeezed atomic states and projection noise in spectroscopy},
  author = {Wineland, D. J. and Bollinger, J. J. and Itano, W. M. and Heinzen, D. J.},
  journal = {Phys. Rev. A},
  volume = {50},
  issue = {1},
  pages = {67--88},
  numpages = {0},
  year = {1994},
  month = {Jul},
  publisher = {American Physical Society},
  doi = {10.1103/PhysRevA.50.67},
  url = {https://link.aps.org/doi/10.1103/PhysRevA.50.67}
}

@article{Bollinger_1996,
  title = {Optimal frequency measurements with maximally correlated states},
  author = {Bollinger, J. J . and Itano, Wayne M. and Wineland, D. J. and Heinzen, D. J.},
  journal = {Phys. Rev. A},
  volume = {54},
  issue = {6},
  pages = {R4649--R4652},
  numpages = {0},
  year = {1996},
  month = {Dec},
  publisher = {American Physical Society},
  doi = {10.1103/PhysRevA.54.R4649},
  url = {https://link.aps.org/doi/10.1103/PhysRevA.54.R4649}
}

@article{Giovannetti_2006,
  title = {Quantum Metrology},
  author = {Giovannetti, Vittorio and Lloyd, Seth and Maccone, Lorenzo},
  journal = {Phys. Rev. Lett.},
  volume = {96},
  issue = {1},
  pages = {010401},
  numpages = {4},
  year = {2006},
  month = {Jan},
  publisher = {American Physical Society},
  doi = {10.1103/PhysRevLett.96.010401},
  url = {https://link.aps.org/doi/10.1103/PhysRevLett.96.010401}
}

@article{Kuzmich_1998,
doi = {10.1209/epl/i1998-00277-9},
url = {https://dx.doi.org/10.1209/epl/i1998-00277-9},
year = {1998},
month = {jun},
publisher = {},
volume = {42},
number = {5},
pages = {481},
author = {A. Kuzmich and  N. P. Bigelow and  L. Mandel},
title = {Atomic quantum non-demolition measurements and squeezing},
journal = {Europhysics Letters}
}

@article{Kuzmich_2000,
  title = {Generation of Spin Squeezing via Continuous Quantum Nondemolition Measurement},
  author = {Kuzmich, A. and Mandel, L. and Bigelow, N. P.},
  journal = {Phys. Rev. Lett.},
  volume = {85},
  issue = {8},
  pages = {1594--1597},
  numpages = {0},
  year = {2000},
  month = {Aug},
  publisher = {American Physical Society},
  doi = {10.1103/PhysRevLett.85.1594},
  url = {https://link.aps.org/doi/10.1103/PhysRevLett.85.1594}
}

@article{Smith_2006,
  title = {Efficient Quantum-State Estimation by Continuous Weak Measurement and Dynamical Control},
  author = {Smith, Greg A. and Silberfarb, Andrew and Deutsch, Ivan H. and Jessen, Poul S.},
  journal = {Phys. Rev. Lett.},
  volume = {97},
  issue = {18},
  pages = {180403},
  numpages = {4},
  year = {2006},
  month = {Oct},
  publisher = {American Physical Society},
  doi = {10.1103/PhysRevLett.97.180403},
  url = {https://link.aps.org/doi/10.1103/PhysRevLett.97.180403}
}

@article{Shah_2010,
  title = {High Bandwidth Atomic Magnetometery with Continuous Quantum Nondemolition Measurements},
  author = {Shah, V. and Vasilakis, G. and Romalis, M. V.},
  journal = {Phys. Rev. Lett.},
  volume = {104},
  issue = {1},
  pages = {013601},
  numpages = {4},
  year = {2010},
  month = {Jan},
  publisher = {American Physical Society},
  doi = {10.1103/PhysRevLett.104.013601},
  url = {https://link.aps.org/doi/10.1103/PhysRevLett.104.013601}
}

@article{Wasilewski_2010,
  title = {Quantum Noise Limited and Entanglement-Assisted Magnetometry},
  author = {Wasilewski, W. and Jensen, K. and Krauter, H. and Renema, J. J. and Balabas, M. V. and Polzik, E. S.},
  journal = {Phys. Rev. Lett.},
  volume = {104},
  issue = {13},
  pages = {133601},
  numpages = {4},
  year = {2010},
  month = {Mar},
  publisher = {American Physical Society},
  doi = {10.1103/PhysRevLett.104.133601},
  url = {https://link.aps.org/doi/10.1103/PhysRevLett.104.133601}
}

@article{Ye_2024,
   title={Essay: Quantum Sensing with Atomic, Molecular, and Optical Platforms for Fundamental Physics},
   volume={132},
   ISSN={1079-7114},
   url={http://dx.doi.org/10.1103/PhysRevLett.132.190001},
   DOI={10.1103/physrevlett.132.190001},
   number={19},
   journal={Physical Review Letters},
   publisher={American Physical Society (APS)},
   author={Ye, Jun and Zoller, Peter},
   year={2024},
   month=may }

@article{Levine_99,
    author = {Levine, Judah},
    title = {Introduction to time and frequency metrology},
    journal = {Review of Scientific Instruments},
    volume = {70},
    number = {6},
    pages = {2567-2596},
    year = {1999},
    month = {06},
    issn = {0034-6748},
    doi = {10.1063/1.1149844},
    url = {https://doi.org/10.1063/1.1149844},
    eprint = {https://pubs.aip.org/aip/rsi/article-pdf/70/6/2567/19022246/2567\_1\_online.pdf},
}

@article{Fraas_16,
author={Fraas, Martin},
year={2016},
title={An Analysis of the Stationary Operation of Atomic Clocks},
journal={Communications in Mathematical Physics},
pages={363-393},
volume={348},
url={https://doi.org/10.1007/s00220-016-2761-1},
doi={10.1007/s00220-016-2761-1},
issn={0003-6951},
}

@article{Colombo_22,
    author = {Colombo, Simone and Pedrozo-Peñafiel, Edwin and Vuletić, Vladan},
    title = {Entanglement-enhanced optical atomic clocks},
    journal = {Applied Physics Letters},
    volume = {121},
    number = {21},
    pages = {210502},
    year = {2022},
    month = {11},
    issn = {0003-6951},
    doi = {10.1063/5.0121372},
    url = {https://doi.org/10.1063/5.0121372},
}

@book{Nielsen_00,
  author = {Nielsen, Michael A. and Chuang, Isaac L.},
  publisher = {Cambridge University Press},
  title = {Quantum Computation and Quantum Information},
  year = {2000}
}

@article{Udem_02,
author={Udem, Th. and Holzwarth, R. and Hänsch, T. W.},
year={2002},
title={Optical frequency metrology},
journal={Nature},
pages={233-237},
volume={416},
issue={6877},
issn={1476-4687},
url={https://doi.org/10.1038/416233a},
doi={10.1038/416233a},
}

@article{Steinel_23,
  title = {Evaluation of a $^{88}{\mathrm{Sr}}^{+}$ Optical Clock with a Direct Measurement of the Blackbody Radiation Shift and Determination of the Clock Frequency},
  author = {Steinel, M. and Shao, H. and Filzinger, M. and Lipphardt, B. and Brinkmann, M. and Didier, A. and Mehlst\"aubler, T. E. and Lindvall, T. and Peik, E. and Huntemann, N.},
  journal = {Phys. Rev. Lett.},
  volume = {131},
  issue = {8},
  pages = {083002},
  numpages = {6},
  year = {2023},
  month = {Aug},
  publisher = {American Physical Society},
  doi = {10.1103/PhysRevLett.131.083002},
  url = {https://link.aps.org/doi/10.1103/PhysRevLett.131.083002}
}

@article{Pelzer_24,
  title = {Multi-ion Frequency Reference Using Dynamical Decoupling},
  author = {Pelzer, Lennart and Dietze, Kai and Mart\'{\i}nez-Lahuerta, V\'{\i}ctor Jos\'e and Krinner, Ludwig and Kramer, Johannes and Dawel, Fabian and Spethmann, Nicolas C. H. and Hammerer, Klemens and Schmidt, Piet O.},
  journal = {Phys. Rev. Lett.},
  volume = {133},
  issue = {3},
  pages = {033203},
  numpages = {7},
  year = {2024},
  month = {Jul},
  publisher = {American Physical Society},
  doi = {10.1103/PhysRevLett.133.033203},
  url = {https://link.aps.org/doi/10.1103/PhysRevLett.133.033203}
}

@article{Hausser_25,
  title = {$^{115}{\mathrm{In}}^{+}\text{\ensuremath{-}}^{172}{\mathrm{Yb}}^{+}$ Coulomb Crystal Clock with $2.5\ifmmode\times\else\texttimes\fi{}{10}^{\ensuremath{-}18}$ Systematic Uncertainty},
  author = {Hausser, H. N. and Keller, J. and Nordmann, T. and Bhatt, N. M. and Kiethe, J. and Liu, H. and Richter, I. M. and von Boehn, M. and Rahm, J. and Weyers, S. and Benkler, E. and Lipphardt, B. and D\"orscher, S. and Stahl, K. and Klose, J. and Lisdat, C. and Filzinger, M. and Huntemann, N. and Peik, E. and Mehlst\"aubler, T. E.},
  journal = {Phys. Rev. Lett.},
  volume = {134},
  issue = {2},
  pages = {023201},
  numpages = {6},
  year = {2025},
  month = {Jan},
  publisher = {American Physical Society},
  doi = {10.1103/PhysRevLett.134.023201},
  url = {https://link.aps.org/doi/10.1103/PhysRevLett.134.023201}
}

@article{Madjarov_19,
  title = {An Atomic-Array Optical Clock with Single-Atom Readout},
  author = {Madjarov, Ivaylo S. and Cooper, Alexandre and Shaw, Adam L. and Covey, Jacob P. and Schkolnik, Vladimir and Yoon, Tai Hyun and Williams, Jason R. and Endres, Manuel},
  journal = {Phys. Rev. X},
  volume = {9},
  issue = {4},
  pages = {041052},
  numpages = {14},
  year = {2019},
  month = {Dec},
  publisher = {American Physical Society},
  doi = {10.1103/PhysRevX.9.041052},
  url = {https://link.aps.org/doi/10.1103/PhysRevX.9.041052}
}

@article{Norcia_2019,
   title={Seconds-scale coherence on an optical clock transition in a tweezer array},
   volume={366},
   ISSN={1095-9203},
   url={http://dx.doi.org/10.1126/science.aay0644},
   DOI={10.1126/science.aay0644},
   number={6461},
   journal={Science},
   publisher={American Association for the Advancement of Science (AAAS)},
   author={Norcia, Matthew A. and Young, Aaron W. and Eckner, William J. and Oelker, Eric and Ye, Jun and Kaufman, Adam M.},
   year={2019},
   month=oct, pages={93–97} }

@article{Young_20,
author={Young, Aaron W. and Eckner, William J. and Milner, William R. and Kedar, Dhruv and Norcia, Matthew A. and Oelker, Eric and Schine, Nathan and Ye, Jun and Kaufman, Adam M.},
year={2020},
title={Half-minute-scale atomic coherence and high relative stability in a tweezer clock},
journal={Nature},
pages={408-413},
volume={588},
issue={7838},
issn={1476-4687},
url={https://doi.org/10.1038/s41586-020-3009-y},
doi={10.1038/s41586-020-3009-y},
}

@article{Shaw_24,
   title={Multi-ensemble metrology by programming local rotations with atom movements},
   volume={20},
   ISSN={1745-2481},
   url={http://dx.doi.org/10.1038/s41567-023-02323-w},
   DOI={10.1038/s41567-023-02323-w},
   number={2},
   journal={Nature Physics},
   publisher={Springer Science and Business Media LLC},
   author={Shaw, Adam L. and Finkelstein, Ran and Tsai, Richard Bing-Shiun and Scholl, Pascal and Yoon, Tai Hyun and Choi, Joonhee and Endres, Manuel},
   year={2024},
   month=jan, pages={195–201} }

@article{Takamoto_15,
title = {Frequency ratios of Sr, Yb, and Hg based optical lattice clocks and their applications},
journal = {Comptes Rendus Physique},
volume = {16},
number = {5},
pages = {489-498},
year = {2015},
note = {The measurement of time / La mesure du temps},
issn = {1631-0705},
doi = {https://doi.org/10.1016/j.crhy.2015.04.003},
url = {https://www.sciencedirect.com/science/article/pii/S1631070515000730},
author = {Masao Takamoto and Ichiro Ushijima and Manoj Das and Nils Nemitz and Takuya Ohkubo and Kazuhiro Yamanaka and Noriaki Ohmae and Tetsushi Takano and Tomoya Akatsuka and Atsushi Yamaguchi and Hidetoshi Katori},
}

@article{Katori_03,
   title={Ultrastable Optical Clock with Neutral Atoms in an Engineered Light Shift Trap},
   volume={91},
   ISSN={1079-7114},
   url={http://dx.doi.org/10.1103/PhysRevLett.91.173005},
   DOI={10.1103/physrevlett.91.173005},
   number={17},
   journal={Physical Review Letters},
   publisher={American Physical Society (APS)},
   author={Katori, Hidetoshi and Takamoto, Masao and Pal’chikov, V. G. and Ovsiannikov, V. D.},
   year={2003},
   month=oct }

@article{Al_Masoudi_15,
   title={Noise and instability of an optical lattice clock},
   volume={92},
   ISSN={1094-1622},
   url={http://dx.doi.org/10.1103/PhysRevA.92.063814},
   DOI={10.1103/physreva.92.063814},
   number={6},
   journal={Physical Review A},
   publisher={American Physical Society (APS)},
   author={Al-Masoudi, Ali and Dörscher, Sören and Häfner, Sebastian and Sterr, Uwe and Lisdat, Christian},
   year={2015},
   month=dec }

@article{Pezze2021,
  title = {Quantum Phase Estimation Algorithm with Gaussian Spin States},
  author = {Pezz\`e, Luca and Smerzi, Augusto},
  journal = {PRX Quantum},
  volume = {2},
  issue = {4},
  pages = {040301},
  numpages = {21},
  year = {2021},
  month = {Oct},
  publisher = {American Physical Society},
  doi = {10.1103/PRXQuantum.2.040301},
  url = {https://link.aps.org/doi/10.1103/PRXQuantum.2.040301}
}

@article{Pezze2020,
  title = {Heisenberg-Limited Noisy Atomic Clock Using a Hybrid Coherent and Squeezed State Protocol},
  author = {Pezz\`e, Luca and Smerzi, Augusto},
  journal = {Phys. Rev. Lett.},
  volume = {125},
  issue = {21},
  pages = {210503},
  numpages = {7},
  year = {2020},
  month = {Nov},
  publisher = {American Physical Society},
  doi = {10.1103/PhysRevLett.125.210503},
  url = {https://link.aps.org/doi/10.1103/PhysRevLett.125.210503}
}

@article{Carrasco2022,
  title = {Extreme Spin Squeezing via Optimized One-Axis Twisting and Rotations},
  author = {Carrasco, Sebastian C. and Goerz, Michael H. and Li, Zeyang and Colombo, Simone and Vuleti\ifmmode \acute{c}\else \'{c}\fi{}, Vladan and Malinovsky, Vladimir S.},
  journal = {Phys. Rev. Appl.},
  volume = {17},
  issue = {6},
  pages = {064050},
  numpages = {6},
  year = {2022},
  month = {Jun},
  publisher = {American Physical Society},
  doi = {10.1103/PhysRevApplied.17.064050},
  url = {https://link.aps.org/doi/10.1103/PhysRevApplied.17.064050}
}

@article{Caves1981,
  title = {Quantum-mechanical noise in an interferometer},
  author = {Caves, Carlton M.},
  journal = {Phys. Rev. D},
  volume = {23},
  issue = {8},
  pages = {1693--1708},
  numpages = {0},
  year = {1981},
  month = {Apr},
  publisher = {American Physical Society},
  doi = {10.1103/PhysRevD.23.1693},
  url = {https://link.aps.org/doi/10.1103/PhysRevD.23.1693}
}

@article{Tse2019,
  title = {Quantum-Enhanced Advanced LIGO Detectors in the Era of Gravitational-Wave Astronomy},
  author = {Tse, M. and \textit{et al.}},
  journal = {Phys. Rev. Lett.},
  volume = {123},
  issue = {23},
  pages = {231107},
  numpages = {8},
  year = {2019},
  month = {Dec},
  publisher = {American Physical Society},
  doi = {10.1103/PhysRevLett.123.231107},
  url = {https://link.aps.org/doi/10.1103/PhysRevLett.123.231107}
}

@article{Levine1973,
url = {https://doi.org/10.1515/nanoph-2019-0209},
title = {Principles and techniques of the quantum diamond microscope},
author = {Edlyn V. Levine and Matthew J. Turner and Pauli Kehayias and Connor A. Hart and Nicholas Langellier and Raisa Trubko and David R. Glenn and Roger R. Fu and Ronald L. Walsworth},
pages = {1945--1973},
volume = {8},
number = {11},
journal = {Nanophotonics},
doi = {doi:10.1515/nanoph-2019-0209},
year = {2019},
lastchecked = {2025-08-11}
}

@article{Barry2020,
  title = {Sensitivity optimization for NV-diamond magnetometry},
  author = {Barry, John F. and Schloss, Jennifer M. and Bauch, Erik and Turner, Matthew J. and Hart, Connor A. and Pham, Linh M. and Walsworth, Ronald L.},
  journal = {Rev. Mod. Phys.},
  volume = {92},
  issue = {1},
  pages = {015004},
  numpages = {68},
  year = {2020},
  month = {Mar},
  publisher = {American Physical Society},
  doi = {10.1103/RevModPhys.92.015004},
  url = {https://link.aps.org/doi/10.1103/RevModPhys.92.015004}
}

@article{Henkel2010,
  title = {Three-Dimensional Roton Excitations and Supersolid Formation in Rydberg-Excited Bose-Einstein Condensates},
  author = {Henkel, N. and Nath, R. and Pohl, T.},
  journal = {Phys. Rev. Lett.},
  volume = {104},
  issue = {19},
  pages = {195302},
  numpages = {4},
  year = {2010},
  month = {May},
  publisher = {American Physical Society},
  doi = {10.1103/PhysRevLett.104.195302},
  url = {https://link.aps.org/doi/10.1103/PhysRevLett.104.195302}
}

@article{Braverman2019,
  title = {Near-Unitary Spin Squeezing in $^{171}\mathrm{Yb}$},
  author = {Braverman, Boris and Kawasaki, Akio and Pedrozo-Pe\~nafiel, Edwin and Colombo, Simone and Shu, Chi and Li, Zeyang and Mendez, Enrique and Yamoah, Megan and Salvi, Leonardo and Akamatsu, Daisuke and Xiao, Yanhong and Vuleti\ifmmode \acute{c}\else \'{c}\fi{}, Vladan},
  journal = {Phys. Rev. Lett.},
  volume = {122},
  issue = {22},
  pages = {223203},
  numpages = {6},
  year = {2019},
  month = {Jun},
  publisher = {American Physical Society},
  doi = {10.1103/PhysRevLett.122.223203},
  url = {https://link.aps.org/doi/10.1103/PhysRevLett.122.223203}
}

@article{Pogorelov2021,
  title = {Compact Ion-Trap Quantum Computing Demonstrator},
  author = {Pogorelov, I. and Feldker, T. and Marciniak, Ch. D. and Postler, L. and Jacob, G. and Krieglsteiner, O. and Podlesnic, V. and Meth, M. and Negnevitsky, V. and Stadler, M. and H\"ofer, B. and W\"achter, C. and Lakhmanskiy, K. and Blatt, R. and Schindler, P. and Monz, T.},
  journal = {PRX Quantum},
  volume = {2},
  issue = {2},
  pages = {020343},
  numpages = {23},
  year = {2021},
  month = {Jun},
  publisher = {American Physical Society},
  doi = {10.1103/PRXQuantum.2.020343},
  url = {https://link.aps.org/doi/10.1103/PRXQuantum.2.020343}
}

@article{Gil2014,
  title = {Spin Squeezing in a Rydberg Lattice Clock},
  author = {Gil, L. I. R. and Mukherjee, R. and Bridge, E. M. and Jones, M. P. A. and Pohl, T.},
  journal = {Phys. Rev. Lett.},
  volume = {112},
  issue = {10},
  pages = {103601},
  numpages = {5},
  year = {2014},
  month = {Mar},
  publisher = {American Physical Society},
  doi = {10.1103/PhysRevLett.112.103601},
  url = {https://link.aps.org/doi/10.1103/PhysRevLett.112.103601}
}

@article{Luo2025,
author={ Luo, Chengyi and Zhang, Haoqing and Chu, Anjun and Maruko, Chitose and Rey, Ana Maria and Thompson, James K.},
year = {2025},
title = {Hamiltonian engineering of collective XYZ spin models in an optical cavity},
journal = {Nature Physics},
pages = {916-923},
volume = {21},
issue = {6},
ISSN = {1745-2481},
url = {https://doi.org/10.1038/s41567-025-02866-0},
doi = {10.1038/s41567-025-02866-0}
}

@article{Bohnet_2016,
   title={Quantum spin dynamics and entanglement generation with hundreds of trapped ions},
   volume={352},
   ISSN={1095-9203},
   url={http://dx.doi.org/10.1126/science.aad9958},
   DOI={10.1126/science.aad9958},
   number={6291},
   journal={Science},
   publisher={American Association for the Advancement of Science (AAAS)},
   author={Bohnet, Justin G. and Sawyer, Brian C. and Britton, Joseph W. and Wall, Michael L. and Rey, Ana Maria and Foss-Feig, Michael and Bollinger, John J.},
   year={2016},
   month=jun, pages={1297–1301} }

@article{Gross_2010,
   title={Nonlinear atom interferometer surpasses classical precision limit},
   volume={464},
   ISSN={1476-4687},
   url={http://dx.doi.org/10.1038/nature08919},
   DOI={10.1038/nature08919},
   number={7292},
   journal={Nature},
   publisher={Springer Science and Business Media LLC},
   author={Gross, C. and Zibold, T. and Nicklas, E. and Estève, J. and Oberthaler, M. K.},
   year={2010},
   month=mar, pages={1165–1169} }

@article{Riedel_2010,
   title={Atom-chip-based generation of entanglement for quantum metrology},
   volume={464},
   ISSN={1476-4687},
   url={http://dx.doi.org/10.1038/nature08988},
   DOI={10.1038/nature08988},
   number={7292},
   journal={Nature},
   publisher={Springer Science and Business Media LLC},
   author={Riedel, Max F. and Böhi, Pascal and Li, Yun and Hänsch, Theodor W. and Sinatra, Alice and Treutlein, Philipp},
   year={2010},
   month=mar, pages={1170–1173} }

@article{Appel_2009,
   title={Mesoscopic atomic entanglement for precision measurements beyond the standard quantum limit},
   volume={106},
   ISSN={1091-6490},
   url={http://dx.doi.org/10.1073/pnas.0901550106},
   DOI={10.1073/pnas.0901550106},
   number={27},
   journal={Proceedings of the National Academy of Sciences},
   publisher={Proceedings of the National Academy of Sciences},
   author={Appel, J. and Windpassinger, P. J. and Oblak, D. and Hoff, U. B. and Kjærgaard, N. and Polzik, E. S.},
   year={2009},
   month=jul, pages={10960–10965} }

@article{Bohnet_2014,
   title={Reduced spin measurement back-action for a phase sensitivity ten times beyond the standard quantum limit},
   volume={8},
   ISSN={1749-4893},
   url={http://dx.doi.org/10.1038/nphoton.2014.151},
   DOI={10.1038/nphoton.2014.151},
   number={9},
   journal={Nature Photonics},
   publisher={Springer Science and Business Media LLC},
   author={Bohnet, J. G. and Cox, K. C. and Norcia, M. A. and Weiner, J. M. and Chen, Z. and Thompson, J. K.},
   year={2014},
   month=jul, pages={731–736} }

@article{Hosten2016,
author = {Hosten, Onur and Engelsen, Nils J. and Krishnakumar, Rajiv and Kasevich, Mark A.},
year = {2016},
title = {Measurement noise 100 times lower than the quantum-projection limit using entangled atoms},
journal = {Nature},
pages = {505-508},
volume = {529},
issue = {7587},
IsSN  = {1476-4687},
url = {https://doi.org/10.1038/nature16176},
doi = {10.1038/nature16176},
}

@article{Bao_2020,
   title={Spin squeezing of 1011 atoms by prediction and retrodiction measurements},
   volume={581},
   ISSN={1476-4687},
   url={http://dx.doi.org/10.1038/s41586-020-2243-7},
   DOI={10.1038/s41586-020-2243-7},
   number={7807},
   journal={Nature},
   publisher={Springer Science and Business Media LLC},
   author={Bao, Han and Duan, Junlei and Jin, Shenchao and Lu, Xingda and Li, Pengxiong and Qu, Weizhi and Wang, Mingfeng and Novikova, Irina and Mikhailov, Eugeniy E. and Zhao, Kai-Feng and Mølmer, Klaus and Shen, Heng and Xiao, Yanhong},
   year={2020},
   month=may, pages={159–163} }

@article{Wildermuth2006,
    author = {Wildermuth, S. and Hofferberth, S. and Lesanovsky, I. and Groth, S. and Krüger, P. and Schmiedmayer, J. and Bar-Joseph, I.},
    title = {Sensing electric and magnetic fields with Bose-Einstein condensates},
    journal = {Applied Physics Letters},
    volume = {88},
    number = {26},
    pages = {264103},
    year = {2006},
    month = {06},
    issn = {0003-6951},
    doi = {10.1063/1.2216932},
    url = {https://doi.org/10.1063/1.2216932},
}

@article{Mao2023,
author = {Mao, Tian-Wei and Liu, Qi and Li, Xin-Wei and Cao, Jia-Hao and Chen, Feng and Xu, Wen-Xin and Tey, Meng Khoon and Huang, Yi-Xiao and You, Li},
year = {2023},
title = {Quantum-enhanced sensing by echoing spin-nematic squeezing in atomic Bose–Einstein condensate},
journal = {Nature Physics},
pages = {1585-1590},
volume = {19},
issue = {11},
ISSN  = {1745-2481},
url = {https://doi.org/10.1038/s41567-023-02168-3},
doi = {10.1038/s41567-023-02168-3}
}

@article{Rudolph2003,
  title = {Quantum Communication Complexity of Establishing a Shared Reference Frame},
  author = {Rudolph, Terry and Grover, Lov},
  journal = {Phys. Rev. Lett.},
  volume = {91},
  issue = {21},
  pages = {217905},
  numpages = {4},
  year = {2003},
  month = {Nov},
  publisher = {American Physical Society},
  doi = {10.1103/PhysRevLett.91.217905},
  url = {https://link.aps.org/doi/10.1103/PhysRevLett.91.217905}
}

@article{Kessler2014,
  title = {Heisenberg-Limited Atom Clocks Based on Entangled Qubits},
  author = {Kessler, E. M. and K\'om\'ar, P. and Bishof, M. and Jiang, L. and S\o{}rensen, A. S. and Ye, J. and Lukin, M. D.},
  journal = {Phys. Rev. Lett.},
  volume = {112},
  issue = {19},
  pages = {190403},
  numpages = {5},
  year = {2014},
  month = {May},
  publisher = {American Physical Society},
  doi = {10.1103/PhysRevLett.112.190403},
  url = {https://link.aps.org/doi/10.1103/PhysRevLett.112.190403}
}

@article{Komar_2014,
   title={A quantum network of clocks},
   volume={10},
   ISSN={1745-2481},
   url={http://dx.doi.org/10.1038/nphys3000},
   DOI={10.1038/nphys3000},
   number={8},
   journal={Nature Physics},
   publisher={Springer Science and Business Media LLC},
   author={Kómár, P. and Kessler, E. M. and Bishof, M. and Jiang, L. and Sørensen, A. S. and Ye, J. and Lukin, M. D.},
   year={2014},
   month=jun, pages={582–587} }

@article{Kaubruegger2021,
  title = {Quantum Variational Optimization of Ramsey Interferometry and Atomic Clocks},
  author = {Kaubruegger, Raphael and Vasilyev, Denis V. and Schulte, Marius and Hammerer, Klemens and Zoller, Peter},
  journal = {Phys. Rev. X},
  volume = {11},
  issue = {4},
  pages = {041045},
  numpages = {21},
  year = {2021},
  month = {Dec},
  publisher = {American Physical Society},
  doi = {10.1103/PhysRevX.11.041045},
  url = {https://link.aps.org/doi/10.1103/PhysRevX.11.041045}
}

@article{Marciniak2022,
author = {Marciniak, Christian D. and Feldker, Thomas and Pogorelov, Ivan and Kaubruegger, Raphael and Vasilyev, Denis V. and van Bijnen, Rick and Schindler, Philipp and Zoller, Peter and Blatt, Rainer and Monz, Thomas},
year = {2022},
title  = {Optimal metrology with programmable quantum sensors},
journal = {Nature},
pages = {604-609},
volume = {603},
issue  = {7902},
ISSN = {1476-4687},
url  = {https://doi.org/10.1038/s41586-022-04435-4},
doi = {10.1038/s41586-022-04435-4}
}

@article{Kaubruegger2023,
  title = {Optimal and Variational Multiparameter Quantum Metrology and Vector-Field Sensing},
  author = {Kaubruegger, Raphael and Shankar, Athreya and Vasilyev, Denis V. and Zoller, Peter},
  journal = {PRX Quantum},
  volume = {4},
  issue = {2},
  pages = {020333},
  numpages = {21},
  year = {2023},
  month = {Jun},
  publisher = {American Physical Society},
  doi = {10.1103/PRXQuantum.4.020333},
  url = {https://link.aps.org/doi/10.1103/PRXQuantum.4.020333}
}

@article{Thurtell2024,
  title = {Optimizing one-axis twists for variational Bayesian quantum metrology},
  author = {Thurtell, Tyler G. and Miyake, Akimasa},
  journal = {Phys. Rev. Res.},
  volume = {6},
  issue = {2},
  pages = {023179},
  numpages = {18},
  year = {2024},
  month = {May},
  publisher = {American Physical Society},
  doi = {10.1103/PhysRevResearch.6.023179},
  url = {https://link.aps.org/doi/10.1103/PhysRevResearch.6.023179}
}

@misc{kielinski2025,
      title={Bayesian Frequency Metrology with Optimal Ramsey Interferometry in Optical Atomic Clocks}, 
      author={Timm Kielinski and Klemens Hammerer},
      year={2025},
      eprint={2505.04287},
      archivePrefix={arXiv},
      primaryClass={quant-ph},
      url={https://arxiv.org/abs/2505.04287}, 
}

@article{Giovannetti_2011,
   title={Advances in quantum metrology},
   volume={5},
   ISSN={1749-4893},
   url={http://dx.doi.org/10.1038/nphoton.2011.35},
   DOI={10.1038/nphoton.2011.35},
   number={4},
   journal={Nature Photonics},
   publisher={Springer Science and Business Media LLC},
   author={Giovannetti, Vittorio and Lloyd, Seth and Maccone, Lorenzo},
   year={2011},
   month=mar, pages={222–229} }

@article{Schindler2011,
author = {Philipp Schindler  and Julio T. Barreiro  and Thomas Monz  and Volckmar Nebendahl  and Daniel Nigg  and Michael Chwalla  and Markus Hennrich  and Rainer Blatt },
title = {Experimental Repetitive Quantum Error Correction},
journal = {Science},
volume = {332},
number = {6033},
pages = {1059-1061},
year = {2011},
doi = {10.1126/science.1203329},
URL = {https://www.science.org/doi/abs/10.1126/science.1203329}
}

@article{Negnevitsky_2018,
   title={Repeated multi-qubit readout and feedback with a mixed-species trapped-ion register},
   volume={563},
   ISSN={1476-4687},
   url={http://dx.doi.org/10.1038/s41586-018-0668-z},
   DOI={10.1038/s41586-018-0668-z},
   number={7732},
   journal={Nature},
   publisher={Springer Science and Business Media LLC},
   author={Negnevitsky, V. and Marinelli, M. and Mehta, K. K. and Lo, H.-Y. and Flühmann, C. and Home, J. P.},
   year={2018},
   month=nov, pages={527–531} }

@article{Ryan_Anderson2021,
  title = {Realization of Real-Time Fault-Tolerant Quantum Error Correction},
  author = {Ryan-Anderson, C. and Bohnet, J. G. and Lee, K. and Gresh, D. and Hankin, A. and Gaebler, J. P. and Francois, D. and Chernoguzov, A. and Lucchetti, D. and Brown, N. C. and Gatterman, T. M. and Halit, S. K. and Gilmore, K. and Gerber, J. A. and Neyenhuis, B. and Hayes, D. and Stutz, R. P.},
  journal = {Phys. Rev. X},
  volume = {11},
  issue = {4},
  pages = {041058},
  numpages = {29},
  year = {2021},
  month = {Dec},
  publisher = {American Physical Society},
  doi = {10.1103/PhysRevX.11.041058},
  url = {https://link.aps.org/doi/10.1103/PhysRevX.11.041058}
}

@article{Egan2021,
author = {Egan, Laird and Debroy, Dripto M. and Noel, Crystal and Risinger, Andrew and Zhu, Daiwei and Biswas, Debopriyo and Newman, Michael and Li, Muyuan and Brown, Kenneth R. and Cetina, Marko and Monroe, Christopher},
year = {2021},
title = {Fault-tolerant control of an error-corrected qubit},
journal = {Nature},
pages = {281-286},
volume = {598},
issue = {7880},
ISSN = {1476-4687},
url = {https://doi.org/10.1038/s41586-021-03928-y},
doi = {10.1038/s41586-021-03928-y}
}

@article{Graham2023,
  title = {Midcircuit Measurements on a Single-Species Neutral Alkali Atom Quantum Processor},
  author = {Graham, T. M. and Phuttitarn, L. and Chinnarasu, R. and Song, Y. and Poole, C. and Jooya, K. and Scott, J. and Scott, A. and Eichler, P. and Saffman, M.},
  journal = {Phys. Rev. X},
  volume = {13},
  issue = {4},
  pages = {041051},
  numpages = {22},
  year = {2023},
  month = {Dec},
  publisher = {American Physical Society},
  doi = {10.1103/PhysRevX.13.041051},
  url = {https://link.aps.org/doi/10.1103/PhysRevX.13.041051}
}

@article{Deist2022,
  title = {Mid-Circuit Cavity Measurement in a Neutral Atom Array},
  author = {Deist, Emma and Lu, Yue-Hui and Ho, Jacquelyn and Pasha, Mary Kate and Zeiher, Johannes and Yan, Zhenjie and Stamper-Kurn, Dan M.},
  journal = {Phys. Rev. Lett.},
  volume = {129},
  issue = {20},
  pages = {203602},
  numpages = {7},
  year = {2022},
  month = {Nov},
  publisher = {American Physical Society},
  doi = {10.1103/PhysRevLett.129.203602},
  url = {https://link.aps.org/doi/10.1103/PhysRevLett.129.203602}
}

@article{Singh_2023,
   title={Mid-circuit correction of correlated phase errors using an array of spectator qubits},
   volume={380},
   ISSN={1095-9203},
   url={http://dx.doi.org/10.1126/science.ade5337},
   DOI={10.1126/science.ade5337},
   number={6651},
   journal={Science},
   publisher={American Association for the Advancement of Science (AAAS)},
   author={Singh, K. and Bradley, C. E. and Anand, S. and Ramesh, V. and White, R. and Bernien, H.},
   year={2023},
   month=jun, pages={1265–1269} }

@article{Lis2023,
  title = {Midcircuit Operations Using the omg Architecture in Neutral Atom Arrays},
  author = {Lis, Joanna W. and Senoo, Aruku and McGrew, William F. and R\"onchen, Felix and Jenkins, Alec and Kaufman, Adam M.},
  journal = {Phys. Rev. X},
  volume = {13},
  issue = {4},
  pages = {041035},
  numpages = {22},
  year = {2023},
  month = {Nov},
  publisher = {American Physical Society},
  doi = {10.1103/PhysRevX.13.041035},
  url = {https://link.aps.org/doi/10.1103/PhysRevX.13.041035}
}

@article{Norcia2023,
  title = {Midcircuit Qubit Measurement and Rearrangement in a $^{171}\mathrm{Yb}$ Atomic Array},
  author = {Norcia, M. A. and \textit{et al.}},
  journal = {Phys. Rev. X},
  volume = {13},
  issue = {4},
  pages = {041034},
  numpages = {12},
  year = {2023},
  month = {Nov},
  publisher = {American Physical Society},
  doi = {10.1103/PhysRevX.13.041034},
  url = {https://link.aps.org/doi/10.1103/PhysRevX.13.041034}
}

@article{Huie2023,
  title = {Repetitive Readout and Real-Time Control of Nuclear Spin Qubits in ${}^{171}\mathrm{Yb}$ Atoms},
  author = {Huie, William and Li, Lintao and Chen, Neville and Hu, Xiye and Jia, Zhubing and Sun, Won Kyu Calvin and Covey, Jacob P.},
  journal = {PRX Quantum},
  volume = {4},
  issue = {3},
  pages = {030337},
  numpages = {28},
  year = {2023},
  month = {Sep},
  publisher = {American Physical Society},
  doi = {10.1103/PRXQuantum.4.030337},
  url = {https://link.aps.org/doi/10.1103/PRXQuantum.4.030337}
}

@article{Leroux2010,
  title = {Implementation of Cavity Squeezing of a Collective Atomic Spin},
  author = {Leroux, Ian D. and Schleier-Smith, Monika H. and Vuleti\ifmmode \acute{c}\else \'{c}\fi{}, Vladan},
  journal = {Phys. Rev. Lett.},
  volume = {104},
  issue = {7},
  pages = {073602},
  numpages = {4},
  year = {2010},
  month = {Feb},
  publisher = {American Physical Society},
  doi = {10.1103/PhysRevLett.104.073602},
  url = {https://link.aps.org/doi/10.1103/PhysRevLett.104.073602}
}

@article{Colombo_2022,
   title={Time-reversal-based quantum metrology with many-body entangled states},
   volume={18},
   ISSN={1745-2481},
   url={http://dx.doi.org/10.1038/s41567-022-01653-5},
   DOI={10.1038/s41567-022-01653-5},
   number={8},
   journal={Nature Physics},
   publisher={Springer Science and Business Media LLC},
   author={Colombo, Simone and Pedrozo-Peñafiel, Edwin and Adiyatullin, Albert F. and Li, Zeyang and Mendez, Enrique and Shu, Chi and Vuletić, Vladan},
   year={2022},
   month=jul, pages={925–930} }

@article{Li_2022,
   title={Collective Spin-Light and Light-Mediated Spin-Spin Interactions in an Optical Cavity},
   volume={3},
   ISSN={2691-3399},
   url={http://dx.doi.org/10.1103/PRXQuantum.3.020308},
   DOI={10.1103/prxquantum.3.020308},
   number={2},
   journal={PRX Quantum},
   publisher={American Physical Society (APS)},
   author={Li, Zeyang and Braverman, Boris and Colombo, Simone and Shu, Chi and Kawasaki, Akio and Adiyatullin, Albert F. and Pedrozo-Peñafiel, Edwin and Mendez, Enrique and Vuletić, Vladan},
   year={2022},
   month=apr }

@article{Derevianko_2014,
   title={Hunting for topological dark matter with atomic clocks},
   volume={10},
   ISSN={1745-2481},
   url={http://dx.doi.org/10.1038/nphys3137},
   DOI={10.1038/nphys3137},
   number={12},
   journal={Nature Physics},
   publisher={Springer Science and Business Media LLC},
   author={Derevianko, A. and Pospelov, M.},
   year={2014},
   month=nov, pages={933–936} }

@article{Safronova2018,
  title = {Search for new physics with atoms and molecules},
  author = {Safronova, M. S. and Budker, D. and DeMille, D. and Kimball, Derek F. Jackson and Derevianko, A. and Clark, Charles W.},
  journal = {Rev. Mod. Phys.},
  volume = {90},
  issue = {2},
  pages = {025008},
  numpages = {106},
  year = {2018},
  month = {Jun},
  publisher = {American Physical Society},
  doi = {10.1103/RevModPhys.90.025008},
  url = {https://link.aps.org/doi/10.1103/RevModPhys.90.025008}
}

@article{Kolkowitz2016,
  title = {Gravitational wave detection with optical lattice atomic clocks},
  author = {Kolkowitz, S. and Pikovski, I. and Langellier, N. and Lukin, M. D. and Walsworth, R. L. and Ye, J.},
  journal = {Phys. Rev. D},
  volume = {94},
  issue = {12},
  pages = {124043},
  numpages = {15},
  year = {2016},
  month = {Dec},
  publisher = {American Physical Society},
  doi = {10.1103/PhysRevD.94.124043},
  url = {https://link.aps.org/doi/10.1103/PhysRevD.94.124043}
}

@article{Mehlstäubler_2018,
doi = {10.1088/1361-6633/aab409},
url = {https://dx.doi.org/10.1088/1361-6633/aab409},
year = {2018},
month = {apr},
publisher = {IOP Publishing},
volume = {81},
number = {6},
pages = {064401},
author = {Mehlstäubler, Tanja E and Grosche, Gesine and Lisdat, Christian and Schmidt, Piet O and Denker, Heiner},
title = {Atomic clocks for geodesy},
journal = {Reports on Progress in Physics}
}

@article{Grotti2018,
author = {Grotti, Jacopo and \textit{et al.}},
year = {2018},
title = {Geodesy and metrology with a transportable optical clock},
journal = {Nature Physics},
pages = {437-441},
volume = {14},
issue = {5},
ISSN = {1745-2481},
url = {https://doi.org/10.1038/s41567-017-0042-3},
doi = {10.1038/s41567-017-0042-3}
}

@misc{rosenband2013,
      title={Exponential scaling of clock stability with atom number}, 
      author={T. Rosenband and D. R. Leibrandt},
      year={2013},
      eprint={1303.6357},
      archivePrefix={arXiv},
      primaryClass={quant-ph},
      url={https://arxiv.org/abs/1303.6357}, 
}

@article{Shaw2024,
author = {Shaw, Adam L. and Finkelstein, Ran and Tsai, Richard Bing-Shiun and Scholl, Pascal and Yoon, Tai Hyun and Choi, Joonhee and Endres, Manuel},
year = {2024},
title = {Multi-ensemble metrology by programming local rotations with atom movements},
journal = {Nature Physics},
pages = {195-201},
volume = {20},
issue = {2},
ISSN = {1745-2481},
url = {https://doi.org/10.1038/s41567-023-02323-w},
doi = {10.1038/s41567-023-02323-w}
}

@article{Borregaard2013,
  title = {Efficient Atomic Clocks Operated with Several Atomic Ensembles},
  author = {Borregaard, J. and S\o{}rensen, A. S.},
  journal = {Phys. Rev. Lett.},
  volume = {111},
  issue = {9},
  pages = {090802},
  numpages = {5},
  year = {2013},
  month = {Aug},
  publisher = {American Physical Society},
  doi = {10.1103/PhysRevLett.111.090802},
  url = {https://link.aps.org/doi/10.1103/PhysRevLett.111.090802}
}

@article{Roussy_2023,
   title={An improved bound on the electron’s electric dipole moment},
   volume={381},
   ISSN={1095-9203},
   url={http://dx.doi.org/10.1126/science.adg4084},
   DOI={10.1126/science.adg4084},
   number={6653},
   journal={Science},
   publisher={American Association for the Advancement of Science (AAAS)},
   author={Roussy, Tanya S. and Caldwell, Luke and Wright, Trevor and Cairncross, William B. and Shagam, Yuval and Ng, Kia Boon and Schlossberger, Noah and Park, Sun Yool and Wang, Anzhou and Ye, Jun and Cornell, Eric A.},
   year={2023},
   month=jul, pages={46–50} }

@article{Leibfried2004,
author = {D. Leibfried  and M. D. Barrett  and T. Schaetz  and J. Britton  and J. Chiaverini  and W. M. Itano  and J. D. Jost  and C. Langer  and D. J. Wineland },
title = {Toward Heisenberg-Limited Spectroscopy with Multiparticle Entangled States},
journal = {Science},
volume = {304},
number = {5676},
pages = {1476-1478},
year = {2004},
doi = {10.1126/science.1097576},
URL = {https://www.science.org/doi/abs/10.1126/science.1097576}}

@article{Leibfried2005,
author = {Leibfried, D. and  Knill, E. and Seidelin, S. and Britton, J. and Blakestad, R. B. and Chiaverini, J. and Hume, D. B. and Itano, W. M. and Jost, J. D. and Langer, C. and Ozeri, R. and Reichle, R. and Wineland, D. J.},
year = {2005},
title = {Creation of a six-atom ‘Schrödinger cat’ state},
journal = {Nature},
pages = {639-642},
volume = {438},
issue = {7068},
ISSN = {1476-4687},
url = {https://doi.org/10.1038/nature04251},
doi = {10.1038/nature04251}
}

@article{Davis2016,
  title = {Approaching the Heisenberg Limit without Single-Particle Detection},
  author = {Davis, Emily and Bentsen, Gregory and Schleier-Smith, Monika},
  journal = {Phys. Rev. Lett.},
  volume = {116},
  issue = {5},
  pages = {053601},
  numpages = {5},
  year = {2016},
  month = {Feb},
  publisher = {American Physical Society},
  doi = {10.1103/PhysRevLett.116.053601},
  url = {https://link.aps.org/doi/10.1103/PhysRevLett.116.053601}
}

@article{Frowis2016,
  title = {Detecting Large Quantum Fisher Information with Finite Measurement Precision},
  author = {Fr\"owis, Florian and Sekatski, Pavel and D\"ur, Wolfgang},
  journal = {Phys. Rev. Lett.},
  volume = {116},
  issue = {9},
  pages = {090801},
  numpages = {5},
  year = {2016},
  month = {Mar},
  publisher = {American Physical Society},
  doi = {10.1103/PhysRevLett.116.090801},
  url = {https://link.aps.org/doi/10.1103/PhysRevLett.116.090801}
}

@article{Marci2016,
  title = {Loschmidt echo for quantum metrology},
  author = {Macr\`{\i}, Tommaso and Smerzi, Augusto and Pezz\`e, Luca},
  journal = {Phys. Rev. A},
  volume = {94},
  issue = {1},
  pages = {010102},
  numpages = {6},
  year = {2016},
  month = {Jul},
  publisher = {American Physical Society},
  doi = {10.1103/PhysRevA.94.010102},
  url = {https://link.aps.org/doi/10.1103/PhysRevA.94.010102}
}

@article{Nolan2017,
  title = {Optimal and Robust Quantum Metrology Using Interaction-Based Readouts},
  author = {Nolan, Samuel P. and Szigeti, Stuart S. and Haine, Simon A.},
  journal = {Phys. Rev. Lett.},
  volume = {119},
  issue = {19},
  pages = {193601},
  numpages = {7},
  year = {2017},
  month = {Nov},
  publisher = {American Physical Society},
  doi = {10.1103/PhysRevLett.119.193601},
  url = {https://link.aps.org/doi/10.1103/PhysRevLett.119.193601}
}

@article{Li_2023,
   title={Improving metrology with quantum scrambling},
   volume={380},
   ISSN={1095-9203},
   url={http://dx.doi.org/10.1126/science.adg9500},
   DOI={10.1126/science.adg9500},
   number={6652},
   journal={Science},
   publisher={American Association for the Advancement of Science (AAAS)},
   author={Li, Zeyang and Colombo, Simone and Shu, Chi and Velez, Gustavo and Pilatowsky-Cameo, Saúl and Schmied, Roman and Choi, Soonwon and Lukin, Mikhail and Pedrozo-Peñafiel, Edwin and Vuletić, Vladan},
   year={2023},
   month=jun, pages={1381–1384} }

@article{Scharnagl_2023,
   title={Optimal Ramsey interferometry with echo protocols based on one-axis twisting},
   volume={108},
   ISSN={2469-9934},
   url={http://dx.doi.org/10.1103/PhysRevA.108.062611},
   DOI={10.1103/physreva.108.062611},
   number={6},
   journal={Physical Review A},
   publisher={American Physical Society (APS)},
   author={Scharnagl, M. S. and Kielinski, T. and Hammerer, K.},
   year={2023},
   month=dec }

@article{Andr__2004,
   title={Stability of Atomic Clocks Based on Entangled Atoms},
   volume={92},
   ISSN={1079-7114},
   url={http://dx.doi.org/10.1103/PhysRevLett.92.230801},
   DOI={10.1103/physrevlett.92.230801},
   number={23},
   journal={Physical Review Letters},
   publisher={American Physical Society (APS)},
   author={André, A. and Sørensen, A. S. and Lukin, M. D.},
   year={2004},
   month=jun }

@article{Huang_2008,
  title = {Optimized Double-Well Quantum Interferometry with Gaussian Squeezed States},
  author = {Huang, Y. P. and Moore, M. G.},
  journal = {Phys. Rev. Lett.},
  volume = {100},
  issue = {25},
  pages = {250406},
  numpages = {4},
  year = {2008},
  month = {Jun},
  publisher = {American Physical Society},
  doi = {10.1103/PhysRevLett.100.250406},
  url = {https://link.aps.org/doi/10.1103/PhysRevLett.100.250406}
}

@article{Liu_2025,
  title = {Enhancing Dynamic Range of Sub-Standard-Quantum-Limit Measurements via Quantum Deamplification},
  author = {Liu, Qi and Xue, Ming and Radzihovsky, Matthew and Li, Xinwei and Vasilyev, Denis V. and Wu, Ling-Na and Vuleti\ifmmode \acute{c}\else \'{c}\fi{}, Vladan},
  journal = {Phys. Rev. Lett.},
  volume = {135},
  issue = {4},
  pages = {040801},
  numpages = {8},
  year = {2025},
  month = {Jul},
  publisher = {American Physical Society},
  doi = {10.1103/25ds-9724},
  url = {https://link.aps.org/doi/10.1103/25ds-9724}
}

\appendix

\begin{widetext}

\section{Phase estimation algorithm for weakly non-Gaussian states}
\label{app:opt}
Here we argue that the scaling behavior of the phase estimation algorithm studied previously~\cite{Pezze2020, Pezze2021, Huang_2008} is optimal not just for GSS states but for a large class of states which we call weakly non-Gaussian. By optimal we mean that if we fix the prior standard deviation $\sigma$, then the $N$-dependence of the $\Delta J_{y}^{2}$ value that minimizes $\overline{\Xi^{2}}$ will depend on $N$ in the same way for all these states. We will also see that optimal $\Delta J_{y}^{2}$ value depends on fourth moments of angular momentum components for these states. We will call a state $|\psi\rangle$ weakly non-Gaussian if it satisfies
\begin{align}
\label{eq:inequal}
1&\gg\left\langle\left(\frac{4J_{y}^{2}}{N^{2}}\right)^{n}\right\rangle>\left\langle\left(\frac{4J_{y}^{2}}{N^{2}}\right)^{n+1}\right\rangle, \\ 
1&\gg\left\langle\left(\frac{4J_{z}^{2}}{N^{2}}\right)^{n}\right\rangle>\left\langle\left(\frac{4J_{z}^{2}}{N^{2}}\right)^{n+1}\right\rangle,
\end{align}
where $n\in\mathbb{N}$, satisfies $\langle J_{y}\rangle=\langle J_{z}\rangle=0$, as implied, for example, by x-parity symmetry $\left(\prod_{j}\sigma_{j}^{(x)}\right)|\psi\rangle=|\psi\rangle$, and exhibits bounded Kurtosis in the y and z-directions, i.e.
\begin{align}
0<W_{1,y}&\leq\textrm{Kurt}[J_{y}]-1\leq W_{2,y} \\
0<W_{1,z}&\leq\textrm{Kurt}[J_{z}]-1\leq W_{2,z}
\end{align}
for some $W_{1/2,y}$ and $W_{1/2,z}$ that do not depend on $N$ and the kurtosis is defined by
\begin{equation}
\textrm{Kurt}[\mathcal{O}]\equiv\bigg\langle\left(\frac{\langle \mathcal{O}-\langle\mathcal{O}\rangle}{\Delta\mathcal{O}}\right)^{4}\bigg\rangle.
\end{equation}
Note that only the leftmost inequality in Eq.~\eqref{eq:inequal} is an assumption while the second inequality is always true for a positive operator whose eigenvalues values are less than one. We use this assumption to keep only the lowest order terms in expectation values of power series of these operators. We call a state a weakly non-Gaussian squeezed state if in addition to these properties it satisfies
\begin{equation}
\Delta J_{y}^{2}\ll\Delta J_{z}^{2}.
\end{equation}
and has vanishing covariance between $J_{y}$ and $J_{z}$. This second is not stringent since it is natural to apply an x-rotation to any probe until $\Delta J_{y}^{2}$ is minimized prior to the $\phi$-dependent z-rotation. The resulting state will satisfy the vanishing covariance condition.

The property of a state being weakly non-Gaussian implies an approximate form the $J_{x}$ expectation value given by 
\begin{equation}
\langle J_{x}\rangle=\left\langle\sqrt{\frac{N}{2}\left(\frac{N}{2}+1\right)-J_{y}^{2}-J_{z}^{2}}\right\rangle\approx\frac{N}{2}\left[1-\frac{2}{N^{2}}(\Delta J_{y}^{2}+\Delta J_{z}^{2})-\frac{2}{N^{4}}\left(\langle J_{y}^{4}\rangle+\langle J_{z}^{4}\rangle+\langle\{J_{y}^{2},J_{z}^{2}\rangle\}\right)\right],
\end{equation}
where the approximation drops smaller terms in the Taylor expansion of the square root. We also know that
\begin{equation}
\langle J_{x}^{2}\rangle\approx\frac{N}{2}-\langle J_{y}^{2}\rangle-\langle J_{z}^{2}\rangle.
\end{equation}
This means that the variance of $J_{x}$ is
\begin{equation}
\Delta J_{x}^{2}\approx\frac{1}{N^{2}}\left[(\langle J_{y}^{4}\rangle-\langle J_{y}^{2}\rangle^{2})+(\langle J_{z}^{4}\rangle-\langle J_{z}^{2}\rangle^{2})+2\textrm{Cov}(J_{y}^{2},J_{z}^{2})\right].
\end{equation}
For the states we are considering
\begin{equation}
\textrm{Kurt}[J_{y}]=\frac{\langle J_{y}^{4}\rangle}{\langle J_{y}^{2}\rangle^{2}}
\end{equation}
so that
\begin{equation}
\label{eq:xvar}
\Delta J_{x}^{2}\approx\frac{1}{N^{2}}\left[\Delta J_{y}^{4}\left(\textrm{Kurt}[J_{y}]-1\right)+\Delta J_{z}^{4}\left(\textrm{Kurt}[J_{z}]-1\right)+2\textrm{Cov}(J_{y}^{2},J_{z}^{2})\right].
\end{equation}
Then the rotated squeezing parameter takes the form
\begin{equation}
\overline{\Xi^{2}}\approx\frac{4\Delta J_{y}^{2}}{N}+\frac{\sigma^{2}}{N^{3}}\left[\Delta J_{y}^{4}\left(\textrm{Kurt}[J_{y}]-1\right)+\Delta J_{z}^{4}\left(\textrm{Kurt}[J_{z}]-1\right)+2\textrm{Cov}(J_{y}^{2},J_{z}^{2})\right]
\end{equation}
We assume that the first term is negligible since $J_{y}$ is the squeezed observable. If we were considering classical probability distributions, the assumption that $\textrm{Cov}(J_{y},J_{z})=0$ would be sufficient to imply that the third term vanished. However, this is not generally the case quantum mechanically. Quantum mechanics does imply
\begin{equation}
\textrm{Cov}(J_{y}^{2},J_{z}^{2})\leq\sqrt{(\Delta J_{y}^{2})^{2}(\Delta J_{z}^{2})^{2}}=\Delta J_{y}^{2}\Delta J_{z}^{2}\sqrt{(\textrm{Kurt}[J_{y}]-1)(\textrm{Kurt}[J_{z}]-1)},
\end{equation}
where $(\Delta J_{y}^{2})^{2}$ is the variance of $J_{y}^{2}$ and the equality is specific to weakly-Gaussian states. Since we assume that the first two terms in Eq.~\eqref{eq:xvar} scale differently with $N$, the third term must itself increase more slowly than the second as $N$ is increased. This means that for large enough system sizes it should be negligible. This reduces the rotated squeezing parameter to
\begin{equation}
\overline{\Xi^{2}}\approx\frac{4\Delta J_{y}^{2}}{N}+\frac{\sigma^{2}}{N^{3}}\left[\Delta J_{z}^{4}\left(\textrm{Kurt}[J_{z}]-1\right)\right]
\end{equation}
Our assumption of bounded kurtoses then implies
\begin{equation}
\frac{4\Delta J_{y}^{2}}{N}+\frac{\sigma^{2}\Delta J_{z}^{4}W_{1,z}}{N^{3}}\leq\overline{\Xi^{2}}\leq\frac{4\Delta J_{y}^{2}}{N}+\frac{\sigma^{2}\Delta J_{z}^{4}W_{2,z}}{N^{3}}
\end{equation}
so again we will focus on optimizing for a general constant $W_{z}$ between $W_{1,z}$ and $W_{2,z}$. If we then assume that the state is minimum uncertainty
\begin{equation}
\Delta J_{y}\Delta J_{z}\approx\frac{N}{4}.
\end{equation}
The rotated squeezing parameter is approximated by (see Eq.~\eqref{eq:rot_avg})
\begin{equation}
\overline{\Xi^{2}}\approx\frac{4\Delta J_{y}^{2}}{N}+\frac{N\sigma^{2}W_{z}}{64\Delta J_{y}^{4}}.
\end{equation}
The value of $\Delta J_{y}^{2}$ that minimizes this expression is
\begin{equation}
\Delta J_{y}^{2}\approx\left(\frac{N^{2}\sigma^{2}W_{z}}{2}\right)^{1/3},
\end{equation}
which is the same relation as the GSS states but adjusted by an amount that depends on the value of $W_{z}$. The associated values of the squeezing parameters are
\begin{align}
\xi^{2}&\approx\left(\frac{32\sigma^{2}W_{z}}{N^{}}\right)^{1/3} \\
\overline{\Xi^{2}}&\approx\left[32^{1/3}+\left(\frac{1}{16\times64^2}\right)^{1/3}\right]\left(\frac{\sigma^{2}W_{z}}{N}\right)^{1/3}.
\end{align}
Assuming that the initial prior standard deviation satisfies $\sigma\sim\frac{1}{\sqrt{N}}$, as will be the case if a coherent spin state is used to obtain a preliminary estimate, and each subsequent ensemble is subject to residual phase uncertainty given by the value of $\overline{\Xi^{2}}$ for the previous ensemble, we again recover the sequence of squeezing parameter $N$-dependences in Eq.~\eqref{eq:s_gss}.

\section{Properties of twisted Gaussian states}
\label{app:tgs}

In this appendix we compute approximation to the spin moments of the state
\begin{equation}
|GSS(N,s,\chi)\rangle=\frac{1}{\sqrt{\mathcal{N}}}\sum_{m=-N/2}^{N/2}e^{-m^{2}/(s^{2}N)-i\chi m^{2}}|J_{z}=m\rangle.
\end{equation}

\subsection{Evaluation of spin moments}
By $x$-parity symmetry we have
\begin{equation}
\langle J_{z}\rangle=\langle J_{y}\rangle=0.
\end{equation}
First, note that
\begin{equation}
J_{z}^{2}\approx\frac{s^{2}N}{4},
\end{equation}
where we have approximated the Gaussian sum for this expectation value by a Gaussian integral and assumed that the sum normalization is identical to the usual Gaussian integral normalization. We will make these types of approximation many times throughout this appendix so we will abbreviate them by INT ad NORM respectively. Next we need to evaluate
\begin{equation}
\langle J_{y}^{2}\rangle=-\frac{1}{4}\langle J_{+}^{2}+J_{-}^{2}-J_{+}J_{-}-J_{-}J_{+}\rangle.
\end{equation}
So we compute
\begin{equation}
\begin{aligned}
\langle J_{+}^{2}\rangle&=\frac{1}{\mathcal{N}}\sum_{m,m'}e^{-\frac{m^{2}}{s^{2}N}-i\chi m^{2}-\frac{m'}{s^{2}N}+i\chi m'^{2}}\langle m'|J_{+}^{2}|m\rangle \\
&\approx\frac{1}{\mathcal{N}}\left(\frac{N}{2}\right)^{2}\int dm e^{-\frac{m^{2}}{s^{2}N}-\frac{(m+2)^{2}}{s^{2}N}-i\chi m^{2}+i\chi(m+2)^{2}} && \text{(INT,HP)} \\
&\approx\left(\frac{N}{2}\right)^{2}e^{-\frac{2}{s^{2}N}-2\chi^{2}s^{2}N} && \text{(NORM)},
\end{aligned}
\end{equation}
where the text in parentheses indicates the approximations made on each line. We have also assumed that
\begin{equation}
\langle m'|J_{+}|m\rangle\approx\left(\frac{N}{2}\right)\delta_{m+1,m'},
\end{equation}
since we  assume that $\langle J_{z}\rangle\ll N$, which can be thought of as a form of the Holstein-Primakoff approximation so we denote it by HP. Since this is real
\begin{equation}
\langle J_{-}^{2}\rangle\approx\left(\frac{N}{2}\right)^{2}e^{-\frac{2}{s^{2}N}-2\chi^{2}s^{2}N}.
\end{equation}
To the level of precision we are currently working at 
\begin{equation}
\langle J_{+}J_{-}\rangle\approx\langle J_{-}J_{+}\rangle\approx\left(\frac{N}{2}\right)^{2},
\end{equation}
where the first approximation is justified by $\langle[J_{+},J_{-}]\rangle=2\langle J_{z}\rangle$ and $\langle J_{z}\rangle\ll N$. Thus
\begin{equation}
\langle J_{y}^{2}\rangle\approx\frac{N^{2}}{8}\left(1-e^{-\frac{2}{s^{2}N}-2\chi^{2}s^{2}N}\right).
\end{equation}
The next expectation value that we need is
\begin{equation}
\langle\{J_{y},J_{z}\}\rangle=\frac{1}{2i}\langle J_{+}J_{z}-J_{-}J_{z}+J_{z}J_{+}-J_{z}J_{-}\rangle.
\end{equation}
So we compute the expectation values
\begin{equation}
\begin{aligned}
\langle J_{+}J_{z}\rangle&=\frac{1}{\mathcal{N}}\sum_{m,m'}e^{-\frac{m^{2}}{s^{2}N}-i\chi m^{2}-\frac{m'}{s^{2}N}+i\chi m'^{2}}\langle m'|J_{+}J_{z}|m\rangle \\
&\approx\frac{1}{\mathcal{N}}\left(\frac{N}{2}\right)\int dm m e^{-\frac{m^{2}}{s^{2}N}-\frac{(m+1)^{2}}{s^{2}N}-i\chi m^{2}+i\chi(m+1)^{2}} && \text{(INT,HP)} \\
&\approx\left(\frac{N}{4}\right)(i\chi s^{2}N-1)e^{\frac{-1}{2s^{2}N}-\frac{1}{2}\chi^{2}s^{2}N} && \text{(NORM)}
\end{aligned}
\end{equation}
and
\begin{equation}
\begin{aligned}
\langle J_{z}J_{+}\rangle&=\frac{1}{\mathcal{N}}\sum_{m,m'}e^{-\frac{m^{2}}{s^{2}N}-i\chi m^{2}-\frac{m'}{s^{2}N}+i\chi m'^{2}}\langle m'|J_{z}J_{+}|m\rangle \\
&\approx\frac{1}{\mathcal{N}}\left(\frac{N}{2}\right)\int dm (m+1) e^{-\frac{m^{2}}{s^{2}N}-\frac{(m+1)^{2}}{s^{2}N}-i\chi m^{2}+i\chi(m+1)^{2}} && \text{(INT,HP)}\\
&\approx\left(\frac{N}{4}\right)(i\chi s^{2}N)e^{\frac{-1}{2s^{2}N}-\frac{1}{2}\chi^{2}s^{2}N} && \text{(NORM)}.
\end{aligned}
\end{equation}
Then altogether
\begin{equation}
\langle\{J_{y},J_{z}\}\rangle=\frac{1}{2}\chi s^{2}N^{2}e^{-\frac{1}{2s^{2}N}-\frac{1}{2}\chi^{2}s^{2}N}.
\end{equation}

It will be useful to also compute $\langle J_{x}\rangle$. To do this we just need to compute
\begin{equation}
\begin{aligned}
\langle J_{+}\rangle&=\frac{1}{\mathcal{N}}\sum_{m,m'}e^{-\frac{m^{2}}{s^{2}N}-i\chi m^{2}-\frac{m'^{2}}{s^{2}N}+i\chi m'^{2}}\langle m'|J_{+}|m\rangle \\
&\approx\frac{1}{\mathcal{N}}\left(\frac{N}{2}\right)\int dme^{-\frac{m^{2}}{s^{2}N}-i\chi m^{2}-\frac{(m+1)^{2}}{s^{2}N}+i\chi(m+1)^{2}} && \text{(INT,HP)} \\
&\approx\left(\frac{N}{2}\right)e^{-\frac{1}{2s^{2}N}-\frac{1}{2}\chi^{2}s^{2}N} && \text{(NORM)}.
\end{aligned}
\end{equation}
This tells us that 
\begin{equation}
\label{eq:ex_x}
\langle J_{x}\rangle\approx\frac{N}{2}e^{-\frac{1}{2s^{2}N}-\frac{1}{2}\chi^{2}s^{2}N}.
\end{equation}
Similarly, we can compute
\begin{equation}
\langle J_{x}^{2}\rangle\approx\frac{N^{2}}{8}\left(1+e^{-\frac{2}{s^{2}N}-2\chi^{2}s^{2}N}\right)
\end{equation}
so that
\begin{equation}
\begin{split}
\label{eq:x_expect}
\Delta J_{x}^{2}&\approx\frac{N^{2}}{8}\left(1+e^{-\frac{2}{s^{2}N}-2\chi^{2}s^{2}N}\right)-\frac{N^{2}}{4}e^{-\frac{1}{s^{2}N}-\chi^{2}s^{2}N} \\
&\approx\frac{N^{2}}{8}\left(\frac{1}{s^{4}N^{2}}+\chi^{4}s^{4}N^{2}+2\chi^{2}\right),
\end{split}
\end{equation}
where in the last approximation we have assumed both $\frac{1}{s^{2}N}\ll1$ and $\chi^{2}s^{2}N\ll1$.

\subsection{Minimum variance angular momentum direction}
The post-rotation form of $J_{y}$ is
\begin{equation}
J(\theta)\equiv e^{i\theta J_{x}}J_{y}e^{-i\theta J_{x}}=\cos(\theta)J_{y}-\sin(\theta)J_{z}.
\end{equation}
For the states we are considering, the expectation value of this is zero $\langle J(\theta)\rangle=0$. The square of this operator is
\begin{equation}
J(\theta)^{2}=\cos^{2}(\theta)J_{y}^{2}+\sin^{2}(\theta)J_{z}^{2}-\frac{1}{2}\sin(2\theta)\{J_{y},J_{z}\}.
\end{equation}
To find the optimal value of $\theta$ we differentiate the expectation value of this w.r.t. $\theta$. This gives
\begin{equation}
\frac{d\langle J(\theta)^{2}\rangle}{d\theta}=-\sin(2\theta)\langle J_{y}^{2}\rangle+\sin(2\theta)\langle J_{z}^{2}\rangle-\cos(2\theta)\langle\{J_{y},J_{z}\}\rangle=0
\end{equation}
so that the optimal value of $\theta$, denoted $\theta_{*}$ satisfies
\begin{equation}
\theta_{*}=\frac{1}{2}\arctan\left(\frac{\langle\{J_{y},J_{z}\}\rangle}{\langle J_{z}^{2}\rangle-\langle J_{y}^{2}\rangle}\right).
\end{equation}
Then the $\theta$ optimized expectation value of $J(\theta)^{2}$ is given by
\begin{equation}
\begin{split}
\langle J(\theta_{*})^{2}\rangle&=\cos^{2}(\theta_{*})\langle J_{y}^{2}\rangle+\sin^{2}(\theta_{*})\langle J_{z}^{2}\rangle-\frac{1}{2}\sin(2\theta_{*})\langle\{J_{y},J_{z}\}\rangle \\
&=\frac{1}{2}\left[\langle J_{z}^{2}\rangle+\langle J_{y}^{2}\rangle+\cos(2\theta_{*})\left(\langle J_{y}^{2}\rangle-\langle J_{z}^{2}\rangle\right)-\sin(2\theta_{*})\langle\{J_{y},J_{z}\}\rangle\right] \\
&=\frac{1}{2}\left[\langle J_{z}^{2}\rangle+\langle J_{y}^{2}\rangle-\sqrt{(\langle J_{y}^{2}\rangle-\langle J_{z}^{2}\rangle)^{2}+\langle\{J_{y},J_{z}\}\rangle^{2}}\right].
\end{split}
\end{equation}
If we then plug in the spin-moments we obtained in the last section, we obtain an approximate form for the minimal spin variance as
\begin{equation}
\begin{split}
\label{eq:expand}
\langle J(\theta_{*})^{2}&\rangle\approx\frac{N}{8}\left\{s^{2}+\frac{N}{2}\left(1-e^{-\frac{2}{s^{2}N}-2\chi^{2}s^{2}N}\right)-\sqrt{\left[\frac{N}{2}\left(1-e^{-\frac{2}{s^{2}N}-2\chi^{2}s^{2}N}\right)-s^{2}\right]^{2}+4\chi^{2}s^{4}N^{2}e^{-\frac{1}{s^{2}N}-\chi^{2}s^{2}N}}\right\} \\
&=\frac{N}{8}\left\{s^{2}+\frac{N}{2}\left(1-e^{-\frac{2}{s^{2}N}-2\chi^{2}s^{2}N}\right)-\left[\frac{N}{2}\left(1-e^{-\frac{2}{s^{2}N}-2\chi^{2}s^{2}N}\right)-s^{2}\right]\sqrt{1+\frac{4\chi^{2}s^{4}N^{2}e^{-\frac{1}{s^{2}N}-\chi^{2}s^{2}N}}{\left(\frac{N}{2}\left(1-e^{-\frac{2}{s^{2}N}-2\chi^{2}s^{2}N}\right)-s^{2}\right)^{2}}}\right\} \\
&\approx\frac{N}{4}\left[s^{2}-\frac{\chi^{2}s^{4}N^{2}e^{-\frac{1}{s^{2}N}-\chi^{2}s^{2}N}}{\frac{N}{2}\left(1-e^{-\frac{2}{s^{2}N}-2\chi^{2}s^{2}N}\right)-s^{2}}\right]\approx\frac{N}{4}\left[s^{2}-\frac{\chi^{2}s^{4}N}{\sinh\left(\frac{1}{s^{2}N}+\chi^{2}s^{2}N\right)}\right] \\
&\approx\frac{s^{2}N}{4}\left[1-\frac{1}{1+\frac{1}{\chi^{2}s^{4}N^{2}}+\frac{1}{6}\chi^{4}s^{4}N^{2}}\right]\approx\frac{s^{2}N}{4}\left[\frac{1}{\chi^{2}s^{4}N^{2}}+\frac{1}{6}\chi^{4}s^{4}N^{2}\right].
\end{split}
\end{equation}
The conditions for this variance to exhibit behavior we are interested in are
\begin{equation}
\frac{1}{N}<\frac{1}{s^{2}N}\ll\chi^{2}s^{2}N\ll1.
\end{equation}
If the first inequality is not satisfied, then the initial GSS state would have $s^{2}>1$ and would actually be anti-squeezed. If the second assumption is not satisfied, then the last term in the square root in the first line of Eq.~\ref{eq:expand} will be asymptotically negligible compared to the first term (due to the polynomial part of the term) and the variance tends to a $\chi$-independent value. If $\chi$ becomes too large the same thing happens (this time due to the exponential part of the term) so the last inequality is also necessary. Additionally, since $s^{2}<\ll$, we have
\begin{equation}
1\ll\chi^{2}s^{2}N^{2}\ll\chi^{2}N^{2}.
\end{equation}

The first approximation in Eq.~\eqref{eq:expand} just accounts for the approximate evaluations of the spin-moments. The second approximation is a binomial expansion of the square root. In order for it to be valid we assume that $\chi^{2}s^{4}N^{2}\ll\chi^{4}s^{4}N^{4}$ which as we've seen, is satisfied in the regime with the interesting physics. The third approximation follows from assuming that $s^{2}\ll\frac{1}{s^{2}}$ and $s^{2}\ll\chi^{2}s^{2}N^{2}$ which are also satisfied in the regime of interest. The fourth approximation is just a Taylor approximation of the hyperbolic sine function where we have assumed as $\frac{1}{s^{2}N}\ll\chi^{2}s^{2}N\ll1$. The final approximation is binomial expansion of the fraction within the brackets and does not make any additional assumptions. Note that if $s^{2}=1$ then this expression reduces to the one obtained by Kitagawa and Ueda~\cite{Kitagawa_1993} for the minimal variance obtainable when beginning from a spin-coherent state. We can now optimize this w.r.t. $\chi$ by differentiating
\begin{equation}
\label{eq:var_dif}
\frac{d\langle J(\theta_{*})^{2}\rangle}{d\chi}\approx\frac{\chi s^{4}N^{2}}{2\sinh\left(\frac{1}{s^{2}N}+\chi^{2}s^{2}N\right)}\left[\chi^{2}s^{2}N\coth\left(\frac{1}{s^{2}N}+\chi^{2}s^{2}N\right)-1\right]\approx\frac{s^{2}N}{4}\left[-\frac{2}{\chi^{3}s^{4}N^{2}}+\frac{2}{3}\chi^{3}s^{4}N^{2}\right]=0
\end{equation}
so that the optimal value of $\chi$, denoted $\chi_{*}$, is given by
\begin{equation}
\label{eq:chi_opt}
\chi_{*}\approx\frac{3^{1/6}}{(s^{2}N)^{2/3}}.
\end{equation}
The first approximation in Eq.~\eqref{eq:var_dif} is due to the use of the last expression on the third line of Eq.~\eqref{eq:expand} and the second approximation denotes the use of the last expression in Eq.~\eqref{eq:expand} as approximations for $\langle J(\theta_{*})\rangle$. The approximation in Eq.~\eqref{eq:chi_opt} is due to it being obtained from Eq.~\eqref{eq:var_dif} which is itself already approximate. We then obtain the minimum possible variance $J_{\textrm{min}}^{2}$ to be
\begin{equation}
\label{eq:j_min}
J_{\textrm{min}}^{2}\approx\frac{s^{2}N}{4}\left[\frac{(s^{4}N^{2})^{2/3}}{3^{1/3}s^{4}N^{2}}+\frac{3^{2/3}s^{4}N^{2}}{6(s^{4}N^{2})^{4/3}}\right]=\frac{(9s^{2}N)^{1/3}}{8},
\end{equation}
where the approximation just indicates the use of the approximate expressions in Eq.~\eqref{eq:expand} and Eq.~\eqref{eq:chi_opt}.

\subsection{Spin-squeezing parameters}

To compute the rotated and unrotated squeezing parameters, first, notice that to lowest order
\begin{equation}
\label{eq:ex_x_approx}
\langle J_{x}\rangle\approx\frac{N}{2},
\end{equation}
where we have used the approximate spin moment obtained above in Eq.~\eqref{eq:ex_x} and assumed that $\frac{1}{s^{2}N}\ll1$ and $\chi^{2}s^{2}N\ll1$. Since this is independent of $\chi$, the unrotated squeezing parameter is minimized at the same value of $\chi$ that minimizes the $\langle J(\theta_{*})^{2}\rangle$ and its minimal value is
\begin{equation}
\xi^{2}_{\textrm{min}}\approx\frac{3^{2/3}s^{2/3}}{2N^{2/3}},
\end{equation}
so that the squeezing decreases more quickly with system size than for the state that can be produced via a applying an OAT operation to an unentangled state if $s^{2}$ decreases with system size. The approximation in this expression is the use the approximation spin moments in Eq.~\eqref{eq:ex_x_approx} and Eq.~\eqref{eq:j_min}.

We can also consider the averaged rotated squeezing parameter, see Eq.~\eqref{eq:rot_avg},
\begin{equation}
\begin{split}
\label{eq:rot_xi_appr}
\overline{\Xi^{2}}&\approx\frac{N\Delta J_{y}^{2}}{\langle J_{x}\rangle^{2}}+\sigma^{2}\frac{N\Delta J_{x}^{2}}{\langle J_{x}\rangle^{2}}\approx\left[s^{2}-\frac{\chi^{2}s^{4}N}{\sinh\left(\frac{1}{s^{2}N}+\chi^{2}s^{2}N\right)}\right]+\frac{N\sigma^{2}}{2}\left(1+e^{-\frac{2}{s^{2}N}-2\chi^{2}s^{2}N}\right)-N\sigma^{2}e^{-\frac{1}{s^{2}N}-\chi^{2}s^{2}N} \\
&\approx\frac{1}{\chi^{2}s^{2}N^{2}}+\frac{1}{6}\chi^{4}s^{6}N^{2}+\frac{\sigma^{2}}{2}\left(\frac{1}{s^{4}N}+\chi^{4}s^{4}N^{3}+2\chi^{2}N\right),
\end{split}
\end{equation}
where the first approximation keeps only the lowest order term in $\sigma^{2}$ and the other two approximations use spin-moment approximations of different coarseness. Differentiating this gives us
\begin{equation}
\label{eq:state2}
\begin{aligned}
\frac{d\overline{\Xi^{2}}}{d\chi}&\approx-\frac{2\chi s^{4}N}{\sinh\left(\frac{1}{s^{2}N}+\chi^{2}s^{2}N\right)}+\frac{2\chi^{3}s^{6}N^{2}\coth\left(\frac{1}{s^{2}N}+\chi^{2}s^{2}N\right)}{\sinh\left(\frac{1}{s^{2}N}+\chi^{2}s^{2}N\right)}-2\sigma^{2}\chi s^{2}N^{2}\left(e^{-\frac{2}{s^{2}N}-2\chi^{2}s^{2}N}-e^{-\frac{1}{s^{2}N}-\chi^{2}s^{2}N}\right) \\
&\approx-\frac{2}{\chi^{3}s^{2}N^{2}}+\frac{2}{3}\chi^{3}s^{6}N^{2}+2\sigma^{2}\chi^{3}s^{4}N^{3},
\end{aligned}
\end{equation}
where the approximations are the same as the ones in Eq.~\eqref{eq:rot_xi_appr} and we have kept only the leading order term proportional to $\sigma^{2}$. This implies that the optimal value of $\chi$ is
\begin{equation}
\label{eq:chi_rot_opt}
\chi_{*}(\sigma)\approx\left(\frac{1}{\left(\frac{1}{3}s^{2}+N\sigma^{2}\right)s^{6}N^{4}}\right)^{1/6},
\end{equation}
where the approximation is using the final expression above. This reduces Eq.~\eqref{eq:rot_xi_appr} to
\begin{equation}
\label{eq:xi_rot_final}
\overline{\Xi_{*}^{2}}\approx\frac{(9s^{2}+27N\sigma^{2})^{1/3}}{2N^{2/3}}+\frac{\sigma^{2}}{2s^{4}N}+\frac{\sigma^{2}}{(\frac{1}{3}s^{2}+N\sigma^{2})^{1/3}s^{2}N^{1/3}}.
\end{equation}

Note that if $\sigma^{2}$ is a constant then the first term in denominator of Eq.~\eqref{eq:chi_rot_opt} can be dropped to give
\begin{equation}
\chi_{*}(\sigma)\approx\left(\frac{1}{\sigma^{2}s^{6}N^{5}}\right)^{1/6}.
\end{equation}
If $s^{2}$ is also a constant, then the corresponding value of Eq.~\eqref{eq:xi_rot_final} is
\begin{equation}
\overline{\Xi^{2}_{*}}\approx\left(\frac{\sigma}{N}\right)^{1/3},
\end{equation}
which is the best possible $N$-dependence for a single squeezed state. This agrees with previous analysis of the available system size dependence in the non-adaptive case~\cite{Andr__2004}.

\subsection{Multi-twist procedures}

Here we derive the twisting angles associated with multi-twist state preparation protocols. These procedures are based on assuming that after each OAT operation and \textit{x}-rotation pair the state returns to a GSS state with squeezing parameter equal to the squeezing parameter of the twisted GSS state. 

First, we consider the sequence based on repeatedly minimizing the value of $J_{\textrm{min}}^{2}$, or equivalently of $\xi^{2}$. If we take 
\begin{equation}
\label{eq:cas}
s^{2}_{j}=\frac{F}{N^{1-\frac{1}{3^{j-1}}}},
\end{equation}
then Eq.~\eqref{eq:j_min} becomes
\begin{equation}
J_{\textrm{min}}^{2}\approx\frac{(9F)^{1/3}N^{\frac{1}{3^{j}}}}{8}.
\end{equation}
Recall that the approximate spin moments imply that for the GSS states
\begin{equation}
\xi^{2}\approx s^{2}.
\end{equation}
This suggests that if we imagine that twist $j$ acts on a GSS state with $s_{j}^{2}$ then after allowing the new twist to proceed to its maximum extent, we should expect the next twist will act on a GSS state with
\begin{equation}
\label{eq:recurs}
s_{j+1}^{2}=\xi^{2}_{j}=\frac{N(J_{\textrm{min}}^{2})_{j}}{(\langle J_{x}\rangle^{2})_{j}}\approx\frac{1}{2}\left(\frac{9s_{j}^{2}}{N^{2}}\right)^{1/3},
\end{equation}
where we have used Eq.~\eqref{eq:j_min}. Suppose again that $s_{j}^{2}$ has the form in Eq.~\eqref{eq:cas}, then plugging this in gives
\begin{equation}
\label{eq:cas_A}
s_{j+1}^{2}\approx\frac{1}{2}\left(\frac{9F}{N^{3-\frac{1}{3^{j-1}}}}\right)^{1/3}=\frac{1}{2}\frac{(9F)^{1/3}}{N^{1-\frac{1}{3^{j}}}},
\end{equation}
which is of the same form as the initial value of $s^{2}$ but with a modified constant and $j+1$ replacing $j$. The coherent spin-state polarized along the \textit{x}-axis is well approximated by a GSS state with $s^{2}=1$. With this in mind, we assume that $s_{1}^{2}=1$ and write out the first several values of the  $s^{2}_{j}$ according to Eq.~\eqref{eq:recurs}: 
\begin{align}
s_{1}^{2}&\approx1 \\
s_{2}^{2}&\approx\frac{1}{2}\left(\frac{3}{N}\right)^{2/3} \\
s_{3}^{2}&\approx\frac{1}{2^{1+\frac{1}{3}}}\left(\frac{9^{1+\frac{1}{3}}}{N^{2+\frac{2}{3}}}\right)^{1/3}=\frac{9^{\frac{1}{3}+\frac{1}{9}}}{2^{1+\frac{1}{3}}}\frac{1}{N^{1-\frac{1}{9}}} \\
s_{4}^{2}&\approx\frac{9^{\frac{1}{3}+\frac{1}{9}+\frac{1}{27}}}{2^{1+\frac{1}{3}+\frac{1}{9}}}\frac{1}{N^{1-\frac{1}{27}}} \\
s_{5}^{2}&\approx\frac{9^{\frac{1}{3}+\frac{1}{9}+\frac{1}{27}+\frac{1}{81}}}{2^{1+\frac{1}{3}+\frac{1}{9}+\frac{1}{27}}}\frac{1}{N^{1-\frac{1}{81}}}.
\end{align}
The pattern is
\begin{equation}
s_{j}^{2}\approx\frac{3^{1-\frac{1}{3^{j-1}}}}{2^{\frac{3-\frac{1}{3^{j-2}}}{2}}}\frac{1}{N^{1-\frac{1}{3^{j-1}}}}=\left(\frac{3}{2^{3/2}N}\right)^{1-\frac{1}{3^{j-1}}}.
\end{equation}
Plugging this into the expression for the optimal value of the twisting angle gives
\begin{equation}
\chi_{j*}\approx\frac{2^{1-\frac{1}{3^{j-1}}}}{3^{\frac{1}{2}-\frac{2}{3^{j}}}N^{\frac{2}{3^{j}}}}.
\end{equation}
The N-dependences here match those required by the phase estimation algorithm.

Now we can consider a multi-twist protocol based on repeatedly minimizing $\Xi^{2}_{*}$. If we begin with a GSS state that satisfies Eq.~\eqref{eq:cas}, then Eq.~\eqref{eq:xi_rot_final} becomes
\begin{equation}
\overline{\Xi^{2}_{*}}\approx\frac{\left(9F+27N^{2-3^{-(j-1)}}\sigma^{2}\right)^{1/3}}{2N^{1-3^{-j}}}+\frac{N^{1-2\times3^{-(j-1)}}\sigma^{2}}{2F^{2}}+\frac{N^{1-4*3^{-j}}\sigma^{2}}{(\frac{1}{3}F+N^{2-3^{-(j-1)}}\sigma^{2})^{1/3}F}.
\end{equation}
Which term dominates depends on the $N$-dependence of $\sigma^{2}.$ Suppose that
\begin{equation}
\label{eq:zeta}
\sigma^{2}=\frac{G}{N^{\zeta}}.
\end{equation}
Then whenever $3^{-(j-1)}>2-\zeta,$ this reduces to a form similar to that of the unrotated squeezing parameter
\begin{equation}
\overline{\Xi^{2}_{*}}\approx\frac{(9F)^{1/3}}{2N^{1-3^{-j}}}.
\end{equation}
However, if this is not the case, then the other two terms become relevant and $\overline{\Xi^{2}_{*}}$ will not decrease as quickly with system size. We begin with a GSS state with $s_{j}^{2}$ given by Eq.~\eqref{eq:cas} and apply a twist given by Eq.~\eqref{eq:chi_rot_opt}, we will treat the resulting state as a GSS state with
\begin{equation}
\begin{aligned}
\label{eq:s_rot}
s^{2}_{j+1}&\approx s^{2}_{j}\left[\frac{(\frac{1}{3}s^{2}_{j}+N\sigma^{2})^{1/3}}{s_{j}^{2}N^{2/3}}+\frac{1}{6N^{2/3}(\frac{1}{3}s^{2}_{j}+N\sigma^{2})^{2/3}}\right] \\
&=\frac{F}{N^{1-3^{-(j-1)}}}\left[\frac{N^{1-3^{-(j-1)}}\left(\frac{F}{3N^{1-3^{-(j-1)}}}+GN^{1-\zeta}\right)^{1/3}}{FN^{2/3}}+\frac{1}{6N^{2/3}\left(\frac{F}{3N^{1-3^{-(j-1)}}}+GN^{1-\zeta}\right)^{2/3}}\right],
\end{aligned}
\end{equation}
which reduces to Eq.~\eqref{eq:cas_A} if the second term in the parentheses can be neglected. This expression was obtained by plugging Eq.~\eqref{eq:chi_rot_opt} into Eq.~\eqref{eq:expand}.

It is illustrative to work through what happens for two initial values of $\zeta.$ First, consider the case where $\zeta=1$ and $G=1$ which is characteristic of the residual phase uncertainty if a spin-coherent state is used to obtain a preliminary estimate of the phase value. If we begin with a spin-coherent state, i.e. $s_{1}^{2}=1$. Then Eq.~\eqref{eq:rot_xi_appr} becomes
\begin{equation}
\label{eq:xi_rot_cas}
\overline{\Xi^{2}}_{1}\approx1+\frac{1}{2}\left(1+e^{-\frac{2}{N}}-2e^{-\frac{1}{N}}\right).
\end{equation}
After the first twist Eq.~\eqref{eq:xi_rot_final} and Eq.~\eqref{eq:s_rot} become
\begin{align}
\overline{\Xi^{2}}_{2}&\approx\frac{\left(36\right)^{1/3}}{2N^{2/3}} \\
s^{2}_{2}&\approx\frac{\frac{3}{2}}{\left(\frac{4}{3}\right)^{2/3}N^{2/3}},
\end{align}
which matches the scaling after applying a single twist to minimize $\langle J(\theta_{*})^{2}\rangle.$ If we apply an additional twist, we obtain
\begin{align}
    \overline{\Xi^{2}}_{3}&\approx\frac{27+4\left(\frac{4}{3}\right)^{4/3}+12\left(\frac{4}{3}\right)^{2/3}}{18N^{2/3}} \\
    s^{2}_{3}&\approx\frac{1}{N^{2/3}}.
\end{align}
So that the $N$-dependence no longer improves and will not improve with additional twists. Note that if an ensemble is prepared under these circumstances we expect that after using it to obtain a preliminary estimate, the residual phase uncertainty will depend on $N$ as $\sim N^{-2/3}.$

Now consider the case where $\zeta=2-3^{-k}$, with $k$ an integer greater than zero, and we again begin from a spin-coherent state. This time Eq.~\eqref{eq:xi_rot_final} and Eq.~\eqref{eq:s_rot} become
\begin{equation}
\overline{\Xi^{2}}_{2}\approx\frac{9^{1/3}}{2N^{2/3}}\approx s^{2}_{2}.
\end{equation}
Which is the same value obtained when we minimizing $\langle J(\theta_{*})\rangle$. As more twists are applied, this pattern will continue until $k$ twists have been applied. At which point, the next twist will give
\begin{align}
\overline{\Xi^{2}}_{k+2}&\approx\frac{\left[9\left(\frac{3}{2^{3/2}}\right)^{1-3^{-k}}+27G\right]^{1/3}}{2N^{1-3^{-(k+1)}}} \\
    s^{2}_{k+2}&\approx\frac{\left(\frac{1}{3}\left(\frac{3}{2^{3/2}}\right)^{1-3^{-k}}+G\right)^{1/3}}{N^{1-3^{-(k+1)}}}+\frac{\left(\frac{3}{2^{3/2}}\right)^{1-3^{-k}}}{6\left(\frac{1}{3}\left(\frac{3}{2^{3/2}}\right)^{1-3^{-k}}+G\right)^{2/3}N^{1-3^{-(k+1)}}}\equiv\frac{F'}{N^{1-3^{-(k+1)}}}.
\end{align}
This is the optimal $N$-dependence that is expected given this residual phase uncertainty $\sigma^{2}$. If an additional twist is applied the result will be 
\begin{align}
\overline{\Xi^{2}}_{k+3}&\approx\frac{\frac{G}{F'^{2}}+3G^{1/3}+2\frac{G^{2/3}}{F'}}{2N^{1-3^{-(k+1)}}} \\
s^{2}_{k+3}&\approx\frac{G^{1/3}}{N^{1-3^{-(k+1)}}},
\end{align}
so that the $N$-dependence no longer improves and this is the maximum number of twists that should usually be considered given this residual phase uncertainty.

\section{Additional numerical results on the non-Gaussianity of twisted-Gaussian states}
\label{app:ng}
In this appendix, we collect some additional results relating to the non-Gaussian features of the probe states considered in this paper. In Fig.~\ref{fig:kurt2}, we show plots similar to those in Fig.~\ref{fig:kurt} but for a two-twist state preparation. The magnitudes of both twists are determined by solving Eq.~\eqref{eq:state1} and then multiplying the result by a $\mathcal{C}=0.7$. In Fig.~\ref{fig:kurt2}(a) we show the kurtoses of $J_{z}$ and $J_{y}$ fit to a sigmoid-exponential function, see Eq.~\eqref{eq:sig_exp}, for $N=750$ and once again the fits are of a high quality. In Fig.~\ref{fig:kurt2}(b), we show the same kurtoses but now for several different system sizes. The turning on behavior again occurs at similar system sizes suggesting a maximum usable value of $\mathcal{C}$. We now also observe that for large values of $\mathcal{C}$ larger values of $\textrm{Kurt}[J_{y}]$ are achieved at larger system sizes but there do appear to be diminishing returns with increasing system size suggesting that this may saturate.

\begin{figure*}
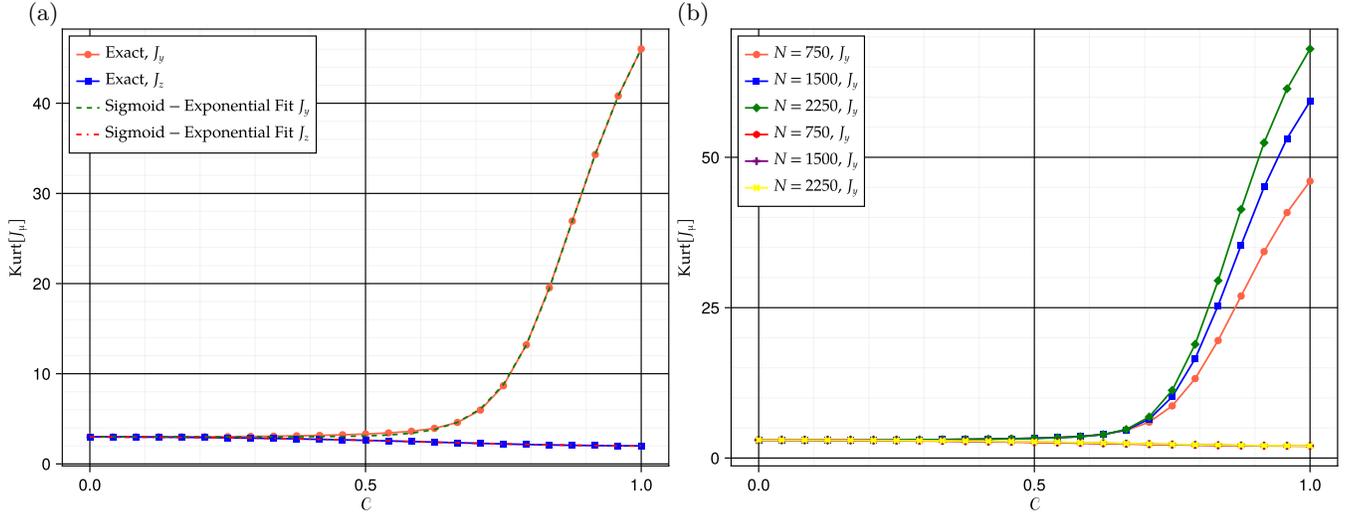

(a) \hspace{8cm} (b) \hspace{8cm} \ \\
\includesvg[width=0.49\linewidth]{kurtfit3_2.svg}
\includesvg[width=0.49\linewidth]{kurtN2_2.svg}
\caption{(a) The kurtoses of $J_{z}$ (orange circles) and $J_{y}$ (blue squares) vs. $\mathcal{C}$ for $N=750$ fit to sigmoid-exponential functions. The fit to $\textrm{Kurt}[J_{y}]$ is the green dashed line and the fit to $\textrm{Kurt}[J_{z}]$ is the red dot-dashed line. (b) Plots of the same kurtoses vs. $\mathcal{C}$ for $N=750$ (orange circles for $\textrm{Kurt}[J_{y}]$ and red hexagons for $\textrm{Kurt}[J_{z}]$), $1500$  (blue squares for $\textrm{Kurt}[J_{y}]$ and purple crosses for $\textrm{Kurt}[J_{z}]$), and $2250$ (green diamonds for $\textrm{Kurt}[J_{y}]$ and yellow x-crosses for $\textrm{Kurt}[J_{z}]$).} 
\label{fig:kurt2}
\end{figure*}

We find that solving the Eq.~\eqref{eq:state3} to fix the final twist actually results in an effective value of $\mathcal{C}$ in the range we set the other $\mathcal{C}$ values to by hand. In Fig.~\ref{fig:Ceff}(a), we plot the effective values of $\mathcal{C}$ obtained for the final twist in this way for the second and third ensembles in a three ensemble setting. In particular, we are plotting the ratio of the solution to Eq.~\eqref{eq:state3} to the solution to Eq.~\eqref{eq:state1}. We find that in both cases the value is less than $0.7$. Additionally, in both cases the values are approximately independent of $N$ with the value for the second ensemble being $\approx0.658$ and the value for the third ensemble being $\approx0.490$. Similar results for the four ensemble setting are shown in Fig.~\ref{fig:Ceff}(b). These effective values of $\mathcal{C}$ suggest the optimal states in this setting are close to being Gaussian. 

\begin{figure*}
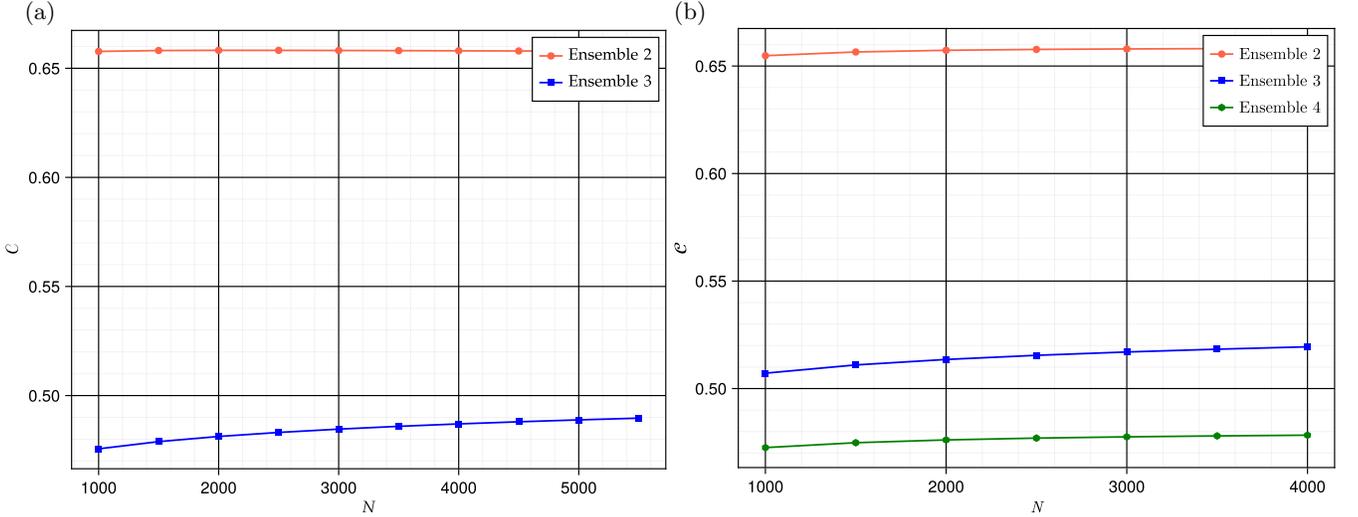

(a) \hspace{8cm} (b) \hspace{8cm} \ \\
\includesvg[width=0.49\linewidth]{Ceff.svg}
\includesvg[width=0.49\linewidth]{Ceff_lvl4.svg}
\caption{(a) Here we plot the effective value of $\mathcal{C}$ for the second and third ensembles of the three ensemble phase estimation protocol vs. system size $N$. The orange circles are for ensemble 2 while the blue squares are for ensemble 3. (b) This is a similar plot for the four ensemble case. The orange circles and blue squares are again for the second and third ensemble respectively while the green hexagons are for the fourth ensemble.} 
\label{fig:Ceff}
\end{figure*}

\section{Details of numerical techniques}
\label{app:numer}

\subsection{Single ensemble simulations}

All of the numerical results in this paper were obtained using code written in the julia programming language. As mentioned in Sec.~\ref{sec:back}, all of the dynamics considered here are constrained to the $J=\frac{N}{2}$ SU(2)-irrep of $N$ spin-1/2 particles. The dimension of this subspace is $N+1$, much less than the $2^{N}$ dimension of the entire subspace allowing for the direct simulation of large ensembles. The numerical results involving only a single ensemble (Fig~\ref{fig:kurt}, Fig.~\ref{fig:xi_collect}, and Fig.~\ref{fig:xi_tild}) were obtained by direct simulation of the state vector restricted to a the symmetric subspace. To evaluate Eq.~\eqref{eq:xi_bar} and produce Fig.~\ref{fig:Nfluct}, we draw 200 samples from $P(N_{s})$ and then directly evaluate $\xi^{2}$ for each sampled system size using the circuit parameters computed for the targeted system size $N$.

\subsection{Multi-ensemble simulations}

The situation becomes slightly more complicated when the there are multiple ensembles under consideration. In this case, direct simulation of the entire state vector is no longer possible for large system sizes. To see why suppose that $|\psi_{1}\rangle$, $|\psi_{2}\rangle$, and $|\psi_{3}\rangle$ are states of $N_{1}$, $N_{2}$, and $N_{3}$ particles respectively. Suppose that, in addition, each of these states is restricted in the symmetric subspace of its state space. The subspace that $|\psi_{1}\rangle\otimes|\psi_{2}\rangle\otimes|\psi_{3}\rangle$ is allowed to occupy has dimension $(N_{1}+1)\times (N_{2}+1)\times (N_{3}+1)$ which rapidly becomes large as the number of ensembles is increased. Luckily, the protocols we consider do not lead to entanglement between the separate ensembles so we can store a separate state vector for each ensemble. Usually in multi-ensemble simulations we are interested in evaluating the averaged estimation error $\Delta\phi^{2}$ associated with a particular protocol.
As mentioned in Sec.~\ref{sec:phase_est}, this quantity can be written as
\begin{equation}
\Delta\phi^{2}=\int d\phi\sum_{\hat{\phi}_{1}}\sum_{\hat\phi_{2}}\dots\sum_{\hat{\phi}_{M}}p(\phi)p(\hat{\phi}_{1}|\phi)p(\hat{\phi}_{2}|\phi,\hat{\phi}_{1})\cdots p(\hat{\phi}_{M}|\phi,\hat{\phi}_{1},\hat{\phi}_{2}\cdots\hat{\phi}_{M-1})(\phi-\hat{\phi})^{2}.
\end{equation}
Our strategy for evaluating this quantity is to first define an effective error operator on the $M$th ensemble by
\begin{equation}
\begin{split}
\label{eq:counter_rot}
W_{M}\equiv\int d\phi\sum_{\hat{\phi}_{1}}\sum_{\hat\phi_{2}}\dots\sum_{\hat{\phi}_{M-1}}&p(\phi)p(\hat{\phi}_{1}|\phi)p(\hat{\phi}_{2}|\phi,\hat{\phi}_{1})\cdots p(\hat{\phi}_{M-1}|\phi,\hat{\phi}_{1},\hat{\phi}_{2}\cdots\hat{\phi}_{M-2}) \\
&\times e^{i\left(\phi-\sum_{j=1}^{M-1}\hat{\phi}_{j}\right)J_{z}^{(M)}}\left(\phi-\sum_{j=1}^{M-1}\hat{\phi}_{j}-\frac{2}{N_{m}}J_{y}^{(M)}\right)^{2} e^{-i\left(\phi-\sum_{j=1}^{M-1}\hat{\phi}_{j}\right)J_{z}^{(M)}},
\end{split}
\end{equation}
where $J_{\mu}^{(M)}$ denotes the $\mu$th angular momentum component on the $M$th ensemble. If the parentheses are expanded this becomes
\begin{equation}
\begin{split}
W_{M}\equiv\int d\phi\sum_{\hat{\phi}_{1}}\sum_{\hat\phi_{2}}\dots\sum_{\hat{\phi}_{M-1}}&p(\phi)p(\hat{\phi}_{1}|\phi)p(\hat{\phi}_{2}|\phi,\hat{\phi}_{1})\cdots p(\hat{\phi}_{M-1}|\phi,\hat{\phi}_{1},\hat{\phi}_{2}\cdots\hat{\phi}_{M-2}) \\
&\times\Bigg\{\left(\phi-\sum_{j=1}^{M-1}\hat{\phi}_{j}\right)^{2} \\
&-\frac{4}{N_{m}}\left(\phi-\sum_{j=1}^{M-1}\hat{\phi}_{j}\right)\left[\cos\left(\phi-\sum_{j=1}^{M-1}\hat{\phi}_{j}\right)J_{y}^{(M)}-\sin\left(\phi-\sum_{j=1}^{M-1}\hat{\phi}_{j}\right)J_{x}^{(M)}\right] \\
&+\frac{4}{N_{M}^{2}}\left[\cos\left(\phi-\sum_{j=1}^{M-1}\hat{\phi}_{j}\right)J_{y}^{(M)}-\sin\left(\phi-\sum_{j=1}^{M-1}\hat{\phi}_{j}\right)J_{x}^{(M)}\right]^{2}\Bigg\}
\end{split}
\end{equation}
At this point
\begin{equation}
\Delta\phi^{2}=\langle\psi_{M}|W_{M}|\psi_{M}\rangle,
\end{equation}
where $|\psi_{M}\rangle$ is the probe state that the $M$th ensemble is prepared in. The advantage of this approach is that once $W_{M}$ has been constructed the circuit parameters, i.e. the twisting and rotation angles, used for the preparation of the $M$th probe state can be optimized by minimizing a simple expectation value. This is done for two of the curves in Fig.~\ref{fig:nu_collect}b: the one labeled ``One Ensemble" and the one labeled ``Three Ensembles (One Twist)" as these scenarios fall outside of the main protocol we are proposing and so we rely on numerical optimization in these cases. We construct this operator even in the cases when we use the protocols we explicitly propose in the main text and no additional optimization is require as this remains a convenient way to evaluate $\Delta\phi^{2}$.

The integral over $\phi$ is evaluated via Hermite-Gauss quadrature. The number of particles in the first ensemble is typically not too large so we exactly evaluate $p(\hat{\phi}_{1}|\phi)$ for all $\hat{\phi}$ and loop over all terms in the sum. We typically do this for all ensembles but there are two exceptions. First, for the Fig.~\ref{fig:nu_collect}b curve labeled ``Three Ensembles (One Twist)" the system sizes at which convergence occurs are so large that this is not practical. Second, for simulations of four ensembles direct evaluation of all sums becomes too costly due to the exponential scaling of the simulation cost with the number of ensembles. In these cases, we utilize Monte-Carlo summation for the sum over $\hat{\phi}_{k}$ when $k\geq2$. We approximately draw $L$ samples from each $p(\hat{\phi}_{k}|\phi,\hat{\phi}_{1},\cdots,\hat{\phi}_{k-1})$. The naive way to do this would be to first draw a number $r$ from the uniform distribution on the interval $(0,1)$. Then the sample is $2m/N_{k}$, where $m$ is the smallest value that satisfies
\begin{equation}
\sum_{j=-N_{2}/2}^{m}p\left(\frac{2j}{N_{2}}\bigg|\phi,\hat{\phi}_{1},\cdots,\hat{\phi}_{k-1}\right)>r.
\end{equation}
This is the method we use for the typical four ensemble simulations. However, due to the large system sizes used to produce the ``Three Ensembles (One Twist)" curve, we are likely to need to check many $m$ before one is selected meaning that this approach is still not practical. Accordingly, we approximate the outcome distribution for the second ensemble in this case by a Gaussian with mean given by 
\begin{equation}
\langle\psi_{2}|e^{i\left(\phi-\hat{\phi}_{1}\right)J_{z}^{(M)}}J_{y}^{(M)}e^{-i\left(\phi-\hat{\phi}_{1}\right)J_{z}^{(M)}}|\psi_{2}\rangle
\end{equation}
and variance given by
\begin{equation}
\langle\psi_{2}|e^{i\left(\phi-\hat{\phi}_{1}\right)J_{z}^{(M)}}\left(J_{y}^{(M)}\right)^{2}e^{-i\left(\phi-\hat{\phi}_{1}\right)J_{z}^{(M)}}|\psi_{2}\rangle-\langle\psi_{2}|e^{i\left(\phi-\hat{\phi}_{1}\right)J_{z}^{(M)}}J_{y}^{(M)}e^{-i\left(\phi-\hat{\phi}_{j}\right)J_{z}^{(M)}}|\psi_{2}\rangle^{2}.
\end{equation}
In other words, we approximate the outcome distribution by a Gaussian that has the same mean and variance and the exact distribution. We numerically verified that this gives good agreement with sampling from the exact distribution at small system sizes. This can be understood as a result of something like a $\mathcal{C}$ parameter keeping the state close to a GSS state. We use $L=10$ to produce the ``Three Ensembles (One Twist)" curve and the four ensemble simulations.

In the two ensemble simulations, we utilize $\lfloor N/5\rfloor$ particles in the first ensemble. In the three ensemble simulations, we utilize $N_{1}=\lfloor N/20\rfloor$ and $N_{2}=4*N_{1}$. Finally, in the four ensemble simulations we use $N_{1}=\lfloor N/50\rfloor$ particles in the first ensemble, $N_{2}=4*N_{1}$ particles in the second ensemble, and $N_{3}=3*N_{2}$ particles in the third ensemble. In all cases, the remaining $N-\sum_{k=1}^{M-1}N_{k}$ particles make up the final ensemble. This pattern roughly follows the pattern used by Pezz\`e and Smerzi~\cite{Pezze2020,Pezze2021} and seems to work well for the states under consideration here.

\subsection{Feedback control simulations}

For simulations that include feedback control fluctuations, we evaluate
\begin{equation}
\begin{split}
\overline{\Delta\phi^{2}}\equiv\int dr_{1}\cdots\int dr_{M-1}\int d\phi\sum_{\hat{\phi}_{1}}\sum_{\hat\phi_{2}}\dots\sum_{\hat{\phi}_{M}}&p(\phi)p(\hat{\phi}_{1}|\phi)P(r_{1})p(\hat{\phi}_{2}|\phi,\hat{\phi}_{1}+r_{1})P(r_{2}) \\
&\times\cdots p(\hat{\phi}_{M}|\phi,\hat{\phi}_{1}+r_{1},\hat{\phi}_{2}+r_{2}\cdots\hat{\phi}_{M-1}+r_{M-1})(\phi-\hat{\phi})^{2}.
\end{split}
\end{equation}
There are multiple reasonable ways of generating an estimator for this quantity. The first, and most straight forward, is to produce $L_{r}$ samples $r_{j}^{(k)}$ for each $r_{j}$. Then the estimator becomes
\begin{equation}
\begin{split}
\overline{\Delta\phi^{2}}_{\textrm{est},1}=\frac{1}{L_{r}^{M-1}}\sum_{k_{1}=1}^{L}\sum_{k_{2}=1}^{L}\cdots\sum_{k_{M-1}=1}^{L}\int d\phi&\sum_{\hat{\phi}_{1}}\sum_{\hat\phi_{2}}\dots\sum_{\hat{\phi}_{M}}p(\phi)p(\hat{\phi}_{1}|\phi)p(\hat{\phi}_{2}|\phi,\hat{\phi}_{1}+r_{1}^{(k_{1})}) \\
&\times\cdots p(\hat{\phi}_{M}|\phi,\hat{\phi}_{1}+r_{1}^{(k_{1})},\hat{\phi}_{2}+r_{2}^{(k_{2})}\cdots\hat{\phi}_{M-1}+r_{M-1}^{(k_{M-1})})(\phi-\hat{\phi})^{2}.
\end{split}
\end{equation}
A closely related approach is to sample the vector $(r_{1},r_{2},\dots,r_{M-1})$ $L_{r}$ times. The associated estimator is
\begin{equation}
\begin{split}
\overline{\Delta\phi^{2}}_{\textrm{est},2}=\frac{1}{L_{r}}\sum_{k=1}^{L_{r}}\int d\phi&\sum_{\hat{\phi}_{1}}\sum_{\hat\phi_{2}}\dots\sum_{\hat{\phi}_{M}}p(\phi)p(\hat{\phi}_{1}|\phi)p(\hat{\phi}_{2}|\phi,\hat{\phi}_{1}+r_{1}^{(k)}) \\
&\times\cdots p(\hat{\phi}_{M}|\phi,\hat{\phi}_{1}+r_{1}^{(k)},\hat{\phi}_{2}+r_{2}^{(k)}\cdots\hat{\phi}_{M-1}+r_{M-1}^{(k)})(\phi-\hat{\phi})^{2}.
\end{split}
\end{equation}
A third option is to draw a different sample for each term in the sums over $\phi$ and the $\hat{\phi}_{j}$ so that the estimator becomes
\begin{equation}
\begin{split}
\overline{\Delta\phi^{2}}_{\textrm{est},3}=\frac{1}{L_{r}^{M-1}}\int d\phi&\sum_{\hat{\phi}_{1}}\sum_{\hat\phi_{2}}\dots\sum_{\hat{\phi}_{M}}\sum_{k_{1}=1}^{L_{r}}\sum_{k_{2}=1}^{L_{r}}\cdots\sum_{k_{M-1}=1}^{L_{r}}p(\phi)p(\hat{\phi}_{1}|\phi)p(\hat{\phi}_{2}|\phi,\hat{\phi}_{1}+r_{1}^{(k_{1})}) \\
&\times\cdots p(\hat{\phi}_{M}|\phi,\hat{\phi}_{1}+r_{1}^{(k_{1}),\hat{\phi}_{2}+r_{2}^{(k_{2})}}\cdots\hat{\phi}_{M-1}+r_{M-1}^{(k_{M-1})})(\phi-\hat{\phi})^{2},
\end{split}
\end{equation}
where $\bar{\phi}_{j}$ in a vector given by $(\phi,\hat{\phi}_{1},\hat{\phi}_{2}\cdots,\hat{\phi}_{j-1})$. These are all unbiased estimators for $\overline{\Delta\phi^{2}}$. However, they do not have the same variances.  In this work we opt for a hybrid approach since we are apriori not certain which estimator will give the lowest variance. In particular, we use
\begin{equation}
\begin{split}
\overline{\Delta\phi^{2}}_{\textrm{est},4}=\frac{1}{L_{O}L_{I}^{M-1}}\sum_{i=j}^{L_{O}}\int d\phi&\sum_{\hat{\phi}_{1}}\sum_{\hat\phi_{2}}\dots\sum_{\hat{\phi}_{M}}\sum_{k_{1,j}=1}^{L_{I}}\sum_{k_{2,j}=1}^{L_{I}}\cdots\sum_{k_{M-1,j}=1}^{L_{I}}p(\phi)p(\hat{\phi}_{1}|\phi)p(\hat{\phi}_{2}|\phi,\hat{\phi}_{1}+r_{1}^{(k_{1,j})}) \\
&\times\cdots p(\hat{\phi}_{M}|\phi,\hat{\phi}_{1}+r_{1}^{(k_{1,j})},\hat{\phi}_{2}+r_{2}^{(k_{2,j})}\cdots\hat{\phi}_{M-1}+r_{M-1}^{(k_{M-1,j})})(\phi-\hat{\phi})^{2},
\end{split}
\end{equation}
with $L_{I}=1$ and $L_{O}=10$.

\subsection{Packages used in this work and some other details}

Optimizations of the circuit parameters used in the ``One Ensemble" and ``Three Ensembles (One Twist)" curves were performed using the implementation of Nelder-Mead in NLopt.jl with stopping criteria of $10^{-15}$ change in the absolute value of $\Delta\phi^{2}$. The zeros and weights used for the Hermite-Gauss quadrature integration were obtained using the FastGaussQuadrature.jl. Sampling and manipulation of standard probability distributions uses Distributions.jl. Matrix manipulations for state vector simulations uses LinearAlgebra.jl. Solutions to Eq.~\eqref{eq:state1} and Eq.~\eqref{eq:state3} are obtained using Roots.jl. Fits to Eq.~\eqref{eq:sig_exp} are obtained with LsqFit.jl. The $Q$ distributions plotted in Fig.~\ref{fig:scheme} were obtained using QuantumOptics.jl. Fits to power law decays are obtained by taking the log of both the x and y variables and then performing a linear least-squares fit.

\end{widetext}

\end{document}